\documentclass[useAMS,usenatbib,usedcolumn,usegraphicx]{mn2e}

\newcommand{\hii}{H\,{\sc ii}}
\newcommand{\hi}{H\,{\sc i}}
\newcommand{\ha}{H{$\alpha$}}
\newcommand{\h}{$^{h}$}
\newcommand{\m}{$^{m}$}
\newcommand{\nodata}{...}
\newcommand{\lum}{erg~s$^{-1}$}
\newcommand{\flux}{erg~s$^{-1}$~cm$^{-2}$}
\newcommand{\cden}{cm$^{-2}$}

\newcommand{\cntr}{cnts~s$^{-1}$}
\newcommand{\msun}{M$_{\sun}$}

\newcommand\phn{\phantom{x}}
\newcommand\phs{\phantom{-}}

\newcommand\aj{AJ}
\newcommand\apj{ApJ}%
\newcommand\apjl{ApJ}%
\newcommand\apjs{ApJS}%
\newcommand\aap{A\&A}%
\newcommand\aaps{A\&AS}%
\newcommand\mnras{MNRAS}%
\newcommand\gca{Geochim.~Cosmochim.~Acta}%
\newcommand\apss{Ap\&SS}%
\newcommand\pasp{PASP}%
\title[{\it Chandra} survey of dwarf starburst galaxies I]
{A {\it Chandra} X-ray survey of nearby dwarf starburst galaxies:
I. Data reduction and results}

\author[J{\"u}rgen Ott, Fabian Walter, and Elias Brinks]{J{\"u}rgen Ott$^{1}$\thanks{Bolton
Fellow, E-mail: Juergen.Ott@csiro.au}, Fabian Walter$^{2}$\thanks{E-mail: walter@mpia.de}, and Elias Brinks\thanks{E-mail:
ebrinks@star.herts.ac.uk}$^{3,4}$\\ 
$^{1}$CSIRO Australia Telescope National Facility, Cnr Vimiera \&
Pembroke Roads, Marsfield NSW 2112, Australia\\ 
$^{2}$Max--Planck--Institut f{\"u}r Astronomie, K{\"o}nigstuhl 17, 69117 Heidelberg, Germany\\
$^{3}$Instituto Nacional de Astrof{\'{\i}}sica {\'O}ptica
y Electr{\'o}nica, Apartado Postal 51 y 216, Puebla, Pue~72000,
Mexico \\
$^{4}$Centre for Astrophysics Research, University of Hertfordshire, College Lane, Hatfield AL10 9AB, England}

\begin{document}


\pagerange{\pageref{firstpage}--\pageref{lastpage}} \pubyear{2005}

\maketitle

\label{firstpage}

\begin{abstract}

We present an analysis of {\it Chandra} X-ray observations of a sample
of eight dwarf starburst galaxies (I\,Zw\,18, VII\,Zw\,403, NGC\,1569,
NGC\,3077, NGC\,4214, NGC\,4449, NGC\,5253, and He\,2--10). Extended,
diffuse X-ray emission is detected in all but two of the
objects. Unresolved sources were found within all dwarf galaxies
(total: 55 sources). These point sources are well fit by power law,
thermal plasma, or black body models. Ten of the point sources exceed
an X-ray luminosity of $10^{39}$\,\lum\ (ultraluminous X-ray
sources). In those galaxies where diffuse X-ray emission is detected,
this emission (with X-ray luminosities ranging from
$4\times10^{38}$\,\lum\ to $2\times10^{40}$\,\lum) contains most
(60-80 per cent) of the X-ray photons. This diffuse emission can be
well fit by MeKaL one-temperature thermal plasma models once the
contribution from the unresolved point sources is subtracted
properly. The diffuse X-ray component is significantly extended,
reaching as far as $0.5-5$\,kpc into the outskirts of their
hosts. Azimuthally averaged X-ray surface brightness profiles are well
approximated by exponential functions. Temperatures of various regions
within the galaxies range from $1.6-7.6\times10^{6}$\,K. With few
exceptions, temperatures of the hot gas are remarkably uniform,
hovering around $2-3\times10^{6}$\,K. Temperatures of the coronal gas
in the outer regions are in general $\sim 2-3$ times lower than those
found in the central regions. Fits to the diffuse emission do not
allow strong constraints to be put on the metallicities of the
emitting plasmas. However, the derived metallicities are compatible
with those determined from their \hii\ regions. An $\alpha/Fe$ ratio
of $\sim 2$ is indicated for the hot gas within at least three objects
(NGC\,1569, NGC\,4449, and He\,2--10). Shadowing of the diffuse X-ray
emission by the cooler disc gas is used to constrain the orientation
of the galaxies.

\end{abstract}

\begin{keywords}

ISM: jets and outflows -- 
galaxies: dwarf -- 
galaxies: individual (I Zw 18, VII Zw 403, NGC 1569, NGC 3077, NGC 4214,
                      NGC 4449, NGC 5253, He 2-10) --  
galaxies: starburst -- 
X-rays: ISM -- 
X-rays: galaxies 

\end{keywords}

\section{Introduction}

Feedback from massive stars is a fundamental process regulating the
energetics and dynamics of the interstellar medium \citep[ISM; e.g.,
][]{mckee77,lei92,kor99,gaz99,wad00}. An extreme form of feedback can
be observed in starburst galaxies which exhibit star formation (SF) at
a level that would consume the total fuel supply of their hosts within
a time frame considerably shorter than a Hubble time. A starburst
typically involves several thousand stars more massive than $\sim
8$\,M$_\odot$ which eventually explode as type II supernovae (SNe),
each releasing a mechanical energy of $\sim 10^{51}$\,erg in to the
ISM. Most of this energy is thermalised, and indirectly can be traced
by X-ray emission emerging from hot, coronal gas
\citep{wea77,str02,hec02}. Theoretically, the influence of a starburst 
on the dynamics of the ISM is expected to be a function of the total
mass of the host galaxy. The most dramatic consequences are expected
to be visible in dwarf starburst galaxies which may lose part or even
all of their metal-enriched gas to the intergalactic medium
\citep[see, e.g.,][]{mac99,fer00,sil01a,sil01b}.

In order to understand the impact of a starburst on the ISM of its
host in more detail, we analyse archival data of a sample of eight
nearby starburst dwarf galaxies obtained with the {\it Chandra} X-ray
observatory. Our study is split into two papers. The description of
the data, the data reduction and the results of the X-ray analysis are
described in Paper\,I (this paper). Paper\,II \citep{ott05} compares
the X-ray data, both unresolved and extended emission, with
observations at other wavelengths and provides a discussion on the
state of the ISM, correlations with star formation tracers, and the
development of superwinds which can lead to outflows.

Some of the galaxies analysed here have recently been published as
individual studies (NGC\,3077: \citealt{ott03}; NGC\,1569:
\citealt{mar02}; NGC\,4449: \citealt{sum03}; NGC\,4214
\citealt{har04}; I\,Zw\,18: \citealt{thu04}). 
Our new analysis, however, is unique in the sense that it provides
{\em uniformly reduced datasets} which allow a direct comparison of
the properties of different dwarf galaxies. This is particularly
important as results from the analysis of X-ray data critically depend
on the binning of the data, the choice of models used for the spectral
fits, and in a subtle way on a host of additional variables, such as
the proper subtraction of unresolved sources and the way absorption
within the object as well as foreground absorption by the Milky Way is
dealt with. The aim of our papers is to provide the best analysis
possible given our current knowledge and techniques and, in case some
systematic effects remain, be able to at least compare results among
objects which have been reduced in the same manner. 

In Sect.\,\ref{sec:dataprep} we describe the sample selection criteria
followed by a description of the data reduction techniques used for
our analysis. The results for the individual galaxies are presented in
Sections\,\ref{sec:diffuse_no} (galaxies without a detected
diffuse X-ray component) and \ref{sec:diffuse_yes} (galaxies with
extended X-ray emission). A summary is provided in
Sect.\,\ref{sec:summary}.

\begin{table*}
\begin{minipage}{170mm}
\centering
\caption{Parameters of the {\it Chandra} observations. Coordinates are given in J2000.}
\begin{tabular}{@{}lcrrrlrrl@{}}
\hline

\multicolumn{1}{l}{Galaxy} & \multicolumn{1}{c}{Right Ascension
} & \multicolumn{1}{c}{Declination
} & \multicolumn{1}{c}{Obs ID.} & \multicolumn{1}{c}{Seq. No.} & \multicolumn{1}{c}{Date} & \multicolumn{1}{c}{$t_{\rm exp}$} & \multicolumn{1}{c}{$T_{\rm fp}$} & \multicolumn{1}{c}{ASCDS Ver.}\\

\multicolumn{1}{l}{} & \multicolumn{1}{c}{$^{h}$~~~$^{m}$~~~$^{s}$~~~~} & \multicolumn{1}{c}{~\degr~~~\arcmin~~~\arcsec } & \multicolumn{1}{c}{} & \multicolumn{1}{c}{} & \multicolumn{1}{c}{} &\multicolumn{1}{c}{[ks]} & \multicolumn{1}{c}{[\degr C]} & \multicolumn{1}{c}{}\\
\multicolumn{1}{l}{(1)}   &\multicolumn{1}{c}{(2)} &\multicolumn{1}{c}{(3)}   &\multicolumn{1}{r}{(4)}  &\multicolumn{1}{c}{(5)}  &\multicolumn{1}{c}{(6)}   &\multicolumn{1}{c}{(7)}   &\multicolumn{1}{c}{(8)} &\multicolumn{1}{c}{(9)} \\

\hline
I\,Zw\,18       & 09 34 02.00    & 55 14 28.0    & 805  &  600108  &2000 Feb 8  & 41.3 & $-120$  &  R4CU5UPD11.2 \\
VII\,Zw\,403    & 11 28 01.30    & 78 59 34.6    & 871  &  700176  &2000 Jan 7  & 10.6 & $-110$  &  R4CU5UPD11.1 \\
NGC\,1569       & 04 30 49.00    & 64 50 54.0    & 782  &  600085  &2000 Apr 11     & 97.1 & $-120$  & R4CU5UPD13.1  \\
NGC\,3077       & 10 03 20.66    & 68 44 03.5    & 2076     &  600210  &2001 Mar 7  & 54.1 & $-120$  & R4CU5UPD14.1  \\
NGC\,4214   &  12 15 38.70      &  36 19 41.9       & 2030     &  600164  &2001 Oct 16     & 26.8 & $-120$  & DS6.3.1 \\ 
NGC\,4449       & 12 28 12.00    & 44 05 41.0    & 2031     &  600165  &2001 Feb 4  & 26.9 & $-120$  & R4CU5UPD14.1  \\
NGC\,5253       & 13 39 56.00    &  $-$31 38 24.4    & 2032     &  600166  &2001 Jan 13     & 57.4 & $-120$  & R4CU5UPD13.3  \\
He\,2-10       & 08 36 15.15    &  $-$26 24 34.0    & 2075     &  600209  &2001 Mar 23     & 20.0 & $-120$  & R4CU5UPD14.1  \\

\hline
\multicolumn{9}{l}{}\\
\end{tabular}

\label{tab:chandra_obs}
\end{minipage}
\end{table*}

\section{{\it Chandra} observations and data reduction}
\label{sec:dataprep}

\subsection{Sample selection}
\label{sec:sample}

As mentioned in the introduction, the effect of a starburst on small
galaxies is expected to be far more devastating than the impact a
similar such event would have if it takes place within a larger
system. We therefore decided to concentrate our efforts on dwarf
starburst galaxies. To obtain a sample of galaxies, we searched the
{\it Chandra} archive for data on actively star forming dwarf galaxies
which were publicly available by November 2002. Only data collected by
the ACIS-S3 CCD were selected as this detector has a high quantum
efficiency ($\sim 0.8$ at 1\,keV), high spectral resolution ($\sim
120$\,eV), high angular resolution ($\sim 1\arcsec$), and a low charge
transfer inefficiency. Given the faint nature of dwarf galaxies,
observations with integration times less than 10\,ks were discarded.

In total, eight {\it Chandra} observations of dwarf starburst galaxies
met the selection criteria mentioned above. These targets are:
I\,Zw\,18, VII\,Zw\,403, NGC\,1569, NGC\,3077, NGC\,4214, NGC\,4449,
NGC\,5253, and He\,2-10. None of the galaxies in the sample is a
member of the Local Group; they fall within a distance range of
$2-13$\,Mpc. Note that all objects were previously detected in X-ray
emission by {\it ROSAT}.

The parameters of the {\it Chandra} observations of the galaxies in
our sample are listed in Table\,\ref{tab:chandra_obs}. In Columns 2
and 3, the coordinates of each object are given. Columns 4 and 5 list
the Observation IDs and the Sequence numbers of the corresponding
entry in the {\it Chandra} archive, respectively. The observation
dates are displayed in Column 6. Column 7 shows the total exposure
time $t_{\rm exp}$. The focal plane temperatures $T_{\rm fp}$ are
listed in Column 8 and the pipeline products version numbers in Column
9.

\subsection{Data preparation and imaging}

\label{sec:ch5:chandra_general}

For all datasets the satellite telemetries were processed at the {\it
{\it Chandra} X-ray Center} (CXC) with the {\sc Standard Data
Processing} (SDP) system, to correct for the motion of the satellite
and to apply instrument calibration. {\sc Order-sorting/Integrated
Probability Tables} were applied to all observations using {\sc CALDB
v.2.9}. All data products were analysed with the CXC {\sc {\it
Chandra} Interactive Analysis of Observations} (CIAO, v.2.2) software,
including {\sc Data Model} tools and the fitting software {\sc
Sherpa}. According to the CXC\footnote{see {\sl
http://cxc.harvard.edu/cal/ASPECT/celmon/index.html}}, the pointing
accuracy of the observations is 1\arcsec\ for each observation. Times
of intermittent strong background rates were discarded at levels
higher than 3$\sigma$ of the quiescent periods.

Broad X-ray bands were constructed in order to separate oxygen line
complexes ({\it soft band (S):} $0.3$\,keV\,$\leqslant E \leqslant
0.7$\,keV) from Fe-L lines ({\it medium band (M):}
$0.7$\,keV\,$\leqslant E \leqslant 1.1$\,keV) and the relatively
line-free continuum emission ({\it hard band (H):}
$1.1$\,keV\,$\leqslant E \leqslant 6.0$\,keV). Also, the sum of the
bands ({\it total band:} $0.3$\,keV\,$\leqslant E \leqslant 6.0$\,keV)
was analysed. For the purpose of imaging we applied the corresponding
bad pixel masks (provided by the CXC) and exposure maps. We also
constructed hardness ratios\footnote{Here we use a three-band
definition which has the advantage of being meaningful even in those
cases where emission is present in only one band.} which are defined
as\\
\begin{eqnarray}
HR1 &=& (S-M-H)/(S+M+H) \,\,\,\,\,\,{\rm and } \nonumber \\
HR2 &=& (S+M-H)/(S+M+H).  
\label{eq:hardnessratios}
\end{eqnarray}
  
In Table\,\ref{tab:chandra_obs_limits} we show the resulting 3$\sigma$
detection limits for all bands (Columns 2--5). The remaining exposure
time $t_{\rm eff}$ after removing the background flaring is given in
Column 6.


\begin{table}
\centering
\caption
{3$\sigma$ detection limits in the different bands (given as flux
count rates in units of $10^{-8}$\,cts~~s$^{-1}$\,cm$^{-2}$). The last
column lists the effective exposure times of the {\it Chandra}
observations in ks.}
\begin{tabular}{@{}lcccccccc@{}}
\hline

\multicolumn{1}{l}{Galaxy} & \multicolumn{1}{c}{Soft}& \multicolumn{1}{c}{Medium} & \multicolumn{1}{c}{Hard} & \multicolumn{1}{c}{Total}& \multicolumn{1}{c}{$t_{\rm eff}$}\\
\multicolumn{1}{l}{(1)} & \multicolumn{1}{c}{(2)}& \multicolumn{1}{c}{(3)} & \multicolumn{1}{c}{(4)} & \multicolumn{1}{c}{(5)}& \multicolumn{1}{c}{(6)}
\\

\hline

I\,Zw\,18       & 2.0  & 0.8  & 2.8  & 2.4 & 18.3    \\  
VII\,Zw\,403    & 2.3  & 1.1  & 4.0  & 3.2 & 10.5    \\
NGC\,1569       & 0.9  & 0.4  & 1.4  & 1.2 & 75.0    \\
NGC\,3077       & 1.2  & 0.5  & 1.8  & 1.4 & 53.2    \\
NGC\,4214       & 2.6  & 1.2  & 3.3  & 2.8 & 11.4   \\
NGC\,4449       & 1.9  & 0.6  & 2.1  & 1.9 & 31.0    \\
NGC\,5253       & 0.9  & 0.7  & 2.3  & 1.9 & 44.3    \\
He\,2-10       & 1.9  & 0.7  & 2.5  & 2.2 & 20.0    \\
\hline

\end{tabular}
\label{tab:chandra_obs_limits}
\end{table}


The deepest observations were performed on NGC\,1569 ($t_{\rm
eff}=75$\,ks). With a distance of only 2.2\,Mpc it is also the nearest
galaxy in the sample and therefore provides the best opportunity to
explore the properties of the hot gas \citep[see][]{mar02}. The
observation of VII\,Zw\,403 has the shortest $t_{\rm eff}$ and the
highest focal point temperature. The detection limits of VII\,Zw\,403
are therefore the highest of our sample.

\begin{table*}
\begin{minipage}{170mm}
\centering
\caption
{X-ray count rates of galaxies where diffuse X-ray
emission is detected. For each band the count rates are given in
units of $10^{-3}$\,cts~s$^{-1}$.}
\begin{tabular}{@{}lcccccc@{}}
\hline

\multicolumn{1}{l}{Band}&\multicolumn{1}{c}{NGC\,1569}&\multicolumn{1}{c}{NGC\,3077}&\multicolumn{1}{c}{NGC\,4214}&\multicolumn{1}{c}{NGC\,4449}&\multicolumn{1}{c}{NGC\,5253}&\multicolumn{1}{c}{He2-10}\\

\hline

&&&&&\\
\multicolumn{7}{c}{X-Ray Emission Including Point Sources}\\
&&&&&\\
\hline

Total    & $112.16\pm 1.64$     & $21.73\pm 0.94$     & $59.82\pm2.65$  &$361.55\pm 4.18$    & $38.57\pm 1.08$    & $69.49\pm 2.03$\\
Soft     & $ \phn13.66\pm 0.68$ & $\phn5.22\pm 0.47$ & $13.13\pm1.30$   &$\phn97.11\pm2.23$ & $\phn8.57\pm 0.53$ & $10.17\pm 0.85$\\
Medium   & $ \phn53.17\pm 0.94$ & $\phn9.02\pm 0.47$  & $18.41\pm1.39$  &$133.46\pm 2.33$    & $19.88\pm 0.66$    &  $34.62 \pm 1.34$\\
Hard     & $ \phn45.33\pm 1.16$ & $\phn7.49\pm 0.67$  & $28.28\pm1.85$  &$130.72\pm 2.65$    & $10.12\pm 0.67$    & $24.70\pm 1.26$\\

\hline
&&&&&\\
\multicolumn{7}{c}{Diffuse X-Ray Emission Only}\\
&&&&&\\
\hline

Total    & $75.95\pm 1.48$ & $13.96\pm 0.86$    & $23.44\pm 1.95$ &$174.70\pm 3.23$     & $26.45\pm 0.96$    & $45.40\pm 1.70$\\
Soft     & $10.49\pm 0.65$ & $\phn4.25\pm 0.45$ & $\phn9.52\pm1.17$ & $\phn68.27\pm 1.98$  & $\phn6.66\pm 0.50$ & $\phn7.90\pm 0.78$\\
Medium   & $41.35\pm 0.85$ & $\phn7.62\pm 0.44$ & $\phn8.90\pm1.05$ & $\phn81.47\pm 1.87$  & $14.67\pm 0.57$    & $ 27.74\pm 1.21$\\
Hard     & $23.96\pm 1.03$ & $\phn1.83\pm 0.59$ & $\phn4.93\pm1.16$ & $\phn23.20\pm 1.73$  & $\phn5.02\pm 0.59$ & $\phn9.46\pm 0.90$\\

\hline

\end{tabular}
\label{tab:common_countrate}
\end{minipage}
\end{table*}

For each galaxy, we separately study the point source population and
the purely diffuse emission. Two CIAO source detection algorithms {\sc
wavdetect} and {\sc celldetect} were applied for the identification of
the point sources using the total band X-ray image. The degradation
of the point spread function (PSF) at large off-axis angles was taken
into account by applying scales of 0\farcs5, 1\farcs0, and 2\farcs0
wavelet radii in {\sc wavdetect}. For {\sc celldetect} we used S/N
ratios of 1.9, 2.1, and 3.2 within off-axis radii $r$ of $0\arcmin < r
< 1\arcmin$, $1\arcmin < r < 2\farcm5$, and $2\farcm5 < r < 3\farcm5$,
respectively. Beyond a radius of $3\farcm5$ the PSF is too degraded to
provide reliable {\sc celldetect} detections.

Only those point sources were considered which were detected within
the general area covered by the \ha, optical or diffuse X-ray emission
of the galaxies under study. All point sources are treated in this
paper as if they reside within the targets. Interlopers, such as stars
in the Milky Way or background active galactic nuclei (AGN) may be
visible as less absorbed, extremely soft or highly absorbed, hard
X-ray point sources, respectively. In particular, those point sources
which are best-fitted by power law models with an index of about 1.4
\citep[cf. the Chandra Deep Field observations, ][]{toz01} and with
large absorbing column densities (due to the dwarf galaxies and the
Galaxy) might be distant AGNs.

In order to obtain images of the diffuse X-ray emission, point sources
over an area corresponding to three times the point spread function
(PSF) were blanked from the data. These areas were subsequently
refilled by interpolated count rates from the ambient diffuse X-ray
emission. As a result we obtained purely diffuse X-ray maps in the
different energy bands. The count rates of the X-ray data with and
without point sources are listed in Table\,\ref{tab:common_countrate}.

In addition to the raw pixel maps, all maps were adaptively smoothed
(task {\sc csmooth}) with a lower and an upper S/N of 3 and 4,
respectively, using a fast Fourier method. The corresponding exposure
maps were smoothed as well, using the same smoothing kernels as
computed for the respective images. Smoothed images, however,
were not used for any quantitative analyses but for visualisation
purposes only.

\subsection{Spectral analysis} 
\label{sec:specana}

In addition to broad band images, we derived spectra of the discrete
and diffuse X-ray emission of the galaxies in our sample (energy
range: 0.3\,keV$\leqslant E \leqslant 8.0$\,keV). For this purpose, we
used {\sc CALDB v.2.9} to calibrate the data. {\sc Redistribution
Matrix Files} (RMFs) and {\sc Auxiliary Response Files} (ARFs) were
constructed for the central positions of the apertures from which
corresponding spectra where extracted.

Finally, the consecutive {\it pulse-invariant} (PI) channels were
binned by a factor of three. This still left the spectra sufficiently
oversampled as the energy width of a single PI channel is 14.6\,eV
whereas the spectral resolution of the ACIS-S3 chip at 1\,keV is $\sim
120$\,eV. The spectra were eventually fitted by different models
within the {\sc Sherpa} v.2.2 software. The models were convolved with
the appropriate RMFs and ARFs and we used the $\chi^{2}$ Gehrels
statistics \citep{geh86}. This statistics is designed to work with low
numbers (accuracy: $\sim 1$\%) and also permits background subtraction
as well as the assessment of the quality of a fit via the
goodness--of--fit $\chi^{2}$ indicator. To obtain the parameters which
best describe the data we used a Monte-Carlo fitting technique which
randomly defines 20 starting values within a physically sensible
parameter space followed by the single-shot Powell optimisation method
to derive the final set of parameters ({\it Monte-Powell} in {\sc
Sherpa}). The apertures for point sources are sufficiently small to
allow the background contribution to be neglected ($\sim 1-5$
counts). The larger apertures used for diffuse emission, however, do
contain significant background emission. This emission was subtracted
using spectra from regions with similar CCD rows as the source
regions.

For all models we fixed the absorbing Galactic foreground column
density to the measurements of atomic neutral hydrogen (\hi) performed
by \citet{har97} (listed in Table\,\ref{tab:sample_new}). A solar
metallicity and element mixture was assigned to this absorbing gas
(model {\sc xsphabs}, \citealt{bal92}; abundances taken from
\citealt{and89}). In addition, we used a second absorption component
with a metal abundance corresponding to the internal metallicity of
the objects (based on \hii\ region oxygen abundances, see
Table\,\ref{tab:sample_new} -- for a justification see
Sect.\,\ref{sec:diffuse_yes}). The data were not corrected for
molecular contamination of the ACIS optical blocking filter which may
introduce another $3-7\times 10^{20}$\,\cden\ of absorbing column with
solar metallicity, dependent on the observing dates. The resulting
model expressions for Sherpa are: $model=xsphabs
(Galactic)~\times~xsvphabs (internal)~\times~EMITTER$, where $EMITTER$
stands for the emission model, e.g., {\sc plaw1d} (power law, PL),
{\sc black} (black body, BB), {\sc xsbremss} (thermal bremsstrahlung,
Brems), {\sc xsmekal}
\citep[MeKaL thermal plasma model;][]{mew85,kaa92,lie95}.  
The normalisations are given as: $10^{-14} (4\pi D^{2})^{-1} \int n_{\rm
e} n_{\rm H} dV$ (MeKaL) and $3.02\times 10^{-15} (4\pi D^{2})^{-1} \int
n_{\rm e} n_{\rm I} dV$ (Brems) where $D$ is the distance to the
source in cm, $n_{\rm e}$, $n_{\rm H}$, and $n_{\rm I}$ are the
electron, proton, and ion densities respectively, and $V$ the volume
of the emission region. The BB model amplitude is given as
$2\pi\,(c^{2}h^{3})^{-1} (R/D)^{2}= 9.884 \times 10^{31} (R/D)^{2}$
($R$: radius of the object; the speed of light, $c$, is given in
cm\,s$^{-1}$, and the Planck constant, $h$, is specified in keV\,s).
The normalisation reference point for the PL model was set to
1\,keV. Confidence regions for the individual fits were produced
using the {\it region--projection} task in {\sc Sherpa}. Whenever the
confidence levels are computed in a plane defined by two parameters,
{\it region--projection} still allows all the other free parameters to
vary, i.e., they are not frozen to the values of the best fit.

\begin{table*}
\begin{minipage}{170mm}
\centering
\caption
{Distances, oxygen abundances $12+\log(O/H)$, the corresponding
  metallicities $Z$ (relative to solar abundances), and Galactic
  \hi\ foreground column densities $N_{\rm H}^{\rm Gal}$ of the sample
  \citep[$N_{\rm H}^{\rm Gal}$ taken from][]{har97}.}
\begin{tabular}{@{}lllllc@{}}
\hline

\multicolumn{1}{c}{Galaxy}&\multicolumn{1}{c}{Distance}&\multicolumn{1}{c}{$12+\log(O/H)$}&\multicolumn{1}{c}{$Z$}&\multicolumn{1}{c}{$N_{\rm H}^{\rm Gal}$}&\multicolumn{1}{c}{Reference}\\
\multicolumn{1}{c}{}&\multicolumn{1}{c}{[Mpc]}&\multicolumn{1}{c}{}&\multicolumn{1}{c}{[solar]}&\multicolumn{1}{c}{[$10^{20}$\,\cden]}&\multicolumn{1}{c}{cols 2, 3}\\
\multicolumn{1}{c}{(1)}&\multicolumn{1}{c}{(2)}&\multicolumn{1}{c}{(3)}&\multicolumn{1}{c}{(4)}&\multicolumn{1}{c}{(5)}&\multicolumn{1}{c}{(6)}\\

\hline

I\,Zw\,18       & 12.6    &    7.16        &   0.02    &  \phn2  &    1, 2\\
VII\,Zw\,403   & 4.5     & $7.73\pm0.01$  &   0.07    &  \phn3  &    3, 4\\
NGC\,1569       & 2.2     & $8.22\pm0.07$  &   0.21    &     12  &    5, 6\\  
NGC\,3077       & 3.6     & 8.90           &   1.00    &  \phn4  &    7, 6\\
NGC\,4214       & 2.9     & $8.28\pm0.08$  &   0.25    &  \phn3  &    8, 9\\
NGC\,4449       & 3.9     & $8.31\pm0.07$  &   0.26    &  \phn1  &    10, 6\\
NGC\,5253       & 3.3     & $8.23\pm0.01$  &   0.21    &  \phn5  &    11, 6\\
He\,2-10       & 9.0     & 8.93           &   1.07    &  \phn9  &    12, 13\\

\hline

\end{tabular}
\label{tab:sample_new}

\end{minipage}
\flushleft {\sc References:} (1) \citet{ost00}; (2) \citet{gus00}; (3) \citet{lyn98}; (4) \citet{itz97}; (5) \citet{isr88}; (6) \citet{mar97}; (7) \citet{fre94}; (8) \citet{mai02}; (9) \citet{kob96}; (10) \citet{hun98}; (11) \citet{gib00}; (12) \citet{vac92}; (13) \citet{kob99}
\end{table*}

\section{Galaxies without diffuse X-ray emission}
\label{sec:diffuse_no}

For two galaxies of our sample, I\,Zw\,18 and VII\,Zw\,403, we did not
detect diffuse X-ray emission. Instead, we calculated upper limits
based on the sizes of their optical bodies. To do so, we simulated
absorbed thermal plasma emission from hot gas that we would be able to
detect. The temperature of the thermal plasma is assumed to be
0.2\,keV (equivalent to $2.3\times 10^{6}$\,K) with a metallicity
corresponding to that measured in their
\hii\ regions (see Table\,\ref{tab:sample_new}). This hot gas temperature 
is similar to that derived by spectral fitting of the other galaxies
in the sample (cf. Sect.\,\ref{sec:diffuse_yes}). We also assumed
photoelectric absorption by gas local to the dwarf starburst galaxy
with a canonical column density of $N_{\rm H}=1.5\times
10^{21}$\,cm$^{-2}$ (see Sect.\,\ref{sec:specana} for details).
Finally, we adjusted the normalisation of the plasma models in a way
such that the peaks of the spectra were at just $\sim 2\sigma$ of the
measured background counts, but visible in $\sim 5$ channels. This
leads to upper limits of the diffuse X-ray fluxes and luminosities
which are summarised in Table\,\ref{tab:plasma_limits}. The results
for the discrete X-ray emission of I\,Zw\,18 and VII\,Zw\,403 are as
follows.

\begin{table}
\centering
\caption
{Sensitivity limits of the {\it Chandra} observations for hot gas in
galaxies, where no diffuse emission is detected (I\,Zw\,18 and
VII\,Zw\,403). The internal absorbing column density is fixed to
$1.5\times 10^{21}$\,cm$^{-2}$ and the plasma temperature to 0.2\,keV
($\simeq 2.3\times 10^{6}$\,K). Absorbed ($F_{\rm X}^{\rm
abs}$[$0.3-8.0$\,keV]) and absorption corrected ($F_{\rm X}$) fluxes
are given in units of $10^{-15}$\,erg~s$^{-1}$\,cm$^{-2}$ and
luminosities ($L_{\rm X}$) in $10^{37}$\,erg~s$^{-1}$.}
\begin{tabular}{@{}lccc@{}}
\hline

\multicolumn{1}{l}{Galaxy}&\multicolumn{1}{c}{$F_{\rm X}^{\rm abs}$}&\multicolumn{1}{c}{$F_{\rm X}$}&\multicolumn{1}{c}{$L_{\rm X}$}\\

\hline

I\,Zw\,18    & 9  & 20 & 37 \\ 
VII\,Zw\,403 & 33     & 120    & 28 \\
\hline

\end{tabular}
\label{tab:plasma_limits}
\end{table}



\subsection[I\,Zw\,18]{{\bf I\,Zw\,18} (Mrk\,116, UGCA\,166)}
\label{sec5:izw18}

\paragraph{Previous X-ray observations and results}

{\it ROSAT} PSPC data (17\,ks) of I\,Zw\,18 were analysed by
\citet{mar96}. They reported the detection of a point source with a
luminosity of $\sim10^{39}$\,erg~s$^{-1}$ and explained its nature
partly with hot thermal gas. Faint, diffuse extensions of this source
are reported by \citet{bom01}, based on {\it ROSAT} HRI observations
(64\,ks), as well as by \citet{bom02} and \citet{thu04} based on the
same {\it Chandra} data as presented here. \citet{thu04} derive the
0.5--10.0\,keV X-ray luminosity of this point source to either
$1.6\times10^{39}$\,\lum or $1.4\times10^{39}$\,\lum\ depending on
whether a PL or an Raymond--Smith \citep[RS; ][]{ray77} thermal plasma
model is applied.

\paragraph{{\it Chandra} observations revisited}
\label{sec:chandra_obs_1}

\begin{figure*}
\centering
\includegraphics[width=\textwidth]{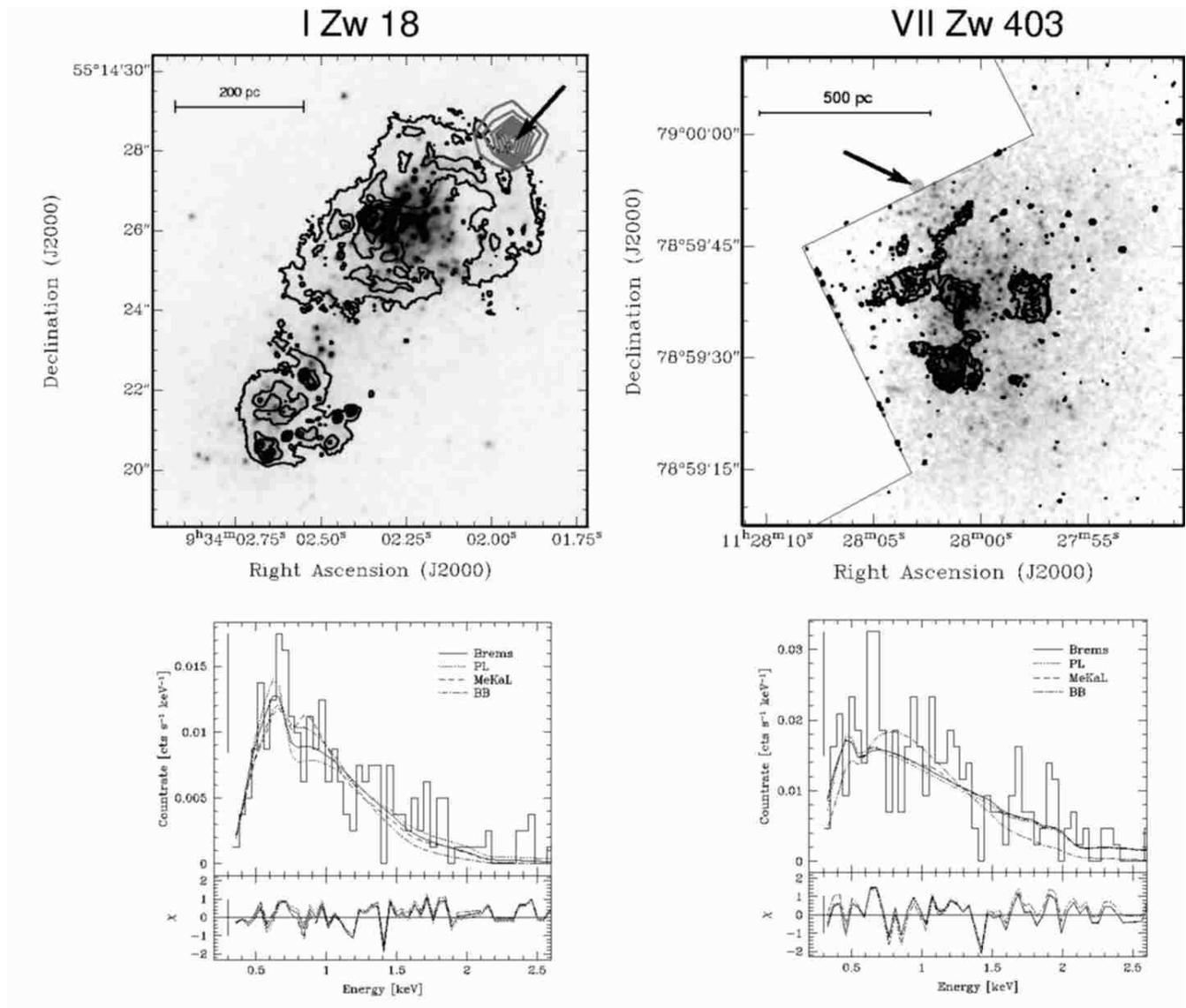}
\caption
{Images and X-ray spectra of I\,Zw\,18 ({\bf left hand panels}) and
VII\,Zw\,403 ({\bf right hand panels}). Displayed are {\it HST}/WFPC2
images of I\,Zw\,18 and VII\,Zw403 in the F675W and F555W bands,
respectively ({\bf top}). \ha\ emission is shown as {\bf black
contours}; the {\bf grey contours} display the locations of the X-ray
point sources (also marked by the arrows). Note that the X-ray point
source in VII\,Zw\,403 is right at the border of the WFPC2 PC CCD
(borders marked by the thin black line). The lower panels show the
spectra of the respective point sources as histograms with the fit
results of each model overlaid ({\bf Brems}: thermal bremsstrahlung,
{\bf PL}: power law, {\bf MeKaL}: MeKaL thermal plasma, {\bf BB}:
black body). The typical errors of the spectra are plotted in the
upper left corners. The lower portions of these graphs display the
errors $\chi=(Data-Model)/Error$ of the corresponding models with the
errorbar to the left. A large $|\chi|$ indicates a bad fit and the
optimum is reached for $|\chi|\sim 1$
\citep[optical images of I\,Zw\,18 taken from][]{can02}.}
\label{fig:nodetect_allfig}
\end{figure*}

In addition to five background sources in the ACIS-S3 field of view,
the {\it Chandra} observations show a source coinciding with the
optical counterpart of I\,Zw\,18 (see Fig.\,\ref{fig:nodetect_allfig},
upper left). This source is located on the rim of a central expanding
H$\alpha$ superbubble \citep[described by][]{mar96}, north-west to the
main optical body. Based on the {\it HST}/WFPC2 data shown in
Fig.\,\ref{fig:nodetect_allfig}, no stellar counterpart can be
assigned to this X-ray point source.

We checked whether or not the source may be slightly extended by
comparing its shape with a model PSF calculated for the same off-axis
angle and the same mean energy as the X-ray source itself. According
to our analysis, the PSF and the X-ray source profiles are
indistinguishable. We do not find evidence for excess emission towards
the extensions claimed by \citet{bom02} and \citet{thu04}.

\begin{table}
\centering
\caption
{Coordinates, net source counts and hardness ratios of the X-ray point sources in I\,Zw\,18 and VII\,Zw\,403. The source counts are given in $10^{-3}$\,cts~s$^{-1}$.}
\begin{tabular}{@{}lcc@{}}
\hline

\multicolumn{1}{c}{}&\multicolumn{1}{c}{I\,Zw\,18}&\multicolumn{1}{c}{VII\,Zw\,403}\\

\hline
Right Ascension (J2000) & 09\h34\m01\fs9            & 11\h28\m03\fs0 \\
Declination (J2000)     & 55\degr14\arcmin28\farcs4 &  78\degr59\arcmin53\farcs3 \\
\hline
Soft   &  \phn$3.22\pm 0.42$ & \phn$7.53\pm 0.90$ \\ 
Medium &  \phn$4.04\pm 0.47$ & \phn$6.31\pm 0.81$ \\
Hard   &  \phn$5.25\pm 0.57$ & $14.55\pm 1.24$ \\
Total  &  $12.51\pm 0.85$& $28.40\pm 1.74$ \\
\hline
HR1    &  $-0.49\pm 0.08$& $-0.47\pm 0.06$\\
HR2    &  $\phs0.16\pm0.07$  & $-0.03\pm 0.06$ \\

\hline

\end{tabular}
\label{tab:source_counts}
\end{table}

Coordinates, net source counts in the different bands, and hardness
ratios of the point source are listed in
Table\,\ref{tab:source_counts}. In addition, we extracted a source
spectrum, which is displayed in Fig.\,\ref{fig:nodetect_allfig} (lower
left). The spectrum peaks at $\sim 0.6$\,keV with a steep decline to
softer energies and a shallower decline to harder energies. We decided
to fit four models to the spectrum: (a) PL, (b) BB, (c) Brems, and (d)
MeKaL. The results are summarised in Table\,\ref{tab:IZw18_point}.
Overlays of these models on the spectrum are shown in
Fig.\,\ref{fig:nodetect_allfig}. All models provide reasonable fits to
the data and their $\chi^{2}_{\rm red}$ are very similar, in the range
of 0.12 (PL fit) to 0.14 (BB fit). The X-ray luminosity of the models
range from $\sim 1.0-5.2\times 10^{39}$\,\lum, which is in good
agreement with previous results. This source falls within the class of
ultraluminous X-ray sources (ULXs, defined as sources having $L_{\rm
X}>10^{39}$\,\lum).


\begin{table*}
\renewcommand{\arraystretch}{1.5}
\begin{minipage}{170mm}
\centering
\caption
{Fits to the point sources in I\,Zw\,18 and VII\,Zw403. Models are
abbreviated as in Fig.\,\ref{fig:nodetect_allfig}. The parameters are:
$N_{\rm H}$: internal absorbing column density, $\gamma$: PL photon
index, $T$: temperature, $F_{\rm X}^{\rm abs}$: absorbed X-ray flux
($0.3-8.0$\,keV), $F_{\rm X}$: unabsorbed X-ray flux, $L_{\rm X}$:
unabsorbed X-ray luminosity, and $\chi^{2}_{\rm red}$: the reduced
$\chi^{2}$ goodness-of-fit parameter. Note that $\gamma$ is only
applicable for the PL model; the normalisation is given for Brems and
MeKaL model fits, whereas the amplitude is used for the PL and BB
models. Errors are given for 68.3 per cent confidence.}
\begin{tabular}{@{}llllll@{}}
\hline

\multicolumn{1}{l}{Parameter}&\multicolumn{1}{l}{Units}&\multicolumn{1}{c}{PL}&\multicolumn{1}{c}{BB}&\multicolumn{1}{c}{Brems}&\multicolumn{1}{c}{MeKaL}\\
\hline

&\multicolumn{5}{c}{I\,Zw\,18}\\
\hline

$N_{\rm H}$                      & [$10^{21}$\,cm$^{-2}$]                & $15.04_{-6.19}^{+5.95}  $       & $6.24_{-5.10}^{+5.09}    $ & $10.45_{-4.93}^{+4.96} $& $8.94_{-3.32}^{+3.39}  $          \\[0.06cm]  
$\gamma/T$                   & [--]/[$10^{6}$\,K]                    & $3.40_{-0.76}^{+0.76}   $       & $2.37_{-0.64}^{+0.56}   $ & $6.75_{-2.20}^{+2.20}  $& $7.35_{-1.64}^{+1.65}   $         \\[0.06cm]  
$Ampl/Norm^{a}$                 & [$10^{-5}$]                           & $3.09_{-1.12}^{+1.11}   $       & $306.2_{-192.9}^{+190.0} $ & $9.36_{-6.71}^{+6.82}  $& $18.17_{-7.57}^{+7.52}  $      \\[0.06cm]    
$F_{\rm X}^{\rm abs}$                & [$10^{-15}$\,erg~s$^{-1}$\,cm$^{-2}$]& $39.40_{-6.05}^{+36.5}$       & $25.95_{-24.74}^{+138}$  & $30.33_{-26.14}^{+79.77}$& $29.97_{-20.14}^{+36.44} $      \\[0.06cm]  
$F_{\rm X}$                      & [$10^{-15}$\,erg~s$^{-1}$\,cm$^{-2}$]& $275_{-160}^{+469}  $       & $52.44_{-47.26}^{+149}$ & $62.61_{-23.04}^{+58.51} $& $84.85_{-43.85}^{+52.57}$         \\[0.06cm]  
$L_{\rm X}$                      & [$10^{37}$\,erg~s$^{-1}$]            & $523_{-304}^{+891}      $       & $99.63_{-89.80}^{+284}$ &$202_{-164}^{+245}       $& $161_{-83}^{+100}        $         \\     
$\chi^{2}_{\rm red}$    &                                                & \multicolumn{1}{c}{0.12}        & \multicolumn{1}{c}{0.14}  & \multicolumn{1}{c}{0.13}& \multicolumn{1}{c}{0.13}        \\

\hline
&\multicolumn{5}{c}{VII\,Zw\,403}\\
\hline

$N_{\rm H}$                      & [$10^{21}$\,cm$^{-2}$]                &  $1.16_{-1.16}^{+1.58}    $       & $0.00_{-0.00}^{+2.08}   $ & $0.00_{-0.00}^{+0.32}   $& $0.12_{-0.12}^{+0.94}  $          \\[0.06cm]                                                                  
$\gamma/T$                   & [--]/[$10^{6}$\,K]                    &  $1.75_{-0.34}^{+0.37}    $       & $3.14_{-0.77}^{+0.77}   $ & $52.4_{-0.1}^{+0.1}   $& $44.0_{-0.2}^{+0.2}  $         \\[0.06cm]                                                                  
$Ampl/Norm^{a}$                  & [$10^{-5}$]                           &  $2.43_{-0.70}^{+0.69}    $       & $110.9_{-56.9}^{+56.8}  $ & $3.07_{-0.85}^{+0.74}   $& $9.60_{-2.37}^{+2.02}  $          \\[0.06cm]                                                                    
$F_{\rm X}^{\rm abs}$                & [$10^{-15}$\,erg~s$^{-1}$\,cm$^{-2}$]&  $151.2_{-79.9}^{+147.8}   $       & $52.61_{-46.7}^{+149.0}$ & $148.7_{-37.2}^{+30.9}  $& $118.5_{-33.8}^{+26.4} $        \\[0.06cm]                                                                 
$F_{\rm X}$                      & [$10^{-15}$\,erg~s$^{-1}$\,cm$^{-2}$]&  $196.9_{-65.0}^{+132.0}   $       & $61.9_{-52.1}^{+163.7}$ & $148.7_{-41.3}^{+35.9}  $& $141.1_{-35.0}^{+30.0} $       \\[0.06cm]                                                               
$L_{\rm X}$                      & [$10^{37}$\,erg~s$^{-1}$]            &  $47.3_{-15.6}^{+31.7}  $       & $14.8_{-12.5}^{+39.3}  $ & $35.7_{-9.9}^{+8.6}$& $33.9_{-8.4}^{+7.2} $        \\                                                                       
$\chi^{2}_{\rm red}$      &                                              &  \multicolumn{1}{c}{0.20}         &  \multicolumn{1}{c}{0.25} & \multicolumn{1}{c}{0.21} & \multicolumn{1}{c}{0.21}      \\

\hline
\footnotetext{$^{a}$ see Sect.\,\ref{sec:specana}} 
\end{tabular}
\label{tab:IZw18_point}
\end{minipage}
\end{table*}

In what follows we will try to derive which of the models best
describes the point source. A PL model would indicate that the source
is likely an X-ray binary (XRB). With a power law photon index of 3.3,
however, it exhibits a very steep spectrum. The Eddington limit of a
source is given by $L_{\rm E}=1.3\times 10^{38}
(M/M_{\odot})$\,erg~s$^{-1}$. If a source exceeds this limit stable
accretion is not possible anymore; the force from radiation pressure
exceeds the force needed to overcome the gravitational potential. If
the accreting object in the XRB has a mass lower than $\sim
40$\,M$_{\odot}$ (PL luminosity of the X-ray point source in
I\,Zw\,18: $L_{\rm X}=5.2\times10^{39}$\,\lum), it is above the
Eddington threshold. This would indicate the presence of a super
Eddington source which would imply that the source should be variable
in time. We therefore checked the point source in I\,Zw\,18 for
possible variability within the observational time frame. To do so, we
binned its spectrum in 60, 200, 600, and 1000\,s time bins and
constructed the corresponding light curves. The light curves do not
suggest temporal variability on the time scales given by the binning
and the length of the observation.

Black body radiation can be emitted by a single star, a helium burning
white dwarf, or an isolated neutron star. The fitted temperatures and
luminosities, however, are too high for any of these objects.

Alternatively, a collisional thermal plasma can be responsible for the
measured spectrum. In addition to the continuum emission (modeled by
thermal bremsstrahlung), line transitions might still play a role
which are described, e.g., by the MeKaL code (which also includes the
bremsstrahlung component). The bremsstrahlung model results in a
higher absorbing column density and a lower temperature as compared to
the MeKaL fits. For temperatures higher than $\sim 10^{7}$\,K, both
models should provide the same results, as line radiation is not a
dominant contribution to the X-ray emission anymore
\citep[see][]{sut93}. For 
lower temperatures, MeKaL models are more appropriate as compared to
pure thermal bremsstrahlung. If the thermal plasma scenario for the
point source in I\,Zw\,18 is realistic, then its nature is probably a
(young) supernova remnant (SNR). However, the X-ray luminosity of this
source is probably too large for this option to be a viable
one. \citet{bla04}, e.g., show that in M\,83 the maximum X-ray
luminosity of an SNR does not exceed $\sim 10^{37}$\,\lum, which is
about two orders of magnitude lower than what we derive for the point
source in I\,Zw\,18.

The source is located on the rim of the H$\alpha$ superbubble
(Fig.\,\ref{fig:nodetect_allfig}, upper left). It may therefore have
formed as a result of induced star formation on the expanding shell
which in turn has probably its origin in the central stellar
population seen in optical bands. Given the extreme luminosity and
small size of this object, we will prefer the XRB scenario and adopt
this for our analysis presented in Paper\,II.


\subsection[VII\,Zw\,403]{{\bf VII\,Zw\,403} (UGC\,6456)}
\label{sec:viizw403}

\paragraph{Previous X-ray observations and results}

\citet{pap94} analysed {\it ROSAT} PSPC (10\,ks) observations of 
VII\,Zw\,403. They detected a point-like source with some extended
faint structures. The total X-ray luminosity of this object was
derived to be $1.9\times 10^{38}$\,erg~s$^{-1}$.

\paragraph{{\it Chandra} observations}
The ACIS-S3 {\it Chandra} observations of VII\,Zw\,403 were performed
with an integration time of $\sim 10$\,ks. One unresolved X-ray source
is detected to coincide with the optical body of VII\,Zw\,403 (see the
upper right panel of Fig.\,\ref{fig:nodetect_allfig} for an overlay,
and Table\,\ref{tab:source_counts} for coordinates, net source counts,
and hardness ratios of the X-ray point source). Comparing our map to
available {\it HST}/WFPC2 data, the X-ray source is located on the
edge of the WFPC2 PC CCD, so unfortunately no cross-identification is
possible. In contrast to the {\it ROSAT} observations, we neither find
evidence for any extension of the discrete source, nor for any
extended, diffuse X-ray emission in VII\,Zw\,403 (the upper limit for
diffuse X-ray emission is given in Table\,\ref{tab:plasma_limits}).

The spectrum of the point source is shown in
Fig.\,\ref{fig:nodetect_allfig} (lower right). We fitted the same
models to the spectrum as for the point source in I\,Zw\,18, but with
7 per cent solar metallicity and the appropriate Galactic foreground
absorption.  From a statistical point of view, all models but the
black body yield fits with about the same $\chi^{2}_{\rm red}$ and are
virtually indistinguishable (see Table\,\ref{tab:IZw18_point}). The
X-ray luminosities of the fits are in the range of $\sim 1.5-4.7\times
10^{38}$\,\lum\ and therefore in agreement with the {\it ROSAT}
observations.

The power law index for VII\,Zw\,403 is lower than for I\,Zw\,18 (1.8
compared to 3.4) and falls within the range common for XRBs
\citep[see, e.g.,][]{irw03}. Thermal plasma models all yield very high
temperatures (a factor of 6--7 higher than the source in
I\,Zw\,18). Therefore, an XRB is the most likely explanation for this
X-ray point source. Lightcurves extracted in the same way as for
I\,Zw\,18 do not indicate any variability at the $3\sigma$ level in
the selected bins.

 
\section{Galaxies with detections of diffuse X-ray emission} 
\label{sec:diffuse_yes}

Diffuse X-ray emission is encountered in the {\it Chandra} data of six
galaxies: NGC\,1569, NGC\,3077, NGC\,4241, NGC\,4449, NGC\,5253, and
He\,2--10. Before discussing the results for these objects one by one,
we first describe in some detail how we analysed the diffuse X-ray
component and how we dealt with the X-ray point sources which fall
within the optical and/or X-ray extent of each target. After
presenting the results on the individual galaxies we justify why we
used the metallicities derived from \hii\ regions for our model
fitting. Lastly we address the issue of deviations from a solar
element mixture of the X--ray emitting gas, in particular a possible
excess of $\alpha$ elements over Fe.

\subsection{Analysis of the diffuse X-ray emission}

Point source identification and subtraction (see
Sect.\,\ref{sec:ch5:chandra_general}) is crucial for the analysis of
the hot gas. The sum of all individual point sources within one
galaxy can not be fitted by a simple function (e.g., PL). This is
illustrated in the total, point source+diffuse X-ray emission spectra
of the galaxies (panels [g] of Figs.\,\ref{fig:olay1569},
\ref{fig:olay3077}, \ref{fig:olay4214},
\ref{fig:olay4449}, \ref{fig:olay5253}, and \ref{fig:olayhe2-10}). 
However, previous analyses of diffuse X-ray components in galaxies, in
particular those based on {\it ROSAT} data, often do not account for
this fact and use a {\em single} power law model to accommodate for
the entire point source population. Incorrect subtraction of the point
sources will inevitably lead to a bias in the fit to the extended
component, the magnitude and nature of which is hard to predict. This
could easily lead one to invoke, for example, two-temperature model
fits where in reality a one-temperature fit would do. The importance
of properly taking into account the contribution from point sources
underscores the need for high angular resolution as provided by the
{\it Chandra} observatory.

Thermal plasma models (see Sect.\,\ref{sec:specana} for reduction and
fitting techniques) are very successful in describing the diffuse
X-ray emission. Several models exist in the literature, such as an RS
model. More recent models are MeKaL \citep{mew85,kaa92,lie95}, APEC
\citep{smi01} and CLOUDY
\citep{fer98}. We decided to use the MeKaL code for our analysis as it
is widely used in the literature and overcomes the shortcomings of the
RS code in the treatment of Fe-L lines. Its atomic data also contains
fluorescence lines which are not available for the other codes. A
detailed comparison of the models is provided on the CXC
webpages\footnote{{\sl
http://cxc.harvard.edu/atomdb/issues\_comparisons.html}}.

The MeKaL models did an excellent job in fitting the X-ray emission
after the unresolved sources had been subtracted. The introduction of
an additional non-thermal component (e.g., in order to account for
faint, unresolved point sources) did not significantly improve the
fits. We used a single temperature to model the emission. As it turns
out, the temperature of the hot, coronal gas varies over the extent of
the targets. This we deal with by defining different regions which are
fit independently by one-temperature models. In general, we define a
central region (R1) and several regions (R2, R3, $\ldots$, Rn) defined
by polygons which we collectively refer to as the 'outer regions'. The
outer regions were selected upon both the \ha\ and X-ray morphology of
each object.

Multi-temperature models do not improve the goodness of the fits
obtained for one-temperature models within an individual, small
region. In Figs.\,\ref{fig:olay1569},
\ref{fig:olay3077}, \ref{fig:olay4214},
\ref{fig:olay4449}, \ref{fig:olay5253}, and \ref{fig:olayhe2-10}
(panels [d] and [e]) these regions are overlaid on \ha\ and unsmoothed
X-ray images (note that for illustration purposes those X-ray images
are not point source subtracted; point sources were removed, however,
for all further analysis). We also combined some of the regions to
study the outskirts of the galaxies. It turns out that a
one-temperature plasma fit is acceptable even for the entire diffuse
X-ray emission. Multi-temperature fits show one component clearly
dominating the emission and all other components to have a very low
normalisation, i.e., X-ray emissivity. Those additional components can
be neglected (cf. panels [g] and [h] of Figs.\,\ref{fig:olay1569},
\ref{fig:olay3077},
\ref{fig:olay4214}, \ref{fig:olay4449}, \ref{fig:olay5253}, and
\ref{fig:olayhe2-10}, where the
single-temperature fits are overlaid on the data). 

The results of the fits are listed in
Tables\,\ref{tab:detect_gas1569},
\ref{tab:detect_gas3077}, \ref{tab:detect_gas4214},
\ref{tab:detect_gas4449}, \ref{tab:detect_gas5253}, and
\ref{tab:detect_gashe2-10}. In these tables $N_{H}$ denotes the internal
absorbing column density, $T$ is the temperature of the hot gas,
$Norm$ the normalisation, $F_{\rm X}^{\rm abs}$ the absorbed flux
($0.3-8.0$\,keV), $F_{\rm X}$ the unabsorbed flux, $L_{\rm X}$ the
unabsorbed luminosity ($0.3-8.0$\,keV), and $\chi^{2}_{red}$ the
reduced $\chi^{2}$ goodness-of-fit parameter. The fits, residuals, and
confidence regions in the $N_{H}-T$ plane (the normalisation was not
fixed in the process) are shown in panels [h] and [i] of
Figs.\,\ref{fig:olay1569},
\ref{fig:olay3077},
\ref{fig:olay4214},
\ref{fig:olay4449}, \ref{fig:olay5253}, and \ref{fig:olayhe2-10}. 
In all our fits we assumed solar abundances for the absorbing
foreground gas from the Milky Way whereas we used metallicities based
on independent measurements of the \hii\ regions when modelling the
emitting X-ray gas and any absorption internal to the objects under
study. This is justified {\em a posteriori} in
Sect.\,\ref{sec:metals}.

\subsection{Analysis of the X-ray Point Sources}

\begin{figure}
\centering
\includegraphics[width=8cm]{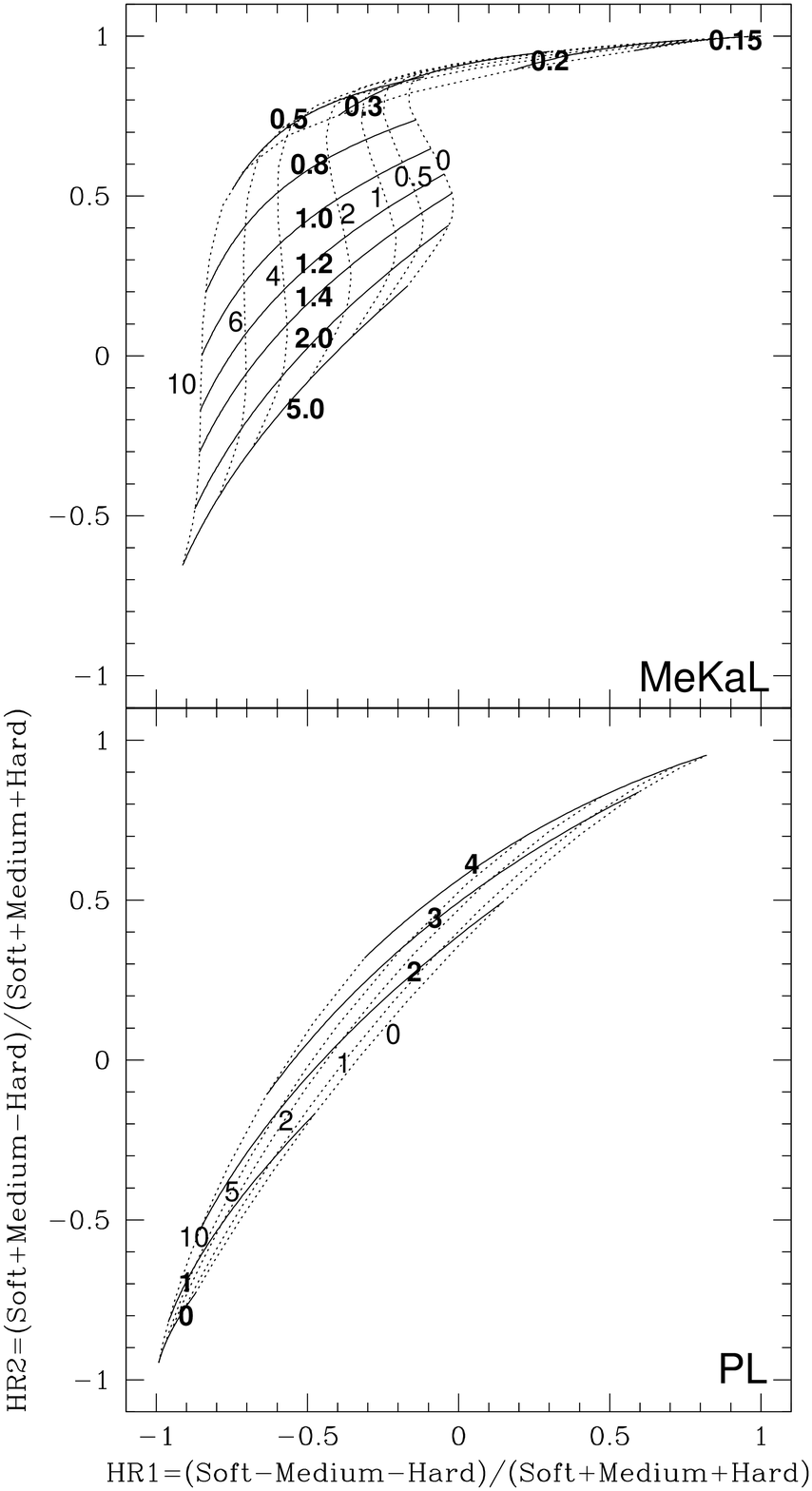}
\caption
{Hardness ratio plots of MeKaL thermal plasma ({\bf top}) and PL ({\bf
bottom}) models. Whereas the PL models are more or less distributed
along a diagonal strip extending from the lower left to the upper
right, the MeKaL models occupy a region in the upper left corner. {\it
Dotted lines} mark models at constant absorbing column densities
$N_{\rm H}$ ({\it normal fonts}, given in units of
$10^{21}$\,\cden). Models at a constant Temperature $T$ (MeKaL) or
power law index $\Gamma$ (PL) are displayed as {\it solid lines}) and
are labelled with {\it bold fonts}. Temperatures in the MeKaL model
are given in units of keV (1\,keV corresponds to
$1.1604\times10^{7}$\,K). The grids presented in this figure are
repeated at a smaller scale in panels (c) of
Figs.\,\ref{fig:point1569}, \ref{fig:point3077},
\ref{fig:point4214}, \ref{fig:point4449}, \ref{fig:point5253}, and
\ref{fig:pointhe2-10} with overplotted the location of the point 
sources in these hardness ratio plots.}
\label{fig:hardnessshow}
\end{figure}

In addition to the diffuse X-ray emission, we study the spectral
properties of the X-ray point source populations of the different
galaxies. As a first step, we extracted the source counts and derived
the hardness ratios (see Sect.\,\ref{sec:ch5:chandra_general}). These
values are listed in Tables\,\ref{tab:point1569a},
\ref{tab:point3077a}, \ref{tab:point4214a}, \ref{tab:point4449a},
\ref{tab:point5253a}, and \ref{tab:pointhe2-10a}. We follow two
approaches for a more detailed analysis of the individual point source
spectra: (a) we compare their hardness ratios to MeKaL and PL models
and (b) we fit different models (Brems, MeKaL, PL, and BB) to
their individual spectra. All models assume a metallicity equal to the
\hii\ oxygen abundances of their host galaxies.
Finally, we derive the best-fitting model for each point source with
sufficient source counts (at least three spectral bins with four
counts minimum) based on (a) a visual inspection of the fitted
spectra, (b) the goodness-of-fit parameter $\chi^{2}_{red}$, and (c)
the source locations in hardness ratio plots (see
Figs.\,\ref{fig:point1569}, \ref{fig:point3077},
\ref{fig:point4214}, \ref{fig:point4449}, \ref{fig:point5253}, and
\ref{fig:pointhe2-10} panels [b] and [c]). 
The grids which are overlaid in the hardness ratio plots are shown in
more detail in Fig.\,\ref{fig:hardnessshow}. They were calculated by
simulating model spectra for a range of thermal plasma and PL
models. In the case of a thermal plasma, we used MeKaL models and
varied the temperature of the plasma and absorbing foreground column
density. For the PL models the PL index and absorbing foreground
column density were varied. By applying the ARFs and RMFs of the
corresponding observations to the artificial spectra, the data were
folded with the ACIS--S3 response. The resulting, convolved spectra
were subsequently used to calculate the hardness ratios. 

The final selection of the source models (based on the criteria
described above) and the corresponding fit parameters are listed in
Tables\,\ref{tab:point1569b}, \ref{tab:point3077b},
\ref{tab:point4214b}, \ref{tab:point4449b}, 
\ref{tab:point5253b}, and \ref{tab:pointhe2-10b}.

\begin{figure*}
\centering
\includegraphics[width=15cm]{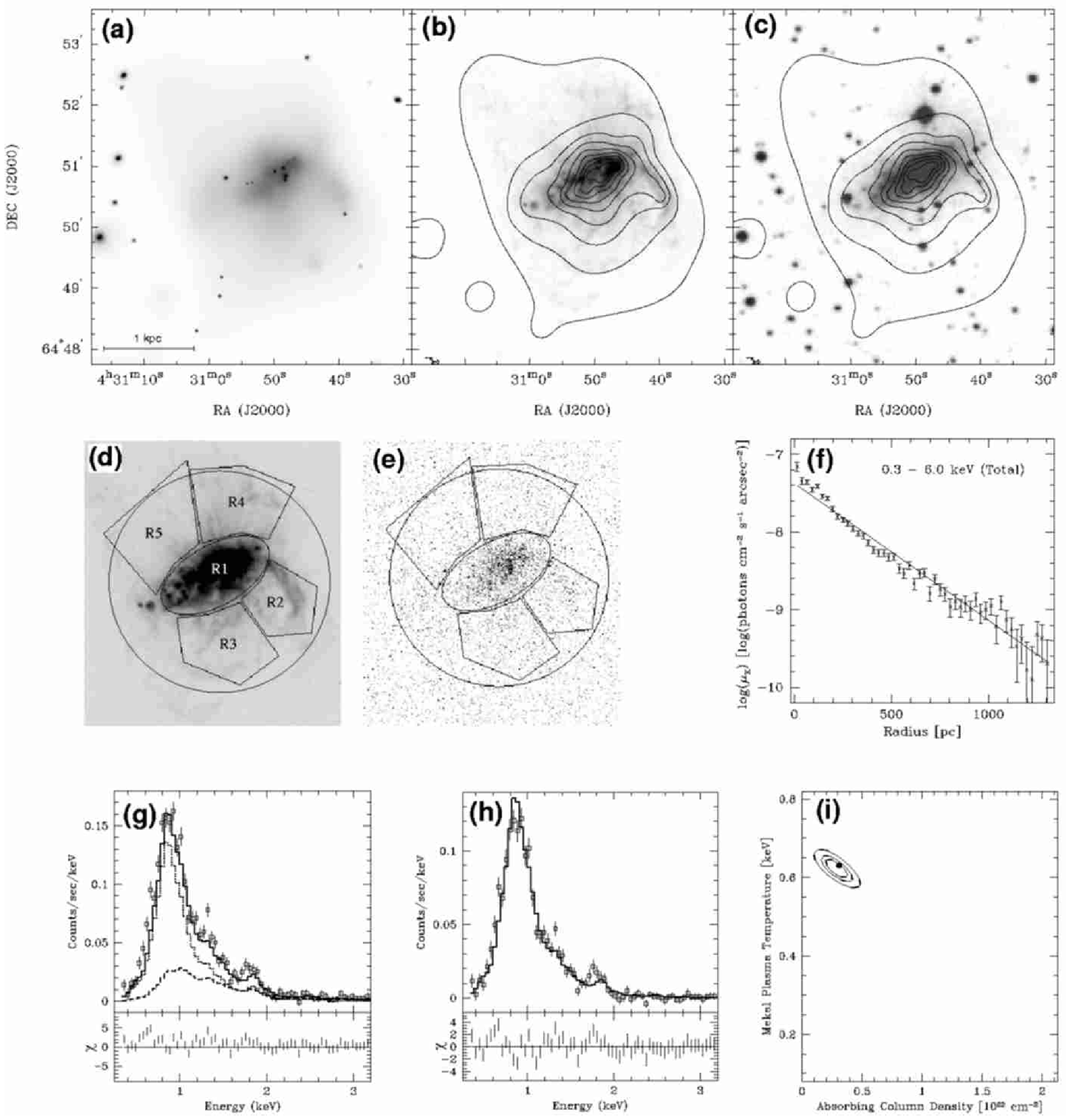}
\caption
{{\bf (a)} The total, adaptively smoothed X-ray emission of NGC\,1569
including point sources on a logarithmic scale. {\bf (b)}
Logarithmically scaled H$\alpha$ image \citep[taken from][]{dce96}
with contours of the diffuse X-ray emission after the removal of the
point sources. Contours of the logarithmic X-ray emission start at and
are spaced by $0.1\log(peak flux)$. {\bf (c)} The same contours as in
(b) overlaid on an optical DSS image. Images (a), (b), and (c) are on
the same scale. {\bf (d)} Definition of the regions used to extract
X-ray spectra overlaid on the \ha\ image. {\bf (e)} The same regions
as in (d) but overlaid on the unsmoothed X-ray image including the
X-ray point sources. {\bf (f)} Logarithmic, azimuthally averaged
surface brightness of the diffuse X-ray emission with the exponential
fit overlaid. The centre for this plot is located on the peak of a NIR
H-band image. {\bf (g)} The total X-ray spectrum of
NGC\,1569. Overlaid are the fits for the diffuse emission ({\it
dotted}), the combined point source models ({\it dashed}), and the
point sources plus diffuse emission ({\it solid}). The lower panels
display the normalised residuals $\chi=(data-model)/error$ (note that
the error is the uncertainty per energy bin) of the combined model as
a function of energy.{\bf (h)} The spectrum of the diffuse X-ray
emission as taken from the region displayed as a large circle in panel
(e) with the fit result on top. Again, the lower panel shows the error
of the fit per bin. {\bf (i)} Confidence regions for the fit of the
diffuse X-ray emission [as displayed in (h)] in the $N_{H}-T$ plane
(the normalisation is allowed to vary in the parameter space). The
best fit is shown by the {\it dot} and the confidence levels are 68.3,
90.0, and 99.0 per cent.}
\label{fig:olay1569}
\end{figure*}

As a last step, we combined the fits for the diffuse and the discrete
sources and overlay the results on the spatially integrated X-ray
spectra of the galaxies (including point sources). These are shown in
panels (g) of Figs.\,\ref{fig:olay1569},
\ref{fig:olay3077},
\ref{fig:olay4214}, \ref{fig:olay4449}, \ref{fig:olay5253}, and 
\ref{fig:olayhe2-10}.

\subsection[NGC\,1569]{{\bf NGC\,1569} (UGC\,3056, Arp\,210, VII\,Zw\,16)}
\label{sec5:1569}


\begin{table*}
\renewcommand{\arraystretch}{1.5}
\begin{minipage}{170mm}
\centering
\caption
{Results of MeKaL collisional thermal plasma model fits applied to the
X-ray spectra of different regions in NGC\,1569 (see panels [d] and
[e] in Fig.\,\ref{fig:olay1569} for the definition of the regions).}
\begin{tabular}{@{}lccccccc@{}}
\hline

\multicolumn{1}{l}{Region}&\multicolumn{1}{c}{$N_{\rm H}$}&\multicolumn{1}{c}{$T$}&\multicolumn{1}{c}{$Norm^{a}$}&\multicolumn{1}{c}{$F_{\rm X}^{\rm abs}$}&\multicolumn{1}{c}{$F_{\rm X}$}&\multicolumn{1}{c}{$L_{\rm X}$}&\multicolumn{1}{c}{$\chi^{2}_{red}$}\\
\multicolumn{1}{c}{}&\multicolumn{1}{c}{[$10^{21}$\,cm$^{-2}$]}&\multicolumn{1}{c}{[$10^{6}$\,K]}&\multicolumn{1}{c}{[$10^{-5}$]}&\multicolumn{1}{c}{[$10^{-15}$\,erg~s$^{-1}$\,cm$^{-2}$]}&\multicolumn{1}{c}{[$10^{-15}$\,erg~s$^{-1}$\,cm$^{-2}$]}&\multicolumn{1}{c}{[$10^{37}$\,erg~s$^{-1}$]}&\\

\hline

Total            & $2.86_{-0.15}^{+0.15}$ & $7.23_{-0.12}^{+0.12}$ & $80.0_{-1.5}^{+1.5}$ & $205_{-10}^{+10}$ & $721_{-15}^{+15}$ & $41.7_{-6.0}^{+6.0}$ & 0.98\\

R1 (Centre)      & $4.99_{-0.19}^{+0.19}$ & $7.23_{-0.13}^{+0.13}$ & $55.2_{-1.2}^{+1.2}$ & $116_{-7}^{+7}$ & $497_{-13}^{+13}$ & $28.8_{-4.1}^{+4.1}$ & 0.68\\

R2               & $4.07_{-0.49}^{+0.53}$ & $5.58_{-0.34}^{+0.34}$ & $9.45_{-0.64}^{+0.64}$ & $16.4_{-3.0}^{+3.4}$ & $78.1_{-7.2}^{+7.5}$ & $4.52_{-0.76}^{+0.76}$ & 0.32\\

R3               & $6.75_{-0.42}^{+0.45}$ & $3.49_{-0.11}^{+0.11}$ & $30.1_{-1.9}^{+1.9}$ & $18.5_{-3.1}^{+3.5}$ & $194_{-15}^{+15}$ & $11.2_{-1.8}^{+1.8}$ & 0.44\\

R4               &  $11.6_{-0.6}^{+0.7}$ & $3.21_{-0.12}^{+0.10}$ & $53.1_{-4.6}^{+4.6}$ & $15.2_{-3.4}^{+4.0}$ & $328_{-33}^{+34}$ & $19.0_{-3.3}^{+3.3}$ & 0.45\\

R5               &  $9.41_{-0.57}^{+0.63}$ & $3.26_{-0.12}^{+0.13}$ & $31.9_{-2.7}^{+2.7}$ & $12.3_{-2.7}^{+3.2}$ & $199_{-20}^{+21}$ & $11.5_{-2.0}^{+2.0}$ & 0.49\\

R2+R3 (South)    &  $6.96_{-0.32}^{+0.34}$ & $3.92_{-0.10}^{+0.10}$ & $45.1_{-2.0}^{+2.0}$ & $33.4_{-4.1}^{+4.5}$ & $308_{-18}^{+18}$ & $17.8_{-2.7}^{+2.7}$ & 0.57\\

R4+R5 (North)    &  $11.2_{-0.4}^{+0.5}$ & $3.21_{-0.08}^{+0.07}$ & $80.3_{-4.7}^{+4.7}$ & $24.2_{-3.8}^{+4.2}$ & $496_{-34}^{+35}$ & $28.7_{-4.5}^{+4.5}$ & 0.61\\

R2 to R5 (Outer Regions)  &  $9.12_{-0.26}^{+0.27}$ & $3.51_{-0.06}^{+0.06}$ & $125_{-4}^{+5}$ & $58.8_{-5.8}^{+6.2}$ & $809_{-36}^{+36}$ & $46.8_{-6.9}^{+6.9}$ & 0.78\\

\hline
\end{tabular}
\footnotetext{$^{a}$ see Sect.\,\ref{sec:specana}} 
\label{tab:detect_gas1569}
\end{minipage}
\end{table*}


\subsubsection{Previous X-ray observations and results}
The first X-ray detection of NGC\,1569 was based on data obtained by
{\it EINSTEIN}. NGC\,1569 appeared as an extended soft source with a
luminosity of $\sim 10^{39}$\,erg~s$^{-1}$ \citep{fab82}. Follow-up
observations with the {\it ROSAT} HRI (11\,ks) led to the detection of
spurs of diffuse X-ray emission on kiloparsec scales which coincide
with \ha\ filaments \citep{hec95}. This emission was interpreted to
originate from hot gas which is heated by supernova events with
mass-loading from the rims of superbubbles. The observations also led
to the speculation that at least part of the hot gas is able to escape
the gravitational potential of NGC\,1569
\citep{hec95}. These claims were supported by \citet{dce96} on the basis of ASCA
(78\,ks) and {\it ROSAT} PSPC (8\,ks) observations. They derive a
total X-ray luminosity of $\sim 3\times 10^{38}$\,erg~s$^{-1}$
emerging from a soft thermal ($T\simeq 7\times10^{6}$\,K) and a hard
component. The hard component was attributed to emission from very hot
gas ($T\simeq 4\times10^{7}$\,K) or alternatively to non-thermal X-ray
emission. The limited quality of their data, however, made a complete
separation of the point source contribution from the diffuse emission
impossible.

\citet{mar02} published a detailed study of NGC\,1569 based on the
same {\it Chandra} data as presented here. They find that they require
a two temperature model for the hot gas in NGC\,1569: a hot component
coinciding with the disc ($T\simeq 7\times10^{6}$\,K) and a cooler
outer component ($T\simeq 3\times10^{6}$\,K) adding up to an X-ray
luminosity of $8.2\times 10^{38}$\,\lum. According to their analysis,
the X-ray emission itself is due to shocks rather than a freely
streaming wind. They also claim that the hot wind contains virtually
all the metals produced by supernova explosions during the starburst
phase which eventually will enrich the intergalactic medium.

\begin{figure*}
\centering
\includegraphics[width=14cm]{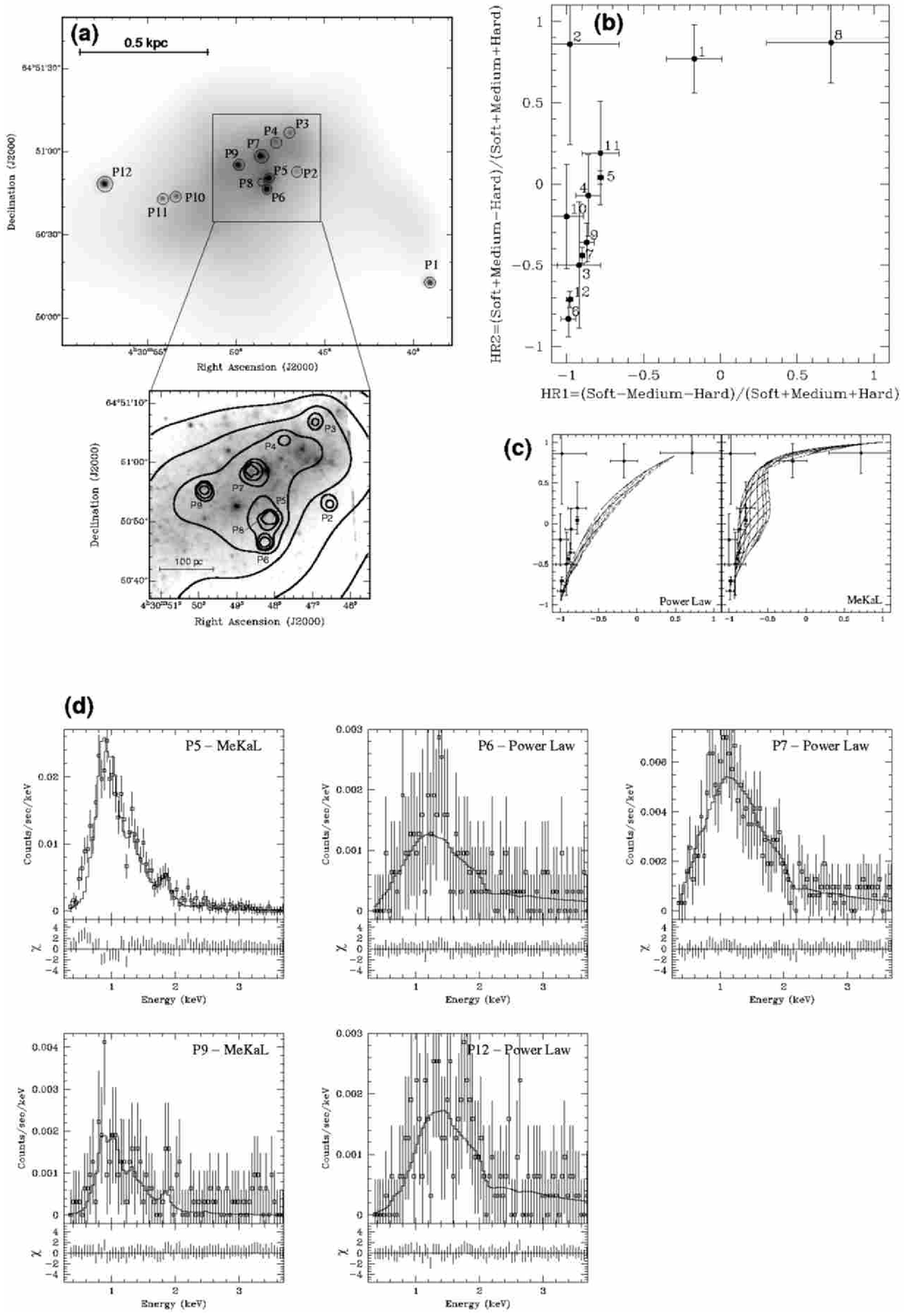}
\caption
{Locations and spectra of the point sources in NGC\,1569. {\bf (a)}
Numbering scheme of the detected sources overlaid on the adaptively
smoothed X-ray image. The blow-up displays an {\it HST}/WFPC2 F555W
(Johnson $V$) band image of the central region of NGC\,1569. Overlaid
are contours from the image above. {\bf (b)} A hardness ratio plot of
these sources. {\bf (c)} The same sources as plotted in (b) but
overlaid on the model grids for a range of power law ({\it left
panel}) and MeKaL thermal plasma ({\it right panel}; using the \hii\
region metallicity listed in Table\,\ref{tab:sample_new}) models
presented in Fig.\,\ref{fig:hardnessshow}. {\bf (d)} X-ray spectra and
the best-fitting models for point sources with counts in at least
three spectral bins with four counts minimum per bin. Normalised
residuals $\chi=(data-model)/error$ (the error is the uncertainty per
energy bin) are plotted below the spectra. For the model selection see
Sect.\,\ref{sec:diffuse_yes} and for the count rates, hardness ratios
and fitting parameters Tables\,\ref{tab:point1569a} and
\ref{tab:point1569b}).}
\label{fig:point1569}
\end{figure*}

\subsubsection{{\it Chandra} observations revisited}

The total X-ray emission of NGC\,1569 extends over $\sim 5\farcm0
\times 3\farcm3$ which corresponds to about $3.2\times 2.1$\,kpc (see
Fig.\,\ref{fig:olay1569}). Twelve point sources were detected (see
Sect.\,\ref{sec:ch5:chandra_general}) within the optical and H$\alpha$
extent. Eight of these sources are located close to the centre of
NGC\,1569 (P2 to P9; see Fig.\,\ref{fig:point1569}[a]), three to the
east (P10, P11, and P12), and one to the south-west (P1). The
extracted count rates of the point sources in the different X-ray
bands and the corresponding hardness ratios are listed in
Table\,\ref{tab:point1569a}. By far the strongest source is P5, which
is located slightly south of the central stellar disc of NGC\,1569. An
overlay with an {\it HST}/WFPC2 F555W (corresponding to Johnson $V$)
band image shows that bright stars coincide with the X-ray sources P7
and P9. Source P1 is not in the field of view of the WFPC2
image. Except for P4 and P12, all other objects are located at
distances $\la 1\arcsec$ to stars and therefore might be related to
them. A hardness ratio plot of all X-ray point sources is shown in
Fig.\,\ref{fig:point1569}(b) and \ref{fig:point1569}(c) and their
spectra with the fitting models are shown in panel (d) of the same
figure. Preliminary thermal plasma emission models provide reasonable
fits to the spectra of seven sources (P1, P2, P4, P5, P9, P10, P11,
see the corresponding hardness ratio plot) and therefore they are
likely SNR candidates or bubbles filled with hot, coronal gas. Four
source spectra are well fit by power law models (likely from XRBs; P3,
P6, P7, P12) and one spectrum (P8) is a supersoft source for which a
black body model is appropriate ($T\simeq 6\times
10^{5}$\,K). However, only five sources have spectra with a minimum of
four counts in at least three spectral bins (P5, P6, P7, P9, P12) --
our limit for which we performed final spectral fits. The resulting
models and their parameters are listed in Table\,\ref{tab:point1569b}
and displayed in Fig.\,\ref{fig:point1569}(d). Note that \citet{mar02}
detect two more point sources in NGC\,1569, based on a somewhat
different detection method. They only fitted models to the two
brightest sources: P5 and P7. Given the different model fitting
parameters and source extraction properties with respect to ours
(metallicity, source extraction region, background subtraction) their
results for those two point sources are in good agreement with the
fits presented here. All X-ray point source luminosities are below the
$10^{39}$\,\lum\ threshold for ULXs.


\begin{table}
\centering
\caption
{Results of fits of exponential functions to the azimuthally 
  averaged X-ray surface brightness profiles in the different bands ($h$:
  exponential scale length in pc).}
\begin{tabular}{@{}lccccc@{}}

\hline

\multicolumn{1}{l}{}&\multicolumn{1}{c}{$h_{\rm Total}$}&\multicolumn{1}{c}{$h_{\rm Soft}$}&\multicolumn{1}{c}{$h_{\rm Medium}$}&\multicolumn{1}{c}{$h_{\rm Hard}$}\\
\hline

NGC\,1569 &  $208\pm  \phn6$ & $229\pm 14$     & $211\pm  \phn6$ & $212\pm 15$     \\
NGC\,3077 & $186\pm 10$     &  $296\pm 49$    &  $200\pm  \phn8$  & $135\pm 21$    \\
NGC\,4214 & $145\pm 12$    & $127\pm 19$   &  $148\pm 21$ &  $187\pm 60$    \\
NGC\,4449 & $579\pm 15$ & $590\pm 22$  &  $505\pm 15$ &  $623\pm 68$  \\
NGC\,5253 & $109\pm \phn9$         & $124\pm 13$         &  $101\pm \phn8$         & $ \phn60\pm \phn6$ \\
He\,2-10 &  $186\pm 15$ & $273\pm 37$  & $166\pm 11$ & $125\pm \phn9$\\

\hline
\end{tabular}
\label{tab:xprofile_fit}
\end{table}

The azimuthally averaged surface brightness ($\mu_{\rm X}$) profiles
of the diffuse X-ray emission in all broad bands are well described by
exponential declines (see Fig.\,\ref{fig:olay1569}[f] for the total
band). As listed in Table\,\ref{tab:xprofile_fit}, the scale lengths of
the profiles of all bands are approximately the same: $\sim
250$\,pc. The X-ray distribution, however, is not spherical and in
general follows the
\ha\ morphology (Fig.\,\ref{fig:olay1569}[b]). In particular, the
prominent \ha\ arm to the west of NGC\,1569 is observed in broad and
narrow optical wavebands as well as in soft X-rays which is slightly
offset to the east. The temperature of the hot gas at the centre
(region R1, see Fig.\,\ref{fig:olay1569} and
Table\,\ref{tab:detect_gas1569}) of NGC\,1569 ($T\simeq 7.2\times
10^{6}$\,K) is clearly higher than toward the outer regions ($T\simeq
3.5 \times 10^{6}$\,K). The second highest temperature is encountered
in the region where the \ha\ arm is located (R2). Based on the
absorbing column density within the regions we can constrain the
orientation of the disc of NGC\,1569. The hot gas emerging from the
southern regions (R2 and R3) is less absorbed than emission coming
from the opposite side. Assuming that the outflow is perpendicular to
the disc, along the z-axis, this implies that the northern half of the
disc of NGC\,1569 lies between us and the outflow and hence is the
near side, whereas the southern part of the disc points away \citep[in
agreement with][]{mar02}. The MeKaL fit to the spectrum of the total
diffuse X-ray emission results in an unabsorbed luminosity of
$4.2\times 10^{38}$\,erg~s$^{-1}$. Adding up the MeKaL X-ray
luminosity of the hot gas and the luminosities of the point sources
with reliable fits (Tables\,\ref{tab:detect_gas1569} and
\ref{tab:point1569b}) we derive a total X-ray luminosity of
$8.3\times 10^{38}$\,\lum\ (see also the total spectrum in
Fig.\,\ref{fig:olay1569}). This value is $\sim 3$ times larger than
the {\it ROSAT} results but lower than the X-ray luminosity obtained
by {\it EINSTEIN} (note that {\it ROSAT} and {\it EINSTEIN} were not
able to provide sufficient angular resolution for an independent
treatment of the individual point sources and the diffuse emission).
The mean temperature of the hot gas ($\sim 7\times 10^{6}$\,K) of the
{\it Chandra} observations as well as the central and outer regions
temperatures are in good agreement with those derived by previous
observations \citep[see][]{dce96,mar02}.


\begin{table*}
\begin{minipage}{170mm}
\centering
\caption
{Positions, count rates and hardness ratios of the point sources in
NGC\,1569. The Right Ascension and Declination are given in $^{h}$
$^{m}$ $^{s}$ and $\degr$ $\arcmin$ $\arcsec$ (J2000), respectively.
The count rates of the total, soft, medium, and hard bands are in
units of $10^{-5}$\,cts\,s$^{-1}$. See Fig.\,\ref{fig:point1569} for
the locations of the point sources coinciding with this galaxy as well
as for their distribution in a hardness ratio plot.}
\begin{tabular}{@{}lllrrrrrr@{}}
\hline

\small

No.&\multicolumn{1}{c}{RA}&\multicolumn{1}{c}{DEC}&\multicolumn{1}{c}{Total}&\multicolumn{1}{c}{Soft}&\multicolumn{1}{c}{Medium}&\multicolumn{1}{c}{Hard}&\multicolumn{1}{c}{HR1}&\multicolumn{1}{c}{HR2}\\

\hline

P1  & 04~30~39.0 & +64~50~12.8 &        $50.40\pm 9.59$ & $20.86\pm 5.39$ & $23.87\pm 6.91$ & $5.66\pm 3.89$ & $-0.17\pm 0.18$ & \multicolumn{1}{c}{$+0.77\pm 0.21$}\\
P2  & 04~30~46.5 & +64~50~52.3 &       $24.28\pm 8.81$ & $0.21\pm 3.86$  & $22.41\pm 6.49$ & $1.65\pm 4.53$ & $-0.98\pm 0.32$ & \multicolumn{1}{c}{$+0.86\pm 0.62$}\\    
P3  & 04~30~46.9 & +64~51~06.9 &       $33.38\pm 10.22$& $1.39\pm 2.41$  & $6.95\pm 6.37$  & $25.03\pm 7.62$& $-0.92\pm 0.14$ & \multicolumn{1}{c}{$-0.50\pm 0.39$}\\     
P4  & 04~30~47.8 & +64~51~03.7 &       $59.54\pm 10.90$& $4.17\pm 2.41$  & $23.47\pm 7.52$ & $31.90\pm 7.51$& $-0.86\pm 0.08$ & \multicolumn{1}{c}{$-0.07\pm 0.25$}\\   
P5   & 04~30~48.1 & +64~50~50.8 &       $1809\pm 52$    & $197\pm 17$     & $742\pm 34$     & $870\pm 35$    & $-0.78\pm 0.02$ & \multicolumn{1}{c}{$+0.04\pm 0.04$}\\
P6   & 04~30~48.3 & +64~50~46.3 &       $192.5\pm 20.4$ & $1.27\pm 5.05$  & $14.90\pm 9.79$ & $176.4\pm 17.2$& $-0.99\pm 0.05$ & \multicolumn{1}{c}{$-0.83\pm 0.11$}\\  
P7   & 04~30~48.6 & +64~50~58.3 &       $784\pm 35$     & $37.55\pm 8.91$ & $183.6\pm 18.2$ & $563\pm 29$    & $-0.90\pm 0.02$ & \multicolumn{1}{c}{$-0.44\pm 0.05$}\\ 
P8  & 04~30~48.6 & +64~50~49.4 &       $26.24\pm 9.05$ & $22.53\pm 6.47$ & $2.04\pm 5.22$  & $1.67\pm 3.58$ & $+0.72\pm 0.42$ & \multicolumn{1}{c}{$+0.87\pm 0.25$}\\     
P9   & 04~30~49.9 & +64~50~55.2 &       $193.4\pm 19.3$ & $12.46\pm 4.63$ & $49.32\pm 11.22$& $131.6\pm 15.0$& $-0.87\pm 0.05$ & \multicolumn{1}{c}{$-0.36\pm 0.12$}\\  
P10 & 04~30~53.4 & +64~50~44.3 &       $34.77\pm 7.99$ & $0.00\pm 1.97$  & $13.91\pm 5.20$ & $20.86\pm 5.73$& $-1.00\pm 0.11$ & \multicolumn{1}{c}{$-0.20\pm 0.32$}\\
P11 & 04~30~54.2 & +64~50~43.0 &       $37.66\pm 8.44$ & $4.17\pm 2.41$  & $18.17\pm 6.35$ & $15.32\pm 5.01$& $-0.78\pm 0.12$ & \multicolumn{1}{c}{$+0.19\pm 0.32$}\\     
P12  & 04~30~57.4 & +64~50~48.6 &       $312.9\pm 21.1$ & $2.78\pm 2.78$  & $43.11\pm 7.99$ & $267.0\pm 19.4$& $-0.98\pm 0.02$ & \multicolumn{1}{c}{$-0.71\pm 0.05$}\\

\hline

\end{tabular}
\label{tab:point1569a}
\end{minipage}
\end{table*}


\begin{table*}
\begin{minipage}{170mm}
\centering
\caption
{Parameters of the fitted models to the point sources in NGC\,1569
($N_{H}$: absorbing column density, $T/\gamma$: temperature (MeKaL
[MEK], black body [BB]) or power law photon index (PL, $I\propto
E^{-\gamma}$), $Norm/Ampl^{a}$: Normalisation (MEK) or Amplitude (PL, BB),
$F_{\rm X}^{\rm abs}$: absorbed flux ($0.3-8.0$\,keV), $F_{\rm X}$:
unabsorbed flux, $L_{\rm X}$: unabsorbed luminosity). Sources which do
not provide sufficient counts (at least three spectral bins with four
counts minimum) are not listed. See Fig.\,\ref{fig:point1569}(d) for
the individual spectra and fits. The last column displays the CXOU
numbering of the sources in \citet{mar02}. Errors in the fitted
parameters are conservatively estimated to be of order 30 per cent.}
\begin{tabular}{@{}lllrrrrrr@{}}
\hline

\small

No.&Model
&\multicolumn{1}{c}{$N_{H}$} &\multicolumn{1}{c}{$T/\gamma$} &\multicolumn{1}{c}{$Norm/Ampl^{a}$} & \multicolumn{1}{c}{$F_{\rm X}^{\rm abs}$ }&\multicolumn{1}{c}{$F_{\rm X}$} &\multicolumn{1}{c}{$L_{\rm X}$}\\
&&\multicolumn{1}{c}{[$10^{21}$\,\cden]} &\multicolumn{1}{c}{[$10^{6}$\,K/--]} &\multicolumn{1}{c}{[$10^{-5}$]} & \multicolumn{2}{c}{[$10^{-15}$\,\flux]} &\multicolumn{1}{c}{[$10^{37}$\,\lum]}&\multicolumn{1}{c}{CXOU}\\

\hline

P5    & MEK & \multicolumn{1}{c}{13.5  }& \multicolumn{1}{c}{6.88 }& \multicolumn{1}{c}{44.5  }& \multicolumn{1}{c}{55.03 }& \multicolumn{1}{c}{397.8 }& \multicolumn{1}{c}{23.03}&\multicolumn{1}{l}{043048.1+645050}\\
P6    & PL  & \multicolumn{1}{c}{8.54  }& \multicolumn{1}{c}{2.02 }& \multicolumn{1}{c}{0.57  }& \multicolumn{1}{c}{19.16 }& \multicolumn{1}{c}{35.13 }& \multicolumn{1}{c}{2.03}&\multicolumn{1}{l}{043048.2+645046}\\
P7    & PL  & \multicolumn{1}{c}{8.82  }& \multicolumn{1}{c}{2.49 }& \multicolumn{1}{c}{2.69  }& \multicolumn{1}{c}{52.03 }& \multicolumn{1}{c}{153.4 }& \multicolumn{1}{c}{8.88}&\multicolumn{1}{l}{043048.6+645058}\\
P9    & MEK & \multicolumn{1}{c}{17.52 }& \multicolumn{1}{c}{6.51 }& \multicolumn{1}{c}{5.64  }& \multicolumn{1}{c}{4.87  }& \multicolumn{1}{c}{49.93 }& \multicolumn{1}{c}{2.89}&\multicolumn{1}{l}{043049.8+645055}\\
P12   & PL  & \multicolumn{1}{c}{17.64 }& \multicolumn{1}{c}{2.33 }& \multicolumn{1}{c}{1.31  }& \multicolumn{1}{c}{25.78 }& \multicolumn{1}{c}{74.56 }& \multicolumn{1}{c}{4.32}&\multicolumn{1}{l}{043057.4+645048}\\

\hline

\end{tabular}
\footnotetext{$^{a}$ see Sect.\,\ref{sec:specana}} 
\label{tab:point1569b}
\end{minipage}
\end{table*}


\subsection[NGC\,3077]{{\bf NGC\,3077} (UGC\,5398)}
\label{sec5:3077}


\begin{table*}
\renewcommand{\arraystretch}{1.5}
\begin{minipage}{170mm}
\centering
\caption
{Results of MeKaL collisional thermal plasma model fits applied to the
X-ray spectra of different regions in NGC\,3077 (see panels [d] and
[e] in Fig.\,\ref{fig:olay3077} for the definition of the regions).}
\begin{tabular}{@{}lccccccc@{}}
\hline

\multicolumn{1}{l}{Region}&\multicolumn{1}{c}{$N_{\rm H}$}&\multicolumn{1}{c}{$T$}&\multicolumn{1}{c}{$Norm^{a}$}&\multicolumn{1}{c}{$F_{\rm X}^{\rm abs}$}&\multicolumn{1}{c}{$F_{\rm X}$}&\multicolumn{1}{c}{$L_{\rm X}$}&\multicolumn{1}{c}{$\chi^{2}_{red}$}\\
\multicolumn{1}{c}{}&\multicolumn{1}{c}{[$10^{21}$\,cm$^{-2}$]}&\multicolumn{1}{c}{[$10^{6}$\,K]}&\multicolumn{1}{c}{[$10^{-5}$]}&\multicolumn{1}{c}{[$10^{-15}$\,erg~s$^{-1}$\,cm$^{-2}$]}&\multicolumn{1}{c}{[$10^{-15}$\,erg~s$^{-1}$\,cm$^{-2}$]}&\multicolumn{1}{c}{[$10^{37}$\,erg~s$^{-1}$]}&\\

\hline

Total            & $4.71_{-0.11}^{+0.11}$ & $2.32_{-0.04}^{+0.04}$ & $50.7_{-2.8}^{+2.8}$ & $33.6_{-5.1}^{+5.6}$ & $1062_{-69}^{+71}$ & $165_{-26}^{+26}$ & 0.43\\

R1 (Centre)      & $0.86_{-0.28}^{+0.32}$ & $6.04_{-0.87}^{+0.72}$ & $0.39_{-0.05}^{+0.05}$ & $6.16_{-1.85}^{+1.95}$& $11.2_{-1.8}^{+1.6}$ & $1.74_{-0.37}^{+0.37}$ & 0.11\\

R2               & $4.36_{-0.15}^{+0.17}$ & $2.18_{-0.04}^{+0.04}$ & $20.9_{-1.8}^{+1.8}$ & $13.0_{-3.0}^{+3.4}$ & $420_{-41}^{+42}$ & $65.1_{-11.3}^{+11.3}$ & 0.24\\

R3               & $3.20_{-0.25}^{+0.29}$ & $3.21_{-0.19}^{+0.20}$ & $2.23_{-0.29}^{+0.28}$ & $6.76_{-2.16}^{+2.63}$& $52.1_{-7.3}^{+7.7}$ & $8.08_{-1.66}^{+1.66}$ & 0.17\\

R4               &  $6.14_{-0.40}^{+0.57}$ & $1.60_{-0.11}^{+0.05}$ & $43.0_{-11.9}^{+11.9}$ & $2.44_{-1.49}^{+2.42}$ & $644_{-224}^{+208}$ & $99.8_{-37.5}^{+37.5}$ & 0.18\\

R2 to R4 (Outer Regions)  & $4.52_{-0.13}^{+0.14}$ & $2.30_{-0.04}^{+0.04}$ & $30.5_{-2.2}^{+2.2}$ & $21.6_{-4.1}^{+4.7}$ & $637_{-53}^{+54}$ & $98.7_{-16.3}^{+16.3}$ & 0.44\\

\hline
\end{tabular}
\label{tab:detect_gas3077}
\footnotetext{$^{a}$ see Sect.\,\ref{sec:specana}} 
\end{minipage}
\end{table*}


\begin{figure*}
\centering
\includegraphics[width=15cm]{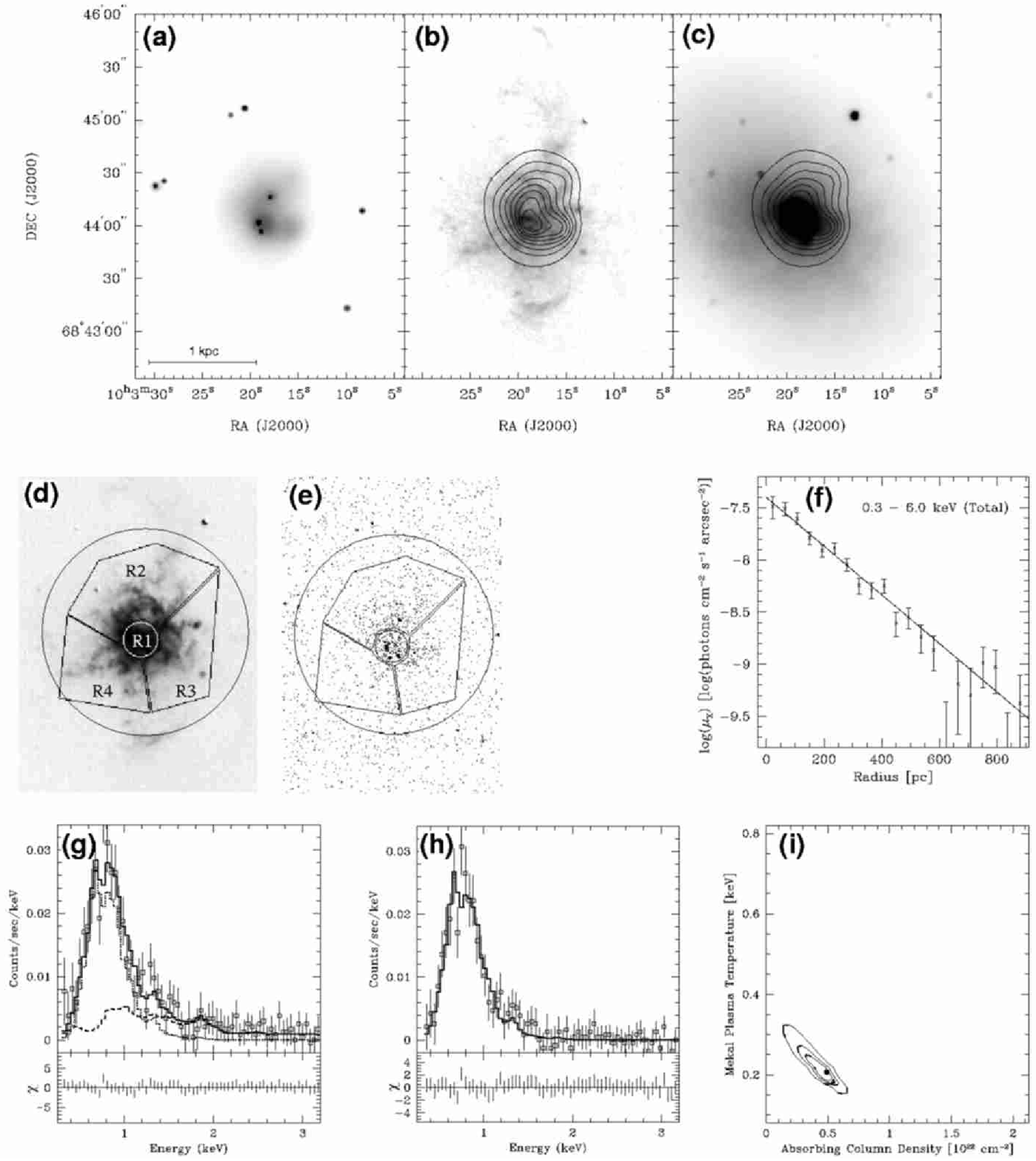}
\caption
{Images and spectra of the X-ray emission of NGC\,3077. See the
caption of Fig.\,\ref{fig:olay1569} for details. The contours of the
logarithmic X-ray emission in panels (b) and (c) are spaced by
$0.1\log(peak flux)$ and start at $0.2\log(peak flux)$ (continuum
subtracted \ha\ image taken from \citealt{mar97}, and optical Calar
Alto 2.2m $R$--band image from \citealt{wal02}).}
\label{fig:olay3077}
\end{figure*}

\subsubsection{Previous X-ray observations and results}
The first X-ray observations of NGC\,3077 were performed by the {\it
EINSTEIN} satellite and resulted in an upper limit for its X-ray
luminosity of $<4\times 10^{38}$\,\lum\ \citep{fab92}. {\it ROSAT}
PSPC data (7\,ks) were analysed by \citet{bi94}. They found evidence
for extended X-ray emission in NGC\,3077. A fit of an RS thermal
plasma model to their X-ray spectrum revealed a temperature of $\sim
7\times 10^{6}$\,K and a luminosity of $2\times 10^{38}$\,\lum.
\citet{ott03} published a paper on the same {\it Chandra} data that 
are presented here. Their main results are that individual,
prominent \ha\ shells appear to be filled with X-ray emitting, hot
thermal plasma (temperatures ranging from $\sim 1$ to $5\times
10^{6}$\,K). The plasma stored in the shells has pressures of $\sim
10^{5-6}$\,K\,cm$^{-3}$ which is most likely what drives their
expansion. The total X-ray luminosity was determined to be in the
range of $2-5\times 10^{39}$\,\lum\ depending on the plasma model
used. About 85 per cent of this emission can be attributed to the
diffuse, hot thermal plasma and the remainder is emitted by six point
sources (two XRBs, three SNR candidates, and a supersoft source).

\subsubsection{{\it Chandra} observations revisited}

\begin{figure*}
\centering
\includegraphics[width=13cm]{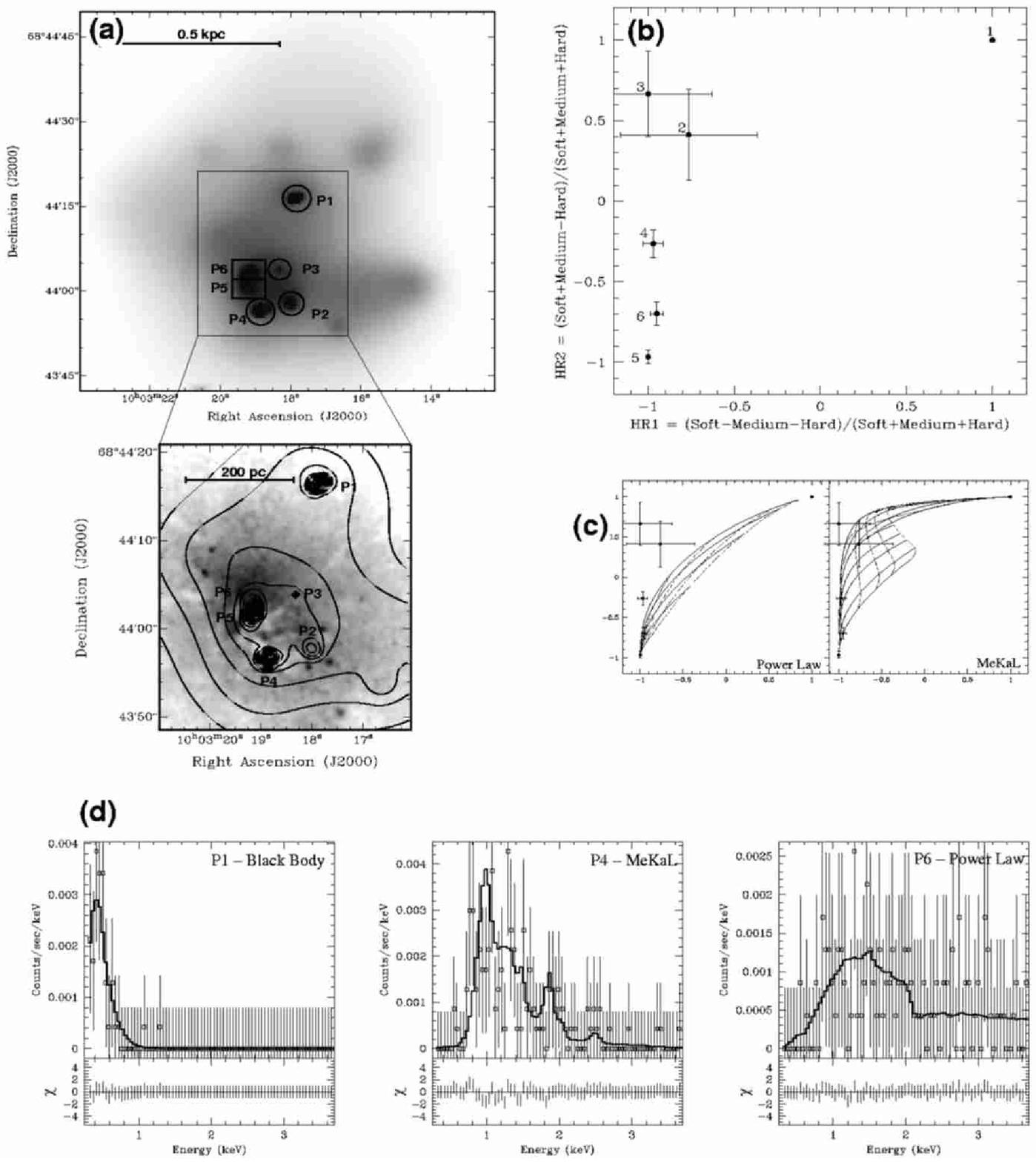}
\caption
{Locations and spectra of the point sources in NGC\,3077. See caption
of Fig.\,\ref{fig:point1569} for details. The {\it HST}/WPFC2 image
shown in the blow-up of panel (a) was obtained in the F814W
filter. The count rates, hardness ratios and model parameters are
listed in Tables\,\ref{tab:point3077a} and \ref{tab:point3077b}.  }
\label{fig:point3077}
\end{figure*}

The diffuse X-ray emission of NGC\,3077 has an extent of $\sim
1\arcmin$ which corresponds to about 1\,kpc (see
(Fig.\,\ref{fig:olay3077}[a]). Six point sources are detected within
the optical and \ha\ extent of NGC\,3077 (Fig.\,\ref{fig:point3077}):
Three sources (P2, P3, and P4) are well modeled by thermal plasmas and
are likely young supernova remnants or small bubbles filled with hot
gas. Two (P5 and P6) show power law spectra and one (P1) is a
supersoft source in the outskirts of NGC\,3077. Only P1, P4, and P6,
however, exceed our threshold for spectral fitting (see
Tables\,\ref{tab:point3077a} and \ref{tab:point3077b} for the point
source properties). A comparison with optical {\it HST}/WFPC2 data
(Fig.\,\ref{fig:point3077}[a]) shows that a bright star is located
just between the two X-ray sources P5 and P6. Given the absolute
positional uncertainties of the {\it HST} and {\it Chandra}
observations, this could be the donor star in the XRB scenario for one
of the sources. P5 and P6, however, are only $\sim 1\arcsec$ apart and
it is therefore not possible to clearly attribute P5 or P6 to this
optical counterpart. \citet{ott03} argue that P5 might also be a
background AGN given the high absorbing column density along this line
of sight. Close to the centre of NGC\,3077 an absorption feature
stretches toward the south-west and the point of strongest absorption
coincides with the X-ray point source P4. No obvious optical features
can be detected at the positions of P1, P2, and P3. A more detailed
analysis of the X-ray point sources is given in
\citet{ott03}.

The morphology of the diffuse X-ray emission of NGC\,3077 is similar
to its \ha\ distribution on a global scale
(Fig.\,\ref{fig:olay3077}[b]). Viewed on smaller scales, however, it
turns out that \ha\ filaments surround certain areas of diffuse X-ray
emission and we can attribute this morphology to expanding
superbubbles filled with hot coronal gas \citep[for a characterisation
of the individual superbubbles, see][]{ott03}. An exponential function
provides a good fit to the azimuthally averaged surface brightness
profiles of the diffuse X-ray emission
(Fig.\,\ref{fig:olay3077}[f]). From Table\,\ref{tab:xprofile_fit} we
derive that the soft emission (scale length $h_{\rm
Soft}\simeq300$\,pc) is about twice as extended as compared to the
hard emission ($h_{\rm Hard}\simeq 140$\,pc). Hardening of the X-ray
emission due to photoelectric absorption can be made responsible for
at least part of this effect (see the prominent optical absorption
features in Fig.\,\ref{fig:point3077}[a]). The fits of thermal plasma
models to spectra which were obtained from individual regions
(Figs.\,\ref{fig:olay3077}[d] and [e]) show some temperature gradient
from the centre toward the outer regions
(Table\,\ref{tab:detect_gas3077}). The orientation of NGC\,3077 as
traced by photoelectric absorption cannot be constrained based on our
analysis. The total, diffuse X-ray luminosity is $1.7\times
10^{39}$\,\lum; adding the point source luminosities result in
$1.9\times10^{39}$\,\lum. The X-ray luminosities as derived from {\it
ROSAT} and {\it EINSTEIN} data are one order of magnitude lower
\citep{bi94}; the likely reason for this discrepancy is an 
approximately seven times lower absorbing column density used in their
fits. In addition, \citet{bi94} use a bremsstrahlung spectrum rather
than a thermal plasma model and were not able to subtract the X-ray
point sources.

\begin{table*}
\begin{minipage}{170mm}
\centering
\caption
{Positions, count rates and hardness ratios of the point sources in
NGC\,3077. The Right Ascension and Declination are given in $^{h}$
$^{m}$ $^{s}$ and $\degr$ $\arcmin$ $\arcsec$ (J2000), respectively.
The count rates of the total, soft, medium, and hard emission are in
units of $10^{-5}$\,cts\,s$^{-1}$. See Fig.\,\ref{fig:point3077} for
the locations of the point sources coinciding with this galaxy as well
as for their distribution in a hardness ratio plot.}
\begin{tabular}{@{}lllrrrrrr@{}}
\hline

\small

No.&\multicolumn{1}{c}{RA}&\multicolumn{1}{c}{DEC}&\multicolumn{1}{c}{Total}&\multicolumn{1}{c}{Soft}&\multicolumn{1}{c}{Medium}&\multicolumn{1}{c}{Hard}&\multicolumn{1}{c}{HR1}&\multicolumn{1}{c}{HR2}\\

\hline
P1  & 10~03~17.8 & +68~44~16.0 &       $78.60\pm 12.97$& $74.86\pm 12.69$  & $1.87\pm 1.87$ & $1.87\pm 1.87$& $+0.90\pm 0.07$ & \multicolumn{1}{c}{$+0.95\pm 0.05$}\\
P2   & 10~03~17.9 & +68~43~57.3 &       $31.81\pm 10.75$    & $3.74\pm 5.92$     & $18.71\pm 7.48$     & $9.35\pm 4.95$    & $-0.76\pm 0.33$ & \multicolumn{1}{c}{$+0.41\pm 0.53$}\\
P3   & 10~03~18.3 & +68~44~03.8 &       $50.53\pm 9.72$ & $5.61\pm 3.24$  & $35.56\pm 8.15$ & $9.36\pm 4.18$& $-0.78\pm 0.12$ & \multicolumn{1}{c}{$+0.63\pm 0.31$}\\  
P4  & 10~03~18.8 & +68~43~56.4 &        $248.9\pm 21.7$ & $3.74\pm 3.74$ & $87.96\pm 12.83$ & $157.2\pm 17.2$ & $-0.97\pm 0.03$ & \multicolumn{1}{c}{$-0.26\pm 0.11$}\\
P5  & 10~03~19.1 & +68~44~01.4 &       $213.3\pm 21.0$ & $0.00\pm 0.00$  & $3.74\pm 4.58$ & $209.6\pm 20.5$ & $-1.00\pm 0.00$ & \multicolumn{1}{c}{$-0.96\pm 0.04$}\\    
P6  & 10~03~19.1 & +68~44~02.3 &       $222.7\pm 21.1$& $5.61\pm 3.24$  & $28.07\pm 8.16$  & $189.0\pm 19.2$& $-0.95\pm 0.02$ & \multicolumn{1}{c}{$-0.70\pm 0.08$}\\

\hline

\end{tabular}
\label{tab:point3077a}
\end{minipage}
\end{table*}


\begin{table*}
\begin{minipage}{170mm}
\centering
\caption
{Parameters of the fitted X-ray emission models to the point sources
in NGC\,3077. See the caption of Table\,\ref{tab:point1569b} for
details and Fig.\,\ref{fig:point3077}(d) for the individual spectra
and fits. Note that contrary to the tables in \citet{ott03} the
absorbing column densities listed here represent internal absorption
of NGC\,3077 only. The last column lists the corresponding source
notation in \citet{ott03}.}

\begin{tabular}{@{}lllrrrrrr@{}}
\hline
\small

No.&Model&\multicolumn{1}{c}{$N_{H}$} &\multicolumn{1}{c}{$T/\gamma$} &\multicolumn{1}{c}{$Norm/Ampl^{a}$} & \multicolumn{1}{c}{$F_{\rm X}^{\rm abs}$ }&\multicolumn{1}{c}{$F_{\rm X}$} &\multicolumn{1}{c}{$L_{\rm X}$}\\
&&\multicolumn{1}{c}{[$10^{21}$\,\cden]} &\multicolumn{1}{c}{[$10^{6}$\,K/--]} &\multicolumn{1}{c}{[$10^{-5}$]} & \multicolumn{2}{c}{[$10^{-15}$\,\flux]} &\multicolumn{1}{c}{[$10^{37}$\,\lum]}&Cross Ident.\\

\hline
P1   & BB & \multicolumn{1}{c}{0.0 }& \multicolumn{1}{c}{0.94 }& \multicolumn{1}{c}{2022  }& \multicolumn{1}{c}{2.88  }& \multicolumn{1}{c}{5.92  }& \multicolumn{1}{c}{0.92}&S4\\
P4   & MEK & \multicolumn{1}{c}{8.77 }& \multicolumn{1}{c}{9.27 }& \multicolumn{1}{c}{3.28  }& \multicolumn{1}{c}{10.44  }& \multicolumn{1}{c}{85.75 }& \multicolumn{1}{c}{13.30}&S1\\
P6   & PL  & \multicolumn{1}{c}{2.93 }& \multicolumn{1}{c}{1.00}& \multicolumn{1}{c}{3.77 }& \multicolumn{1}{c}{58.05}& \multicolumn{1}{c}{65.14 }& \multicolumn{1}{c}{10.10}&S3\\

\hline

\normalsize

\end{tabular}
\footnotetext{$^{a}$ see Sect.\,\ref{sec:specana}} 
\label{tab:point3077b}
\end{minipage}
\end{table*}


\subsection[NGC\,4214]{{\bf NGC\,4214} (NGC\,4228, UGC\,7278)}
\label{sec5:4214}

\subsubsection{Previous X-ray observations and results}
{\it EINSTEIN} was the first X-ray telescope to observe
NGC\,4214 and \citet{fab92} derived an upper limit to the X-ray
luminosity of $1.5\times 10^{40}$\,\lum. {\it ROSAT} observations were
performed with the HRI instrument for $\sim 43$\,ks \citep{rob00} as
part of a large survey. They detected a point-like nuclear source with
an X-ray luminosity of $1.8\times10^{38}$\,\lum. Recently,
\citet{har04} published a more detailed analysis of the same 
Chandra data as discussed here. They fitted a two-temperature thermal
plasma with $T=1.6$ and $6.0\times10^{6}$\,K to the diffuse X-ray
emission.


\begin{table*}
\renewcommand{\arraystretch}{1.5}
\begin{minipage}{170mm}
\centering
\caption
{Results of MeKaL collisional thermal plasma model fits applied to the
X-ray spectra of different regions in NGC\,4214 (see panels [d] and
[e] in Fig.\,\ref{fig:olay4214} for the definition of the regions).
}
\begin{tabular}{@{}lccccccc@{}}
\hline

\multicolumn{1}{l}{Region}&\multicolumn{1}{c}{$N_{\rm H}$}&\multicolumn{1}{c}{$T$}&\multicolumn{1}{c}{$Norm^{a}$}&\multicolumn{1}{c}{$F_{\rm X}^{\rm abs}$}&\multicolumn{1}{c}{$F_{\rm X}$}&\multicolumn{1}{c}{$L_{\rm X}$}&\multicolumn{1}{c}{$\chi^{2}_{red}$}\\
\multicolumn{1}{c}{}&\multicolumn{1}{c}{[$10^{21}$\,cm$^{-2}$]}&\multicolumn{1}{c}{[$10^{6}$\,K]}&\multicolumn{1}{c}{[$10^{-5}$]}&\multicolumn{1}{c}{[$10^{-15}$\,erg~s$^{-1}$\,cm$^{-2}$]}&\multicolumn{1}{c}{[$10^{-15}$\,erg~s$^{-1}$\,cm$^{-2}$]}&\multicolumn{1}{c}{[$10^{37}$\,erg~s$^{-1}$]}&\\

\hline

Total            & $9.87_{-0.52}^{+0.59}$ & $2.17_{-0.06}^{+0.06}$ & $129_{-14}^{+14}$ & $37.9_{-10.6}^{+12.9}$ & $759_{-95}^{+98}$ & $75.9_{-14.5}^{+14.5}$& 0.14\\

R1 (Centre)           & $10.1_{-1.1}^{+1.4}$ & $2.80_{-0.28}^{+0.27}$ & $22.7_{-5.3}^{+5.3}$ & $12.5_{-6.7}^{+9.8}$ & $155_{-41}^{+44}$ & $15.5_{-4.9}^{+4.9}$ & 0.04\\

R2 (North)              & $14.4_{-0.9}^{+1.2}$ & $1.64_{-0.08}^{+0.07}$ & $300_{-71}^{+71}$ & $11.7_{-6.5}^{+10.0}$ & $1321_{-377}^{+397}$ & $132_{-44}^{+44}$ & 0.10\\

R3 (South)              & $2.99_{-0.85}^{+1.13}$ & $2.60_{-0.33}^{+0.33}$ & $6.11_{-1.45}^{+1.45}$ & $12.2_{-6.7}^{+9.2}$ & $40.8_{-12.5}^{+12.1}$ & $4.08_{-1.38}^{+1.38}$ & 0.07\\

R2+R3 (Outer Regions)  & $2.91_{-0.56}^{+0.67}$ & $2.68_{-0.24}^{+0.21}$ & $12.2_{-1.8}^{+1.8}$ & $25.8_{-9.6}^{+11.3}$ & $82.3_{-15.6}^{+14.9}$ & $8.23_{-1.95}^{+1.95}$ & 0.14\\

\hline
\end{tabular}
\footnotetext{$^{a}$ see Sect.\,\ref{sec:specana}} 
\label{tab:detect_gas4214}
\end{minipage}
\end{table*}

\begin{figure*}
\centering
\includegraphics[width=15cm]{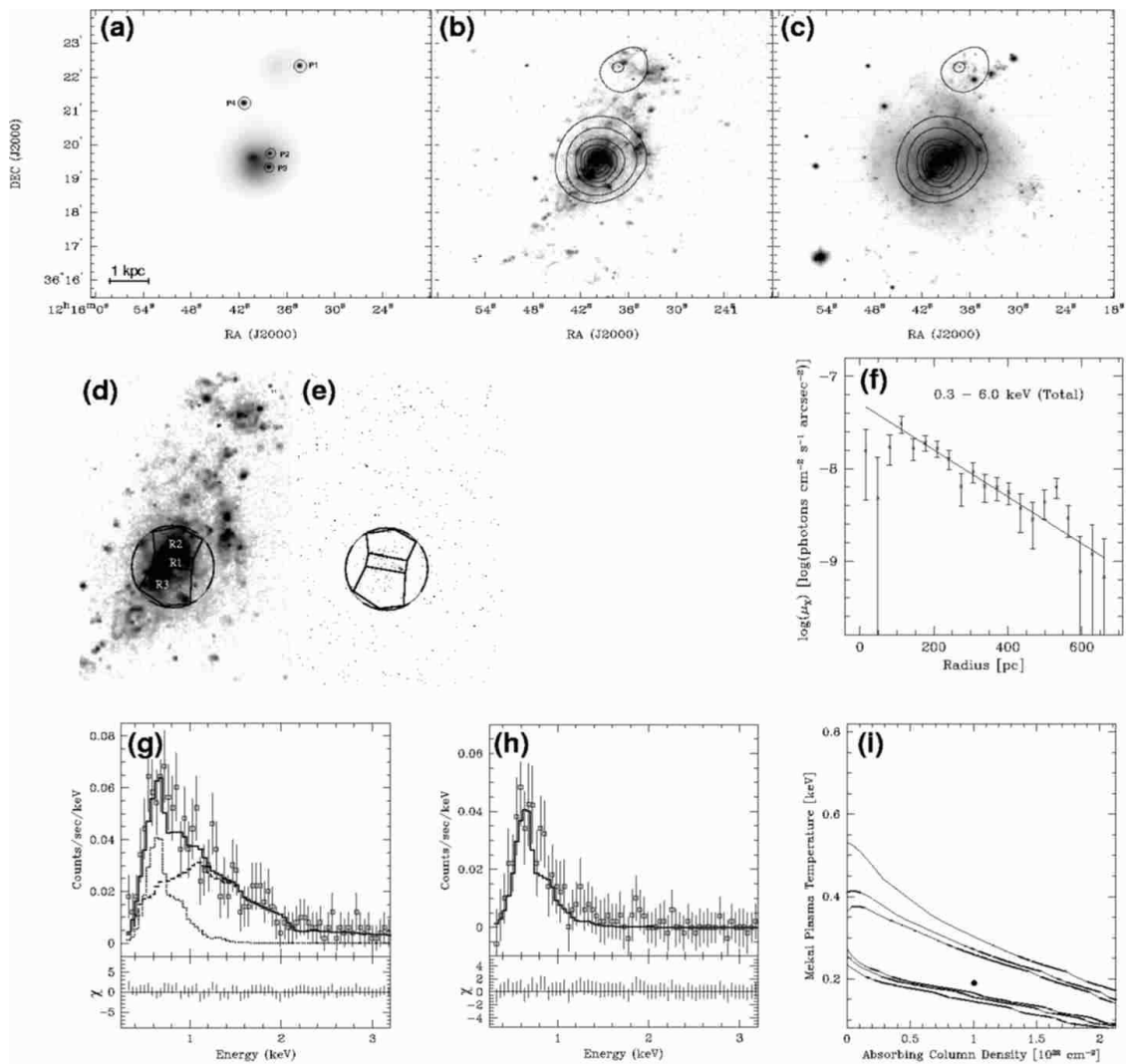}
\caption
{Images and spectra for the X-ray emission of NGC\,4214. See the
caption of Fig.\,\ref{fig:olay1569} for details. The outermost contour
of the logarithmic X-ray emission displays the $0.15\log(peak
flux)$. The other contours are spaced by $0.1\log(peak flux)$ starting
at a $0.2\log(peak flux)$ level \citep[$B$--band and continuum
subtracted \ha\ images taken from][]{wal01}.}
\label{fig:olay4214}
\end{figure*}

\begin{figure*}
\centering
\includegraphics[width=15cm]{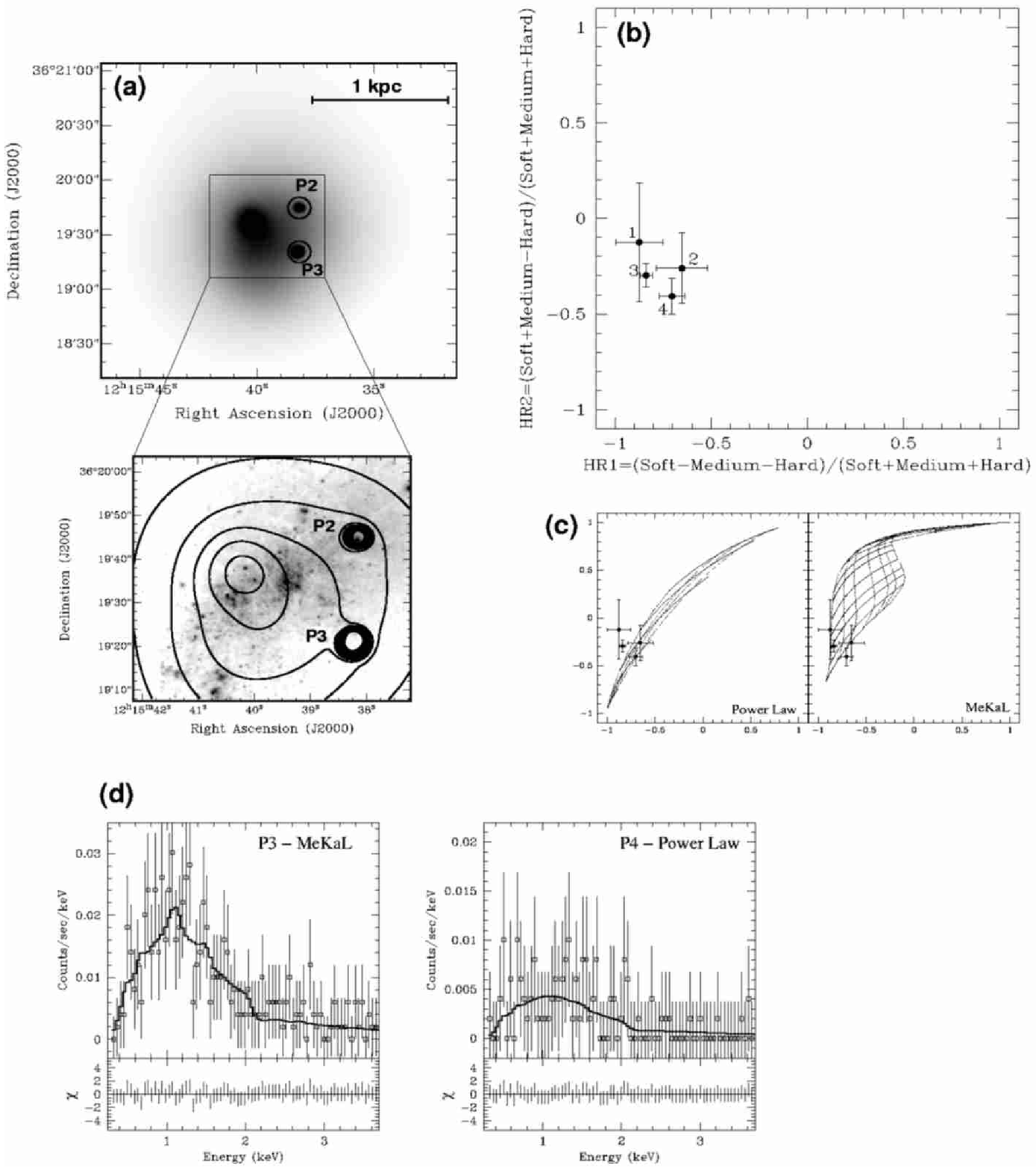}
\caption
{Locations and spectra of the point sources in NGC\,4214. See caption
  of Fig.\,\ref{fig:point1569} for details. The count rates,
  hardness ratios and fitted parameters are listed in
  Tables\,\ref{tab:point4214a} and \ref{tab:point4214b}. }
\label{fig:point4214}
\end{figure*}

\subsubsection{{\it Chandra} observations revisited}

The extent of the diffuse X-ray emission centred on NGC\,4214 is about
$1\farcm3$ (corresponding to 1\,kpc) and slightly elongated along the
north-south axis (see Fig.\,\ref{fig:olay4214}[a]). A second, smaller
X-ray emitting region is found close to an \hii\ region at the
northern tip of this galaxy (Fig.\,\ref{fig:olay4214}[a] and [b]),
which, however, is not significant and might be an artefact from the
adaptive smoothing process (note that this feature cannot be seen when
smoothed with a Gaussian kernel of fixed size). The X-ray emitting
region is much smaller than the \ha\ distribution ($6\farcm8 \times
3\farcm2$).

Four point sources are detected coinciding with the diffuse X-ray and
\ha\ emission of NGC\,4214. A comparison with {\it HST}/WFPC2 F555W
data shows that P2 and P4 coincide with the location of bright
stars. No optical counterpart can be found for P3 (with a count rate
of $1.2\times 10^{-2}$\,cts\,s$^{-1}$ it is by far the strongest
source) and P1 is not in the WFPC2 field of view (see
Fig.\,\ref{fig:point4214}). All point sources, however, are located on
or close to \hii\ regions. In particular, P2 and P3 are near the
centre of NGC\,4214 where the current star formation rate is
highest. The positions, count rates and hardness ratios of the point
sources are listed in Table\,\ref{tab:point4214a}. For P3 and P4,
where enough counts were accumulated, the models which provided the
best fits are a thermal plasma and a power law, respectively (see
Fig.\,\ref{fig:point4214}[b], [c], and [d]). Their parameters are
listed in Table\,\ref{tab:point4214b}. The strongest source, P3, is
most likely identical to the source observed by {\it ROSAT} and its
luminosity is with $2.6\times 10^{38}$\,\lum\ in good agreement with
the {\it ROSAT} measurements.

The azimuthally averaged, diffuse X-ray emission is well fit by an
exponential law with similar scale lengths in all bands ($\sim145$\,pc,
see Table\,\ref{tab:xprofile_fit} and Fig.\,\ref{fig:olay4214}[f]) but
shows the X-ray surface brightness flattening off toward the centre
and a local maximum at large radii. The temperature of the hot gas
($T\simeq 2.2\times10^{6}$\,K -- somewhat higher as compared to the
low temperature component fitted by \citealt{har04}) does not show any
variation over the field of view. The northern portion of the diffuse
X-ray emission, however, is about an order of magnitude more absorbed
than the south (see Table\,\ref{tab:detect_gas4214}). Therefore, the
north of the disc of NGC\,4214 is directed toward the observer and
shadows the X-ray emitting superwind whereas the southern part of the
disc points away and is situated behind the hot gas. The total X-ray
luminosity of the diffuse component is determined to $7.7\times
10^{39}$\,\lum\ (see also Fig.\,\ref{fig:olay4214}(h); note that the
errors of the fits are quite large as can be seen by the size of the
confidence regions in Fig.\,\ref{fig:olay4214}[i]). The total X-ray
emission (luminosity: $8.5\times 10^{39}$\,\lum; point sources plus
MeKaL diffuse emission) is shown in Fig.\,\ref{fig:olay4214}(g) and is
dominated by the flux of the point sources, in contrast to the other
galaxies in our sample.

\begin{table*}
\begin{minipage}{170mm}
\centering
\caption
{Positions, count rates and hardness ratios of the point sources in
NGC\,4214. The Right Ascension and Declination are given in $^{h}$
$^{m}$ $^{s}$ and $\degr$ $\arcmin$ $\arcsec$ (J2000), respectively.
The count rates of the total, soft, medium, and hard emission are in
units of $10^{-5}$\,cts\,s$^{-1}$. See Fig.\,\ref{fig:point4214} for
the locations of the point sources coinciding with this galaxy as well
as for their distribution in a hardness ratio plot.}
\begin{tabular}{@{}lllrrrrrr@{}}
\hline

No.&\multicolumn{1}{c}{RA}&\multicolumn{1}{c}{DEC}&\multicolumn{1}{c}{Total}&\multicolumn{1}{c}{Soft}&\multicolumn{1}{c}{Medium}&\multicolumn{1}{c}{Hard}&\multicolumn{1}{c}{HR1}&\multicolumn{1}{c}{HR2}\\

\hline

P1  & 12~15~34.4 & +36~22~20.5 & $140.9\pm 39.4$ & $8.81\pm 8.81$ & $52.8\pm 21.6$ &  $79.3\pm 31.8$ & $-0.88\pm 0.12$ & $-0.13\pm 0.31$\\
P2  & 12~15~38.2 & +36~19~45.1 & $405.1\pm 64.7$ & $70.5\pm 30.5$ & $79.3\pm 31.8 $ & $255.4\pm 47.4$ & $-0.65\pm 0.13$ & $-0.26\pm 0.18$\\
P3  & 12~15~38.2 & +36~19~21.4 & $3162\pm 171 $ & $255.4\pm 55.0$ & $854.3\pm 87.6$  &  $2052\pm 137$ & $-0.84\pm 0.03$ & $-0.30\pm 0.06$\\
P4  & 12~15~41.4 & +36~21~14.9 & $951.2\pm 91.5$ & $140.9\pm 35.2$   & $140.9\pm 35.2$  &  $669.3\pm 76.8$ & $-0.70\pm 0.07$ & $-0.41\pm 0.09$\\

\hline

\end{tabular}
\label{tab:point4214a}
\end{minipage}
\end{table*}


\begin{table*}
\begin{minipage}{170mm}
\centering
\caption
{Parameters of the fitted X-ray emission models to the point sources
in NGC\,4214. See the caption of Table\,\ref{tab:point1569b} for
details and Fig.\,\ref{fig:point4214}(d) for the individual spectra
and fits. The last column lists the source number allocated by
\citet{har04}.}
\begin{tabular}{@{}lllrrrrrr@{}}
\hline

No.&Model &\multicolumn{1}{c}{$N_{H}$} &\multicolumn{1}{c}{$T/\gamma$} &\multicolumn{1}{c}{$Norm/Ampl^{a}$} & \multicolumn{1}{c}{$F_{\rm X}^{\rm abs}$ }&\multicolumn{1}{c}{$F_{\rm X}$} &\multicolumn{1}{c}{$L_{\rm X}$}&\\
&&\multicolumn{1}{c}{[$10^{21}$\,\cden]} &\multicolumn{1}{c}{[$10^{6}$\,K/--]} &\multicolumn{1}{c}{[$10^{-5}$]} & \multicolumn{2}{c}{[$10^{-15}$\,\flux]} &\multicolumn{1}{c}{[$10^{37}$\,\lum]}&Cross Ident.\\

\hline


P3   & MEK  & \multicolumn{1}{c}{4.36 }& \multicolumn{1}{c}{39.4 }& \multicolumn{1}{c}{18.82   }& \multicolumn{1}{c}{186}& \multicolumn{1}{c}{262}& \multicolumn{1}{c}{26.36}&10\\ 
P4   & PL  & \multicolumn{1}{c}{5.87 }& \multicolumn{1}{c}{1.93}& \multicolumn{1}{c}{1.39}& \multicolumn{1}{c}{157.62 }& \multicolumn{1}{c}{89.70 }& \multicolumn{1}{c}{9.03}&15\\

\hline
\end{tabular}
\footnotetext{$^{a}$ see Sect.\,\ref{sec:specana}} 
\label{tab:point4214b}
\end{minipage}
\end{table*}



\subsection[NGC\,4449]{{\bf NGC\,4449} (UGC\,7592)}
\label{sec5:4449}


\begin{table*}
\renewcommand{\arraystretch}{1.5}
\begin{minipage}{170mm}
\centering
\caption
{Results of MeKaL collisional thermal plasma model fits applied to the
X-ray spectra of different regions in NGC\,4449 (see panels [d] and
[e] in Fig.\,\ref{fig:olay4449} for the definition of the regions).
}
\begin{tabular}{@{}lccccccc@{}}
\hline

\multicolumn{1}{l}{Region}&\multicolumn{1}{c}{$N_{\rm H}$}&\multicolumn{1}{c}{$T$}&\multicolumn{1}{c}{$Norm^{a}$}&\multicolumn{1}{c}{$F_{\rm X}^{\rm abs}$}&\multicolumn{1}{c}{$F_{\rm X}$}&\multicolumn{1}{c}{$L_{\rm X}$}&\multicolumn{1}{c}{$\chi^{2}_{red}$}\\
\multicolumn{1}{c}{}&\multicolumn{1}{c}{[$10^{21}$\,cm$^{-2}$]}&\multicolumn{1}{c}{[$10^{6}$\,K]}&\multicolumn{1}{c}{[$10^{-5}$]}&\multicolumn{1}{c}{[$10^{-15}$\,erg~s$^{-1}$\,cm$^{-2}$]}&\multicolumn{1}{c}{[$10^{-15}$\,erg~s$^{-1}$\,cm$^{-2}$]}&\multicolumn{1}{c}{[$10^{37}$\,erg~s$^{-1}$]}&\\

\hline

Total            & $7.19_{-0.10}^{+0.10}$ & $2.98_{-0.03}^{+0.03}$ & $353_{-7}^{+7}$ & $393_{-20}^{+21}$ & $2481_{-55}^{+55}$ & $452_{-65}^{+65}$ & 1.07\\

R1 (Centre)      & $7.87_{-0.22}^{+0.23}$ & $3.12_{-0.07}^{+0.07}$ & $71_{-3}^{+3}$ & $75.9_{-8.5}^{+9.2}$ & $510_{-26}^{+26}$ & $92.8_{-14.0}^{+14.0}$ & 0.26\\

R2               & $1.08_{-0.23}^{+0.25}$ & $4.43_{-0.22}^{+0.23}$ & $9.93_{-0.51}^{+0.51}$ & $56.8_{-7.9}^{+8.7}$ & $84.1_{-6.5}^{+6.8}$ & $15.3_{-2.5}^{+2.5}$ & 0.23 \\

R3               & $9.66_{-0.24}^{+0.25}$ & $2.91_{-0.06}^{+0.06}$ & $99.7_{-4.6}^{+4.6}$ & $68.5_{-8.5}^{+9.1}$ & $694_{-37}^{+38}$ & $126_{-19}^{+19}$ & 0.40\\

R4               & $7.89_{-0.30}^{+0.32}$ & $2.68_{-0.07}^{+0.07}$ & $46.7_{-3.0}^{+3.0}$ & $37.7_{-6.3}^{+7.0}$ & $315_{-23}^{+24}$ & $57.3_{-9.2}^{+9.2}$ & 0.21\\

R5 (South)       & $3.27_{-0.21}^{+0.23}$ & $3.26_{-0.13}^{+0.10}$ & $25.4_{-1.2}^{+1.2}$ & $68.8_{-9.2}^{+8.9}$ & $185_{-12}^{+11}$ & $33.7_{-5.2}^{+5.2}$ & 0.43\\

R2 to R4 (North) & $5.52_{-0.15}^{+0.15}$ & $3.35_{-0.06}^{+0.06}$ & $89.3_{-2.6}^{+2.6}$ & $161_{-12}^{+13}$ & $660_{-24}^{+24}$ & $120_{-17}^{+17}$ & 0.64\\

R2 to R5 (Outer Regions)  & $4.88_{-0.12}^{+0.13}$ & $3.33_{-0.05}^{+0.05}$ & $111_{-3}^{+3}$ & $225_{-15}^{+15}$ & $821_{-25}^{+26}$ & $149_{-22}^{+22}$ & 0.83\\

\hline
\end{tabular}
\footnotetext{$^{a}$ see Sect.\,\ref{sec:specana}} 
\label{tab:detect_gas4449}
\end{minipage}
\end{table*}


\begin{figure*}
\centering
\includegraphics[width=15cm]{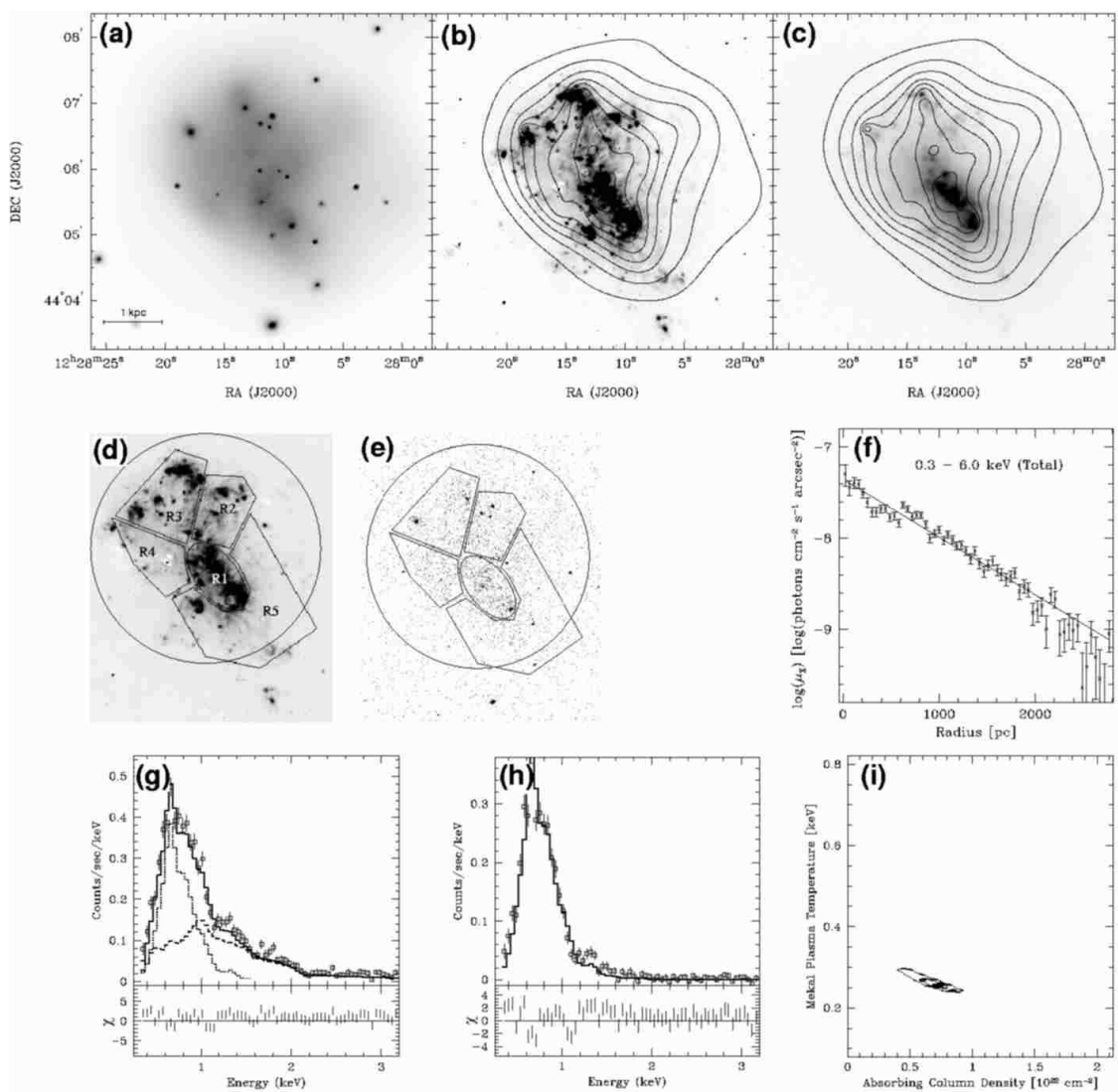}
\caption
{Images and spectra for the X-ray emission of NGC\,4449. See the
caption of Fig.\,\ref{fig:olay1569} for details. The contours of the
logarithmic X-ray emission in panels (b) and (c) are spaced by
$0.1\log(peak flux)$ starting at $0.2\log(peak flux)$ (continuum
subtracted \ha\ image taken from \citealt{bom97}; Optical Lowell 1.1m
$R$--band image from \citealt{frei96}).}
\label{fig:olay4449}
\end{figure*}

\subsubsection{Previous X-ray observations and results}

\citet{bla83} were the first to report an {\it EINSTEIN} X-ray detection of
NGC\,4449. They concentrated their study on a strong point source (P9,
see below) which, based on radio continuum measurements, was
previously classified as a SNR \citep[][]{big83}. They derived an
extraordinary high X-ray luminosity of $\sim 10^{39}$\,erg~s$^{-1}$.
Based on {\it ROSAT} HRI (15\,ks) and PSPC (8\,ks) data, \citet{vog97}
detected seven point sources embedded in diffuse X-ray emission
(combined X-ray luminosity of $2.5\times 10^{39}$\,\lum). They
attributed the diffuse emission to hot thermal gas ($T\simeq 3\times
10^{6}$\,K) and concluded that the global X-ray emission resembles
that of the Large Magellanic Cloud after scaling for the different
galaxy masses. \citet{bom97} assigned the diffuse thermal emission to
active star forming regions, to outflows, and to the interior of a
supergiant shell which was detected via \ha\ observations (PSPC data:
2\,ks, HRI data: 32\,ks). \citet{dce97} combined the {\it ROSAT} PSPC
data with observations from ASCA (45\,ks). In addition to the hard
emission from the point source population they postulate a soft
($T\simeq 8\times 10^{6}$\,K) and a very soft ($T\simeq 3\times
10^{6}$\,K) diffuse thermal X-ray component. They concluded that
mass-loading plays an important role in NGC\,4449. In addition, they
inferred that the gas is hot enough the be blown-out, i.e., to leave
the gravitational potential of NGC\,4449. More recently, \citet{sum03}
published a detailed analysis of the {\it Chandra} data also presented
here. They fit a two-temperature thermal plasma model with $T=3.2$ and
$10\times10^{6}$\,K to the diffuse component and estimate its mass to
$\sim 10^{7}$\,\msun. Some of the hot gas was detected in the interior
of an expanding superbubble visible in \ha. They also attribute 28
X-ray point sources to NGC\,4449.

\subsubsection{{\it Chandra} observations revisited}

\begin{figure*}
\centering
\includegraphics[width=13cm]{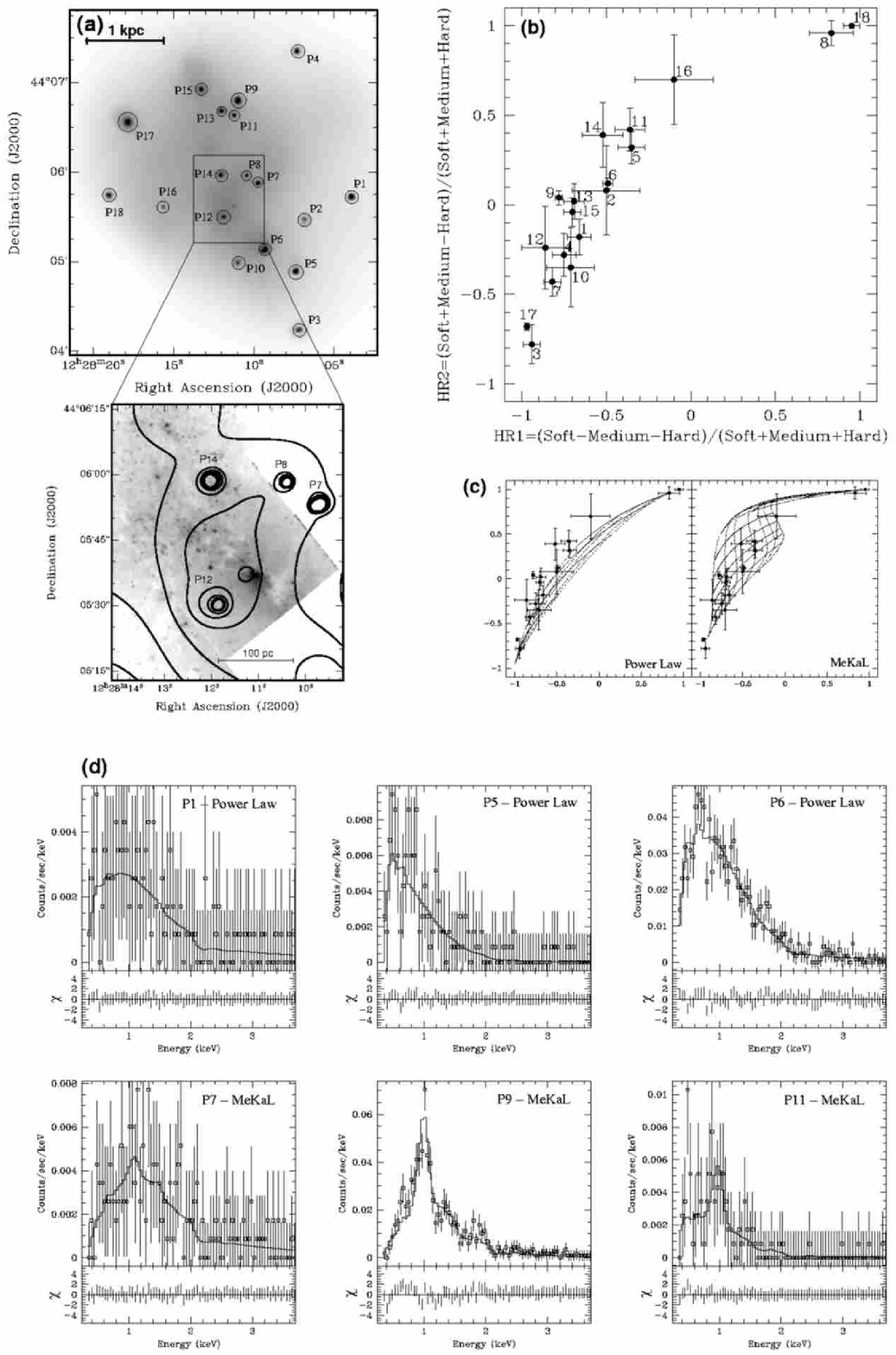}
\caption
{Locations and spectra of the point sources in NGC\,4449. See caption
  of Fig.\,\ref{fig:point1569} for details. The count rates,
  hardness ratios and fit parameters are listed in
  Tables\,\ref{tab:point4449a} and \ref{tab:point4449b}. }
\label{fig:point4449}
\end{figure*}
\addtocounter{figure}{-1}
\begin{figure*}
\centering
\includegraphics[width=13cm]{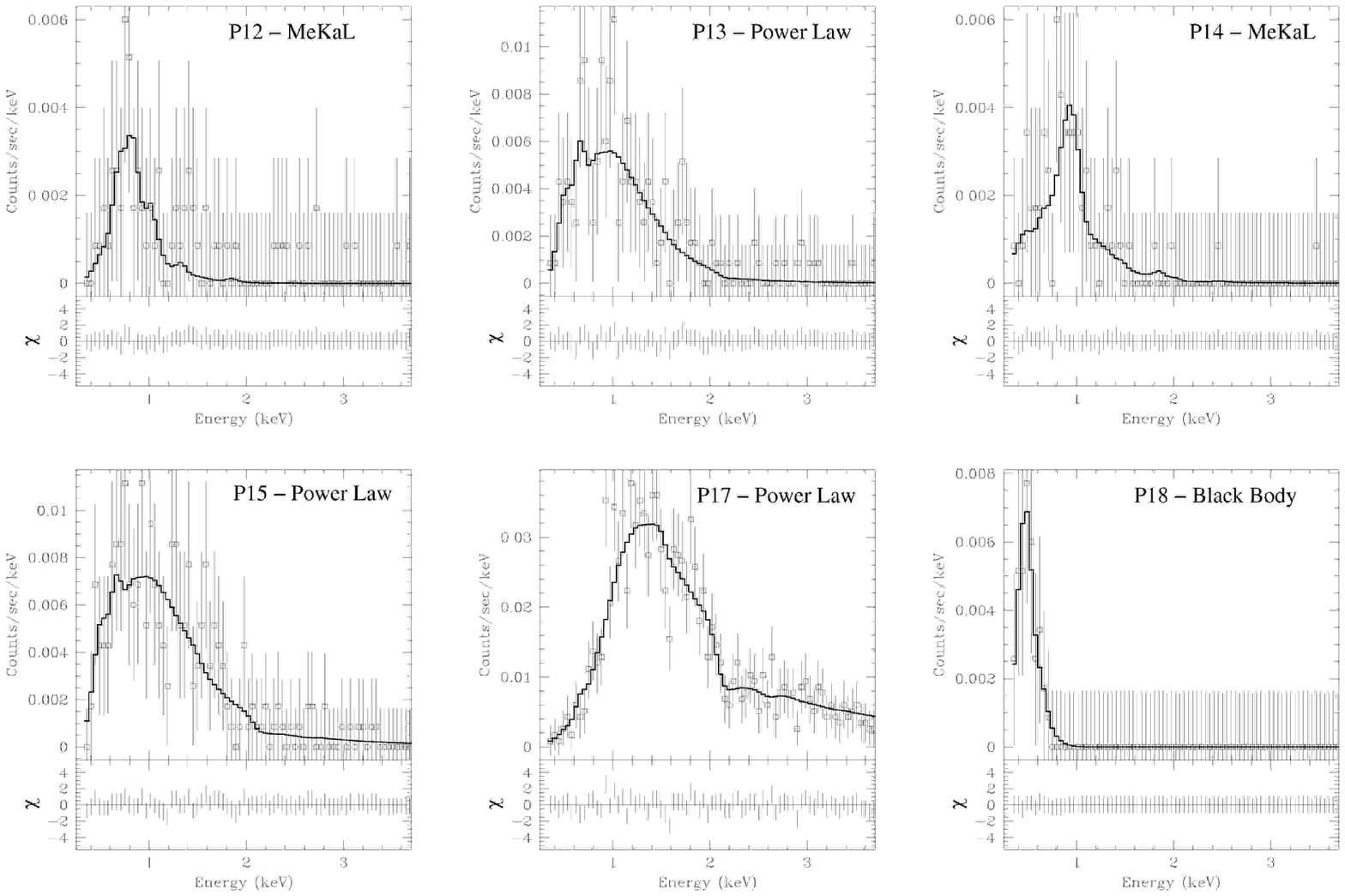}
\caption
{continued.}
\end{figure*}

\begin{table*}
\begin{minipage}{170mm}
\centering
\caption
{Positions, count rates and hardness ratios of the point sources in
NGC\,4449. The Right Ascension and Declination are given in $^{h}$
$^{m}$ $^{s}$ and $\degr$ $\arcmin$ $\arcsec$ (J2000), respectively.
The count rates of the total, soft, medium, and hard emission are in
units of $10^{-5}$\,cts\,s$^{-1}$. See Fig.\,\ref{fig:point4449} for
the locations of the point sources coinciding with this galaxy as well
as for their distribution in a hardness ratio plot. }
\begin{tabular}{@{}lllrrrrrr@{}}
\hline

No.&\multicolumn{1}{c}{RA}&\multicolumn{1}{c}{DEC}&\multicolumn{1}{c}{Total}&\multicolumn{1}{c}{Soft}&\multicolumn{1}{c}{Medium}&\multicolumn{1}{c}{Hard}&\multicolumn{1}{c}{HR1}&\multicolumn{1}{c}{HR2}\\

\hline

P1  & 12~28~03.9 & +44~05~43.7 & $511.3\pm 44.5$ & $86.47\pm 18.80$ & $124.1\pm 21.6$  & $300.8\pm 34.0$ & $-0.66\pm 0.07$ & $-0.18\pm 0.10$\\
P2  & 12~28~06.8 & +44~05~27.5 & $90.23\pm 19.89$& $22.56\pm 10.63$ & $26.32\pm 9.95$  & $41.35\pm 13.55$& $-0.50\pm 0.20$ & $+0.08\pm 0.25$\\
P3  & 12~28~07.2 & +44~04~14.7 & $135.3\pm 22.6$ & $3.76\pm 3.76$   & $11.28\pm 6.51$  & $120.3\pm 21.3$ & $-0.94\pm 0.05$ & $-0.78\pm 0.11$\\ 
P4  & 12~28~07.3 & +44~07~21.1 & $300.8\pm 34.0$ & $37.59\pm 11.89$ & $71.43\pm 17.23$ & $191.7\pm 26.8$ & $-0.75\pm 0.07$ & $-0.28\pm 0.12$\\
P5  & 12~28~07.3 & +44~04~53.8 & $594.0\pm 48.1$ & $191.7\pm 27.4$  & $199.3\pm 28.4$  & $203.0\pm 27.6$ & $-0.35\pm 0.08$ & $+0.32\pm 0.09$\\
P6   & 12~28~09.3 & +44~05~08.1 & $4417\pm 129$   & $1132\pm 65$     & $1338\pm 71$     & $1947\pm 86$    & $-0.49\pm 0.03$ & $+0.12\pm 0.03$\\  
P7   & 12~28~09.7 & +44~05~53.1 & $736.9\pm 55.3$ & $67.67\pm 19.89$ & $142.9\pm 23.8$  & $526.3\pm 45.7$ & $-0.82\pm 0.05$ & $-0.43\pm 0.08$\\
P8   & 12~28~10.4 & +44~05~58.0 & $172.9\pm 29.1$ & $157.9\pm 26.6$  & $11.28\pm 9.95$  & $3.76\pm 6.51$  & $+0.83\pm 0.13$ & $+0.96\pm 0.07$\\
P9   & 12~28~11.0 & +44~06~48.4 & $3793\pm 120$   & $424.8\pm 40.0$  & $1545\pm 77$     & $1823\pm 83$    & $-0.78\pm 0.02$ & $+0.04\pm 0.04$\\
P10 & 12~28~11.0 & +44~04~59.4 & $127.8\pm 24.4$ & $18.80\pm 9.95$  & $22.56\pm 13.02$ & $86.47\pm 18.03$& $-0.71\pm 0.14$ & $-0.35\pm 0.22$\\
P11  & 12~28~11.2 & +44~06~38.1 & $402.3\pm 40.7$ & $127.8\pm 22.6$  & $157.9\pm 26.0$  & $116.5\pm 21.6$ & $-0.36\pm 0.09$ & $+0.42\pm 0.12$\\ 
P12  & 12~28~11.9 & +44~05~29.7 & $218.0\pm 36.1$ & $15.04\pm 15.95$ & $67.67\pm 21.27$ & $135.3\pm 24.4$ & $-0.86\pm 0.14$ & $-0.24\pm 0.23$\\
P13  & 12~28~12.0 & +44~06~41.5 & $699.3\pm 53.4$ & $109.0\pm 22.2$  & $248.1\pm 32.3$  & $342.1\pm 36.3$ & $-0.69\pm 0.06$ & $+0.02\pm 0.10$\\
P14  & 12~28~12.1 & +44~05~58.5 & $233.1\pm 32.3$ & $56.39\pm 15.50$ & $105.3\pm 23.2$  & $71.43\pm 16.39$& $-0.52\pm 0.12$ & $+0.39\pm 0.18$\\
P15  & 12~28~13.3 & +44~06~55.6 & $954.9\pm 61.8$ & $142.9\pm 25.5$  & $315.8\pm 36.1$  & $496.2\pm 43.2$ & $-0.70\pm 0.05$ & $-0.04\pm 0.08$\\
P16 & 12~28~15.6 & +44~05~36.8 & $75.19\pm 17.63$& $33.84\pm 11.29$ & $30.08\pm 11.89$ & $11.29\pm 6.51$ & $-0.10\pm 0.23$ & $+0.70\pm 0.25$\\
P17  & 12~28~17.8 & +44~06~33.6 & $5222\pm 141$   & $86.47\pm 20.25$ & $755.6\pm 54.1$  & $4380\pm 129$   & $-0.97\pm 0.01$ & $-0.68\pm 0.02$\\
P18  & 12~28~19.1 & +44~05~44.4 & $161.7\pm 24.7$ & $157.9\pm 24.4$  & $3.76\pm 3.76$   & $0.00\pm 0.00$  & $+0.95\pm 0.05$ & $+1.00\pm 0.01$\\

\hline

\end{tabular}
\label{tab:point4449a}
\end{minipage}
\end{table*}

\begin{table*}
\begin{minipage}{170mm}
\centering
\caption
{Parameters of the fitted X-ray emission models to the point sources
in NGC\,4449. See the caption of Table\,\ref{tab:point1569b} for
details and Fig.\,\ref{fig:point4449}(d) for the individual spectra
and fits. The last column displays the numbering scheme of
\citet{sum03}.}
\begin{tabular}{@{}lllrrrrrr@{}}
\hline

No.&Model &\multicolumn{1}{c}{$N_{H}$} &\multicolumn{1}{c}{$T/\gamma$} &\multicolumn{1}{c}{$Norm/Ampl^{a}$} & \multicolumn{1}{c}{$F_{\rm X}^{\rm abs}$ }&\multicolumn{1}{c}{$F_{\rm X}$} &\multicolumn{1}{c}{$L_{\rm X}$}&Cross Ident.\\
&&\multicolumn{1}{c}{[$10^{21}$\,\cden]} &\multicolumn{1}{c}{[$10^{6}$\,K/--]} &\multicolumn{1}{c}{[$10^{-5}$]} & \multicolumn{2}{c}{[$10^{-15}$\,\flux]} &\multicolumn{1}{c}{[$10^{37}$\,\lum]}& \multicolumn{1}{c}{No.}\\

\hline


P1   & PL  & \multicolumn{1}{c}{3.44 }& \multicolumn{1}{c}{1.85 }& \multicolumn{1}{c}{0.66   }& \multicolumn{1}{c}{32.66 }& \multicolumn{1}{c}{44.38 }& \multicolumn{1}{c}{8.08}&\multicolumn{1}{c}{5}\\ 
P5   & PL  & \multicolumn{1}{c}{5.77 }& \multicolumn{1}{c}{3.72 }& \multicolumn{1}{c}{1.02   }& \multicolumn{1}{c}{14.58 }& \multicolumn{1}{c}{119.5 }& \multicolumn{1}{c}{21.75}&\multicolumn{1}{c}{9}\\
P6   & PL  & \multicolumn{1}{c}{6.63 }& \multicolumn{1}{c}{3.08 }& \multicolumn{1}{c}{11.05  }& \multicolumn{1}{c}{170.6 }& \multicolumn{1}{c}{791.6 }& \multicolumn{1}{c}{144.1}&\multicolumn{1}{c}{10}\\
P7   & MEK & \multicolumn{1}{c}{5.34 }& \multicolumn{1}{c}{40.5 }& \multicolumn{1}{c}{4.33   }& \multicolumn{1}{c}{42.67 }& \multicolumn{1}{c}{61.01 }& \multicolumn{1}{c}{11.10}&\multicolumn{1}{c}{11}\\
P8   & BB  & \multicolumn{1}{c}{11.36}& \multicolumn{1}{c}{0.32 }& \multicolumn{1}{c}{2e10   }& \multicolumn{1}{c}{2.67  }& \multicolumn{1}{c}{35000 }& \multicolumn{1}{c}{6370}&\multicolumn{1}{c}{12}\\
P9   & MEK & \multicolumn{1}{c}{6.20 }& \multicolumn{1}{c}{12.78}& \multicolumn{1}{c}{26.54  }& \multicolumn{1}{c}{122.6 }& \multicolumn{1}{c}{265.8 }& \multicolumn{1}{c}{48.38}&\multicolumn{1}{c}{15}\\
P11  & MEK & \multicolumn{1}{c}{0.79 }& \multicolumn{1}{c}{12.53}& \multicolumn{1}{c}{1.39   }& \multicolumn{1}{c}{11.19 }& \multicolumn{1}{c}{13.97 }& \multicolumn{1}{c}{2.54}&\multicolumn{1}{c}{17}\\
P12  & MEK & \multicolumn{1}{c}{6.22 }& \multicolumn{1}{c}{4.69 }& \multicolumn{1}{c}{1.70   }& \multicolumn{1}{c}{4.09  }& \multicolumn{1}{c}{14.84 }& \multicolumn{1}{c}{2.70}&\multicolumn{1}{c}{19}\\
P13  & PL  & \multicolumn{1}{c}{15.45}& \multicolumn{1}{c}{5.04 }& \multicolumn{1}{c}{5.11   }& \multicolumn{1}{c}{18.78 }& \multicolumn{1}{c}{2361  }& \multicolumn{1}{c}{429.7}&\multicolumn{1}{c}{20}\\
P14  & MEK & \multicolumn{1}{c}{1.58 }& \multicolumn{1}{c}{10.01}& \multicolumn{1}{c}{0.87   }& \multicolumn{1}{c}{6.24  }& \multicolumn{1}{c}{8.95  }& \multicolumn{1}{c}{1.63 }&\multicolumn{1}{c}{21}\\
P15  & PL  & \multicolumn{1}{c}{9.32 }& \multicolumn{1}{c}{3.35 }& \multicolumn{1}{c}{3.32   }& \multicolumn{1}{c}{36.22 }& \multicolumn{1}{c}{285.9 }& \multicolumn{1}{c}{52.03}&\multicolumn{1}{c}{23}\\
P17  & PL  & \multicolumn{1}{c}{16.07}& \multicolumn{1}{c}{2.16 }& \multicolumn{1}{c}{20.42  }& \multicolumn{1}{c}{519.9 }& \multicolumn{1}{c}{1195  }& \multicolumn{1}{c}{217.5}&\multicolumn{1}{c}{27}\\
P18  & BB  & \multicolumn{1}{c}{9.67 }& \multicolumn{1}{c}{0.58 }& \multicolumn{1}{c}{5.4e6  }& \multicolumn{1}{c}{3.83  }& \multicolumn{1}{c}{1024  }& \multicolumn{1}{c}{186.4}&\multicolumn{1}{c}{28}\\

\hline
\end{tabular}
\footnotetext{$^{a}$ see Sect.\,\ref{sec:specana}} 
\label{tab:point4449b}
\end{minipage}
\end{table*}


With a size of $8\farcm9\times 9\farcm2$ (corresponding to 10.0\,kpc
$\times 10.4$\,kpc) the diffuse X-ray distribution of NGC\,4449 is
impressively large. Its morphology is similar to the \ha\ distribution
of NGC\,4449 (see Fig.\,\ref{fig:olay4449}[b]). To the west, however,
the X-ray emission is more extended. We detect 18 point sources in
NGC\,4449 (see Fig.\,\ref{fig:point4449}[a]). With a count rate of
$\sim 0.05$\,\cntr\ P17 is the strongest, followed by P6 and P9 (see
Table\,\ref{tab:point4449a}). The latter source was previously
classified as a SNR (see above). In Figs.\,\ref{fig:point4449}(b) and
\ref{fig:point4449}(c) we plot a hardness ratio diagram of all the
point sources. Seven of the point source spectra are well fit by
provisional thermal plasma MeKaL models (P2, P7, P9, P11, P12, P14,
P16) and nine by power laws (P1, P3, P4, P5, P6, P10, P13, P15,
P17). To select which emission model provides the best fit, the
location of the sources in the hardness ratio plots was equally taken
into account. In addition, we detect two supersoft sources (P8 and
P18), one near the centre of NGC\,4449 and another one further
out. Individual, final fits to the spectra with sufficient counts are
shown in Fig.\,\ref{fig:point4449}(d) and the fitted parameters are
given in Table\,\ref{tab:point4449b}. Due to crowding and absorption
effects, an unambiguous cross identification of X-ray point sources
with optical counterparts within the {\it HST}/WFPC2 field (see
Fig.\,\ref{fig:point4449}[a]) is not possible.

The azimuthally averaged X-ray surface brightness profiles of
NGC\,4449 are well described by an exponential function with small
deviations only (Fig.\,\ref{fig:olay4449}[f]). The scale length of the
soft X-ray emission is with $\sim 700$\,pc slightly smaller compared
to that of the hard emission ($\sim 740$\,pc). This is not expected,
since hardening of the emission due to photoelectric absorption would
restrict the hard emission more to the centre. Regarding the large
error of typically 90\,pc in the scale length, however, this is only a
tentative result. The temperature of the hot, diffuse gas in the
centre of NGC\,4449 is equal to the temperature in the outer regions
($\sim 3.2\times 10^{6}$\,K; see Figs.\,\ref{fig:olay4449}(d) and
\ref{fig:olay4449}(e) for the definition of the regions and
Fig.\,\ref{fig:olay4449}(h) and \ref{fig:olay4449}(i) for the fits and
confidence regions of the total diffuse X-ray emission; see also
Table\,\ref{tab:detect_gas4449} for the parameters derived). Following
our analysis, we do not need to introduce a second temperature
component to find a good fit to the spectra. This result suggests that
the hot gas has been heated more uniformly over the disc or that the
cooling time exceeds the age of the starburst. The absorbing column
density of the northern half of NGC\,4449 is higher than that toward
the south. This leads to the conclusion that the z-axis pointing to
the north is tipped towards us and hence that the disc gas absorbs the
X-rays emerging from the hot gas. The southern part of the gaseous
disc is located behind the hot outflow. The total diffuse unabsorbed
X-ray luminosity derived from the MeKaL fit is $4.5\times
10^{39}$\,\lum. The X-ray luminosity and temperature are in good
agreement with the {\it ROSAT} measurements (within a factor of $\sim
2$). Note that the {\it ROSAT} result was determined within an energy
range of 0.1--2.0\,keV and that the point source confusion of the {\it
ROSAT} data can contribute to the difference. Adding up point sources
and the the diffuse emission, the total X-ray luminosity of NGC\,4449
results in $7.9\times10^{40}$\,\lum.


\subsection[NGC\,5253]{{\bf NGC\,5253} (UGCA\,369, Haro\,10, ESO\,445-G\,4)}
\label{sec5:5253}


\begin{table*}
\renewcommand{\arraystretch}{1.5}
\begin{minipage}{170mm}
\centering
\caption
{Results of MeKaL collisional thermal plasma model fits applied to the
X-ray spectra of different regions in NGC\,5253 (see panels [d] and
[e] in Fig.\,\ref{fig:olay5253} for the definition of the regions).
}
\begin{tabular}{@{}lccccccc@{}}
\hline

\multicolumn{1}{l}{Region}&\multicolumn{1}{c}{$N_{\rm H}$}&\multicolumn{1}{c}{$T$}&\multicolumn{1}{c}{$Norm^{a}$}&\multicolumn{1}{c}{$F_{\rm X}^{\rm abs}$}&\multicolumn{1}{c}{$F_{\rm X}$}&\multicolumn{1}{c}{$L_{\rm X}$}&\multicolumn{1}{c}{$\chi^{2}_{red}$}\\
\multicolumn{1}{c}{}&\multicolumn{1}{c}{[$10^{21}$\,cm$^{-2}$]}&\multicolumn{1}{c}{[$10^{6}$\,K]}&\multicolumn{1}{c}{[$10^{-5}$]}&\multicolumn{1}{c}{[$10^{-15}$\,erg~s$^{-1}$\,cm$^{-2}$]}&\multicolumn{1}{c}{[$10^{-15}$\,erg~s$^{-1}$\,cm$^{-2}$]}&\multicolumn{1}{c}{[$10^{37}$\,erg~s$^{-1}$]}&\\

\hline

Total              & $5.67_{-0.22}^{+0.23}$ & $3.51_{-0.08}^{+0.08}$ & $55.5_{-2.1}^{+2.0}$ & $81.6_{-8.1}^{+8.6}$ & $359_{-17}^{+18}$ & $46.6_{-7.0.}^{+7.0}$ & 0.64\\
R1 (Centre)        & $10.1_{-0.3}^{+0.3}$ & $3.79_{-0.09}^{+0.09}$ & $44.6_{-2.0}^{+2.0}$ & $39.3_{-4.6}^{+5.0}$ & $299_{-17}^{+17}$ & $38.9_{-5.9}^{+5.9}$ & 0.23\\

R2 (West)          & $11.4_{-0.9}^{+1.1}$ & $2.62_{-0.17}^{+0.16}$ & $22.0_{-3.8}^{+3.8}$ & $7.79_{-3.27}^{+4.36}$ & $124_{-25}^{+25}$ & $16.1_{-4.0}^{+4.0}$ & 0.34\\

R3 (North)         & $0.90_{-0.74}^{+0.88}$ & $5.48_{-1.01}^{+0.89}$ & $2.09_{-0.29}^{+0.29}$ & $10.4_{-3.9}^{+4.4}$ & $17.1_{-3.9}^{+3.7}$ & $2.23_{-0.59}^{+0.59}$ & 0.40\\

R4 (South)         & $6.69_{-0.55}^{+0.61}$ & $3.10_{-0.16}^{+0.16}$ & $13.6_{-1.4}^{+1.4}$ & $14.2_{-3.6}^{+4.2}$ & $82.5_{-9.9}^{+10.4}$ & $10.7_{-2.1}^{+2.1}$ & 0.36\\

R3+R4              & $6.69_{-0.60}^{+0.67}$ & $3.28_{-0.18}^{+0.20}$ & $18.6_{-2.0}^{+2.0}$ & $21.2_{-5.6}^{+6.8}$ & $116_{-15}^{+16}$ & $15.1_{-3.0}^{+3.0}$ & 0.45\\

R2 to R4 (Outer Regions)    & $7.58_{-0.75}^{+0.88}$ & $3.21_{-0.19}^{+0.23}$ & $25.6_{-3.4}^{+3.4}$ & $24.6_{-7.8}^{+10.1}$ & $158_{-25}^{+27}$ & $20.6_{-4.6}^{+4.6}$ & 0.49\\

\hline
\end{tabular}
\footnotetext{$^{a}$ see Sect.\,\ref{sec:specana}} 
\label{tab:detect_gas5253}
\end{minipage}
\end{table*}


\begin{figure*}
  \centering \includegraphics[width=15cm]{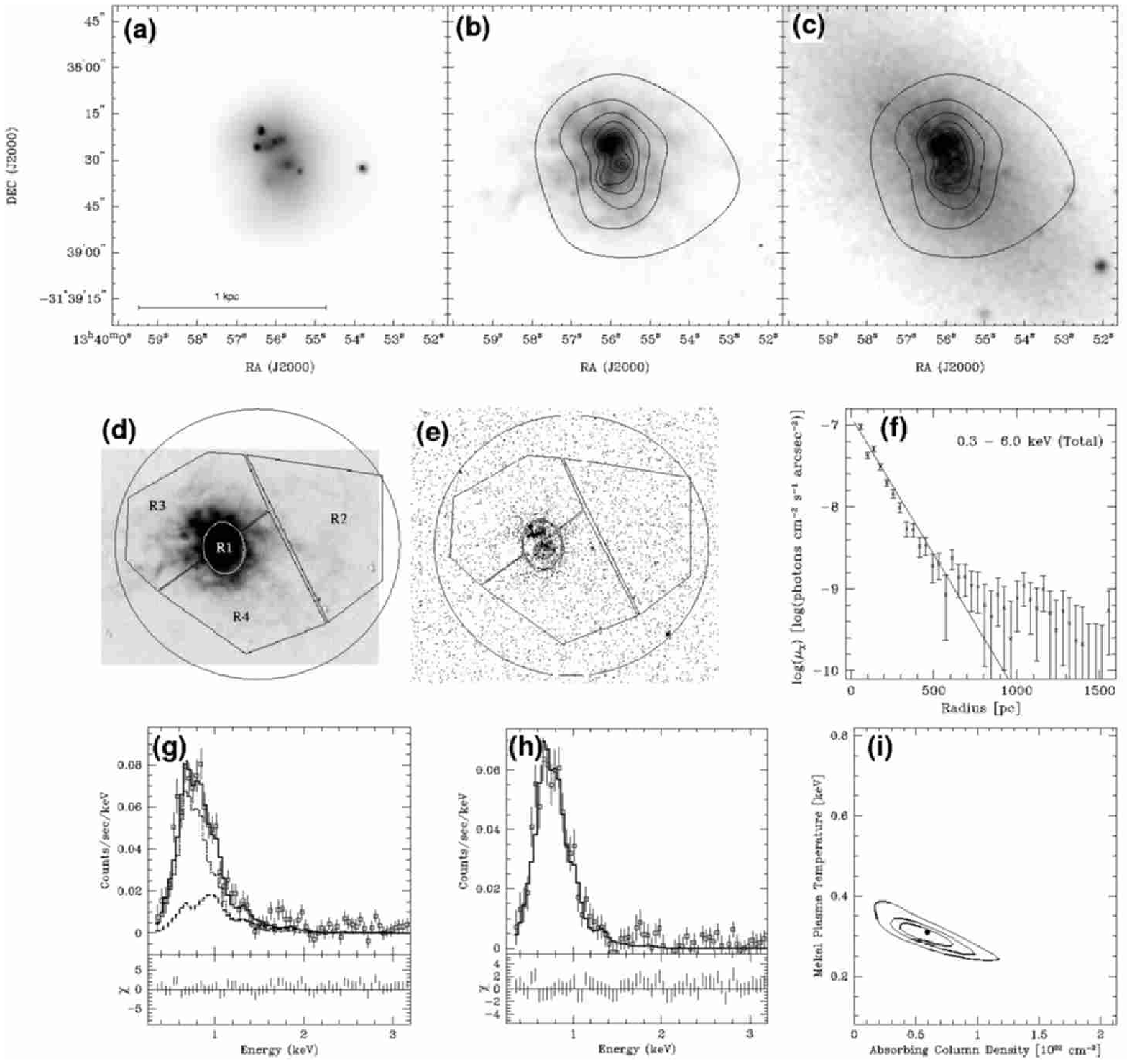}
\caption
{Images and spectra for the X-ray emission of NGC\,5253. See the
caption of Fig.\,\ref{fig:olay1569} for details (continuum subtracted
\ha\ image taken from \citealt{mar98}, the optical $R$--band 1.1m ESO
Schmidt image from \citealt{lau89}).}
\label{fig:olay5253}
\end{figure*}

\subsubsection{Previous X-ray observations and results}

{\it EINSTEIN} was the first X-ray observatory to detect NGC\,5253 and
\citet{fab92} derived a luminosity of $7\times 10^{38}$\,\lum.
\citet{mar95b} analysed follow-up observations with the {\it ROSAT} PSPC
instrument (35\,ks). They detected an extended soft source coinciding
with the galaxy's centre which they fit with a thermal plasma with a
temperature of $4\times 10^{6}$\,K (luminosity: $6.5\times
10^{38}$\,erg~s$^{-1}$). The X-ray emission was interpreted as
emerging from the interior of an \ha\ superbubble. In order to resolve
the soft X-ray source, \citet{str99} observed NGC\,5253 with the {\it
ROSAT} HRI (71\,ks). They detected five sources which they identified
with young massive stellar clusters at the centre. Three of the
sources were found to be extended and they concluded that X-rays
originate in sources which are commonly associated with star forming
regions, i.e., SNRs and massive XRBs (point sources) as well as
superbubbles (extended sources). More recently, \citet{sum04}
presented an analysis of {\it Chandra} and {\it XMM-Newton} data of
NGC\,5253. They examine in detail 17 X-ray point sources and the
diffuse X-ray emission and derive an X-ray luminosity of the hot gas
of $\sim 2\times10^{38}$\,\lum\ emitted by hot gas with temperatures of
2.6 and 8.2$\times 10^{6}$\,K.

\subsubsection{{\it Chandra} observations revisited}

\begin{figure*}
\centering
\includegraphics[width=14cm]{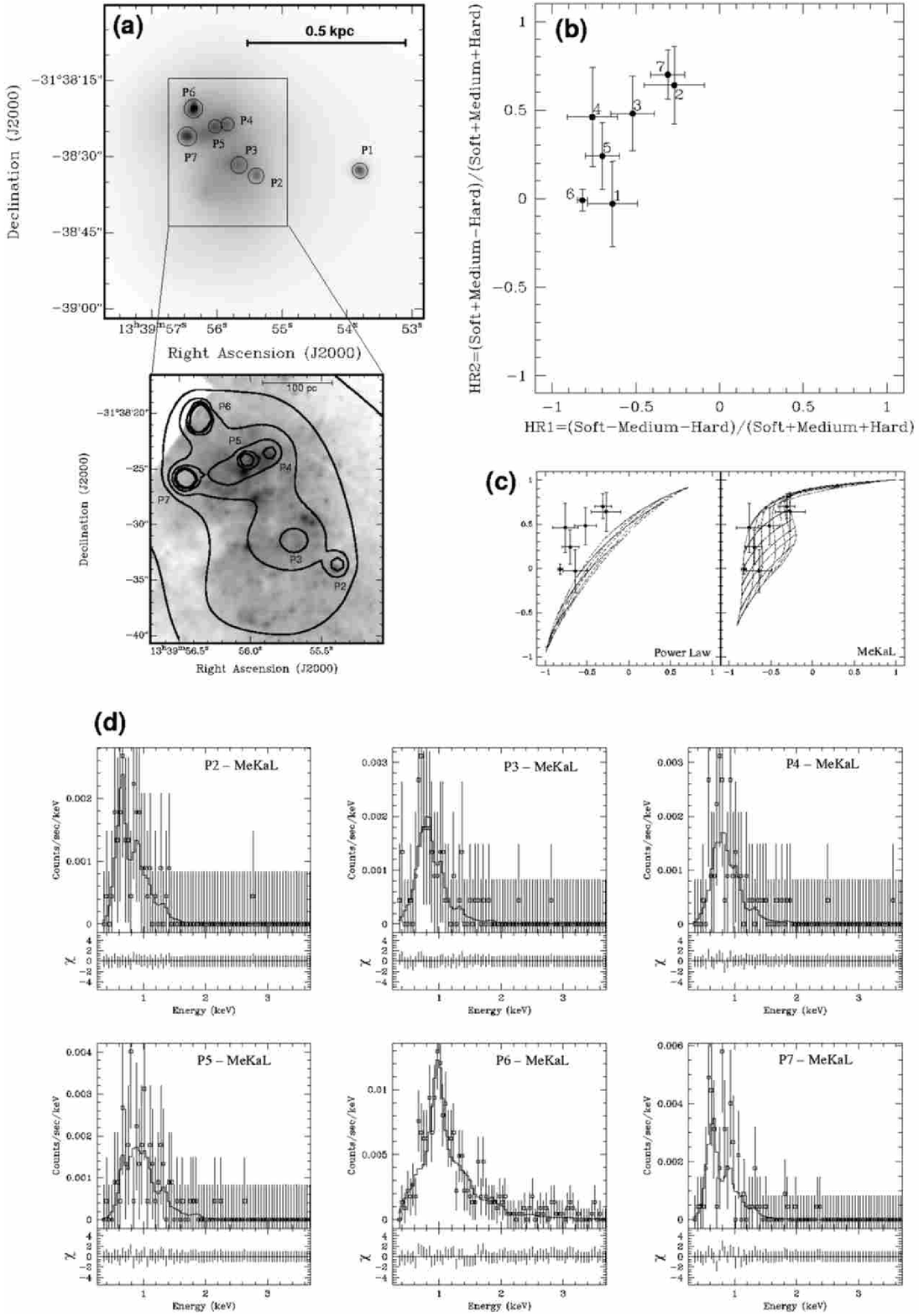}
\caption
{Locations and spectra of the point sources in NGC\,5253. See caption
  of Fig.\,\ref{fig:point1569} for details. The count rates,
  hardness ratios and fitted parameters are listed in
  Tables\,\ref{tab:point5253a} and \ref{tab:point5253b}. }
\label{fig:point5253}
\end{figure*}

The {\it Chandra} data of NGC\,5253 show a total of seven point
sources embedded in an extended region, $1\farcm8 \times 1\farcm3$ in
size (corresponding to $1.7 \times 1.2$\,kpc). The area of high
surface brightness, however, is less extended in the east-west
direction ($1\farcm0$ or $\sim 1.0$\,kpc) as compared to the faint
emission. The distribution of the diffuse X-ray emission is similar to
the \ha\ morphology, which also exhibits a low surface brightness
extension to the west (see Fig.\,\ref{fig:olay5253}[b]). Except for
the point source P1 (Fig.\,\ref{fig:point5253}[a]), which is located
toward the south-west, all X-ray point sources lie close to the centre
of NGC\,5253. The count rates of all point sources and their hardness
ratios are listed in Table\,\ref{tab:point5253a} and the corresponding
hardness ratio plot is shown in Figs.\,\ref{fig:point5253}(b) and
(c). The fits to the spectra for the six point sources with sufficient
counts are displayed in Fig.\,\ref{fig:point5253} and the resulting
parameters are listed in Table\,\ref{tab:point5253b}. Of the seven
point sources, only P1 is best modeled by a power law. All the others
appear to emit line and continuum radiation due to a hot thermal
plasma with temperatures of $\sim 2-4\times 10^{6}$\,K (P6, the
strongest point source, exhibits a temperature which is about a factor
of 4 higher). A highly obscured, very young super-stellar cluster
resides between P4 and P5. The X-ray data are therefore in good
agreement with the scenario that those X-ray point sources are members
of this or neighbouring stellar clusters. No clear optical counterparts
can be detected on an {\it HST}/WFPC2 F555W (Johnson $V$ band) image
(Fig.\,\ref{fig:point5253}[a]).

The azimuthally averaged, X-ray surface brightness profiles of
NGC\,5253 in the different X-ray bands are well described by an
exponential decline at least toward smaller radii; at large radii the
profiles level off (Fig.\,\ref{fig:olay5253}[f]). We derive the soft
emission to be more extended ($h_{Soft}\simeq150$\,pc, see
Table\,\ref{tab:xprofile_fit}) compared to the hard emission ($\sim
70$\,pc). The strong optical absorption feature to the east of
NGC\,5253 is in agreement with the detection of dust lanes and
molecular complexes at that position
\citep[see, e.g.,][]{mei02}. This absorption is also evident in
diffuse X-rays; not only the surface brightness is lower at this
location but the emission is also considerably harder than the
emission toward the north and south. Fits of thermal plasma models to
individual regions in NGC\,5253 (see Fig.\,\ref{fig:olay5253}[d] and
[e]) reveal that temperatures of the hot gas are very similar for all
areas ($T\simeq 3\times 10^{6}$\,K; Table\,\ref{tab:detect_gas5253}
and Fig.\,\ref{fig:olay5253}[h] and [i] for the model fits and the
confidence regions to the overall emission). The absorbing column
density in the centre ($N_{\rm H}\simeq 10^{22}$\,cm$^{-2}$), however,
is larger than that of the other regions (in agreement with the
observed dust lanes and molecular complexes). The fit to the northern
part of NGC\,5253 results in a very low value of $N_{\rm H}\la
10^{20}$\,cm$^{-2}$, whereas the southern part is absorbed by a column
density which is about a factor of 10 higher. We conclude that the
south of the disc of NGC\,5253 is located between the hot gas and the
observer, whereas the northern part of the cool disc gas lies behind
the hot wind. The total diffuse X-ray luminosity is $4.7\times
10^{38}$\,erg~s$^{-1}$. The temperature of the hot gas agrees well
with previous {\it ROSAT} observations. Adding up the X-ray
luminosities of the point sources and of the hot gas (total X-ray
luminosity: $1.9\times10^{39}$\,\lum) the {\it ROSAT} and {\it
Chandra} results are almost identical (see
Fig.\,\ref{fig:olay5253}[g]).


\begin{table*}
\begin{minipage}{170mm}
\centering
\caption
{Positions, count rates and hardness ratios of the point sources in
NGC\,5253. The Right Ascension and Declination are given in $^{h}$
$^{m}$ $^{s}$ and $\degr$ $\arcmin$ $\arcsec$ (J2000), respectively.
The count rates of the total, soft, medium, and hard emission are in
units of $10^{-5}$\,cts\,s$^{-1}$. See Fig.\,\ref{fig:point5253} for
the locations of the point sources coinciding with this galaxy as well
as for their distribution in a hardness ratio plot.}
\begin{tabular}{@{}lllrrrrrr@{}}
\hline

No.&\multicolumn{1}{c}{RA}&\multicolumn{1}{c}{DEC}&\multicolumn{1}{c}{Total}&\multicolumn{1}{c}{Soft}&\multicolumn{1}{c}{Medium}&\multicolumn{1}{c}{Hard}&\multicolumn{1}{c}{HR1}&\multicolumn{1}{c}{HR2}\\

\hline

P1 & 13~39~53.8 & $-$31~38~32.9 & $64.45\pm 12.51$ & $11.72\pm 5.52$ & $19.53\pm 7.31$  & $33.20\pm 8.51$ & $-0.64\pm 0.15$ & $-0.03\pm 0.24$\\
P2  & 13~39~55.4 & $-$31~38~33.7 & $85.94\pm 16.12$ & $31.25\pm 9.57$ & $39.06\pm 10.70$ & $15.63\pm 7.31$ & $-0.27\pm 0.18$ & $+0.64\pm 0.22$\\
P3  & 13~39~55.7 & $-$31~38~31.6 & $97.21\pm 15.68$ & $23.37\pm 7.32$ & $48.58\pm 11.26$ & $25.26\pm 8.08$ & $-0.52\pm 0.13$ & $+0.48\pm 0.21$\\
P4  & 13~39~55.9 & $-$31~38~23.7 & $91.80\pm 17.04$ & $11.12\pm 7.47$ & $55.89\pm 12.32$ & $24.79\pm 9.09$ & $-0.76\pm 0.15$ & $+0.46\pm 0.28$\\
P5  & 13~39~56.0 & $-$31~38~24.3 & $131.3\pm 18.3$  & $19.73\pm 6.71$ & $61.96\pm 12.90$ & $49.64\pm 11.09$& $-0.70\pm 0.10$ & $+0.24\pm 0.19$\\
P6  & 13~39~56.4 & $-$31~38~20.7 & $830.8\pm 41.3$  & $74.43\pm 12.92$& $338.2\pm 26.40$ & $418.2\pm 29.0$ & $-0.82\pm 0.03$ & $-0.01\pm 0.06$\\
P7  & 13~39~56.5 & $-$31~38~25.9 & $180.5\pm 20.3$  & $62.34\pm 11.42$& $91.31\pm 14.29$ & $26.86\pm 8.85$ & $-0.31\pm 0.10$ & $+0.70\pm 0.14$\\
\hline

\end{tabular}
\label{tab:point5253a}
\end{minipage}
\end{table*}


\begin{table*}
\begin{minipage}{170mm}
\centering
\caption
{Parameters of the fitted X-ray emission models to the point
  sources in NGC\,5253. See the caption of Table\,\ref{tab:point1569b}
  for details and Fig.\,\ref{fig:point5253}(d) for the individual
  spectra and fits.}
\begin{tabular}{@{}lllrrrrrr@{}}
\hline

No.&Model &\multicolumn{1}{c}{$N_{H}$} &\multicolumn{1}{c}{$T/\gamma$} &\multicolumn{1}{c}{$Norm/Ampl^{a}$} & \multicolumn{1}{c}{$F_{\rm X}^{\rm abs}$ }&\multicolumn{1}{c}{$F_{\rm X}$} &\multicolumn{1}{c}{$L_{\rm X}$}\\
&&\multicolumn{1}{c}{[$10^{21}$\,\cden]} &\multicolumn{1}{c}{[$10^{6}$\,K/--]} &\multicolumn{1}{c}{[$10^{-5}$]} & \multicolumn{2}{c}{[$10^{-15}$\,\flux]} &\multicolumn{1}{c}{[$10^{37}$\,\lum]}&\\
\hline
P2   & MEK & \multicolumn{1}{c}{19.78}& \multicolumn{1}{c}{2.03 }& \multicolumn{1}{c}{61.22}& \multicolumn{1}{c}{2.46 }& \multicolumn{1}{c}{285.2 }& \multicolumn{1}{c}{37.08}\\
P3   & MEK & \multicolumn{1}{c}{6.79 }& \multicolumn{1}{c}{4.73 }& \multicolumn{1}{c}{1.42 }& \multicolumn{1}{c}{2.55 }& \multicolumn{1}{c}{10.74 }& \multicolumn{1}{c}{1.40}\\
P4   & MEK & \multicolumn{1}{c}{9.37 }& \multicolumn{1}{c}{4.00 }& \multicolumn{1}{c}{2.20 }& \multicolumn{1}{c}{2.27 }& \multicolumn{1}{c}{15.20 }& \multicolumn{1}{c}{1.98}\\
P5   & MEK & \multicolumn{1}{c}{22.23}& \multicolumn{1}{c}{2.87 }& \multicolumn{1}{c}{27.89}& \multicolumn{1}{c}{3.17 }& \multicolumn{1}{c}{163.6 }& \multicolumn{1}{c}{21.27}\\
P6   & MEK & \multicolumn{1}{c}{5.35 }& \multicolumn{1}{c}{12.04}& \multicolumn{1}{c}{5.90 }& \multicolumn{1}{c}{24.97}& \multicolumn{1}{c}{54.77 }& \multicolumn{1}{c}{7.12}\\
P7   & MEK & \multicolumn{1}{c}{19.77}& \multicolumn{1}{c}{1.80 }& \multicolumn{1}{c}{136.2}& \multicolumn{1}{c}{3.47 }& \multicolumn{1}{c}{557.6 }& \multicolumn{1}{c}{72.49}\\
\hline

\end{tabular}
\footnotetext{$^{a}$ see Sect.\,\ref{sec:specana}} 
\label{tab:point5253b}
\end{minipage}
\end{table*}


\subsection[He\,2-10]{{\bf He\,2-10} (ESO\,495-G\,21)}
\label{sec5:he2-10}

\begin{table*}
\renewcommand{\arraystretch}{1.5}
\begin{minipage}{170mm}
\centering
\caption
{Results of MeKaL collisional thermal plasma model fits applied to the
X-ray spectra of different regions in He\,2--10 (see panels [d] and
[e] in Fig.\,\ref{fig:olayhe2-10} for the definition of the regions).
}
\begin{tabular}{@{}lccccccc@{}}
\hline

\multicolumn{1}{l}{Region}&\multicolumn{1}{c}{$N_{\rm H}$}&\multicolumn{1}{c}{$T$}&\multicolumn{1}{c}{$Norm^{a}$}&\multicolumn{1}{c}{$F_{\rm X}^{\rm abs}$}&\multicolumn{1}{c}{$F_{\rm X}$}&\multicolumn{1}{c}{$L_{\rm X}$}&\multicolumn{1}{c}{$\chi^{2}_{red}$}\\
\multicolumn{1}{c}{}&\multicolumn{1}{c}{[$10^{21}$\,cm$^{-2}$]}&\multicolumn{1}{c}{[$10^{6}$\,K]}&\multicolumn{1}{c}{[$10^{-5}$]}&\multicolumn{1}{c}{[$10^{-15}$\,erg~s$^{-1}$\,cm$^{-2}$]}&\multicolumn{1}{c}{[$10^{-15}$\,erg~s$^{-1}$\,cm$^{-2}$]}&\multicolumn{1}{c}{[$10^{37}$\,erg~s$^{-1}$]}&\\

\hline

Total              & $4.44_{-0.29}^{+0.35}$ & $2.82_{-0.14}^{+0.14}$ & $93.6_{-13.0}^{+13.0}$ & $103_{-36}^{+46}$ & $2101_{-313}^{+319}$ & $2038_{-423}^{+423}$ & 0.32\\

R1 (Centre)        &$1.23_{-0.20}^{+0.22}$ & $7.60_{-0.44}^{+0.44}$ & $3.38_{-0.27}^{+0.27}$ & $41.9_{-6.5}^{+6.9}$ & $95.6_{-6.4}^{+6.3}$ & $92.7_{-14.5}^{+14.5}$ & 0.09\\

R2 (West)          & $2.11_{-0.25}^{+0.29}$ & $3.21_{-0.21}^{+0.22}$ & $3.91_{-0.53}^{+0.53}$ & $15.4_{-5.2}^{+6.2}$ & $91.4_{-13.7}^{+14.5}$ & $88.7_{-18.2}^{+18.2}$ & 0.06\\

R3 (South)         & $3.19_{-0.31}^{+0.38}$ & $2.89_{-0.20}^{+0.20}$ & $5.60_{-0.95}^{+0.95}$ & $11.0_{-4.5}^{+6.0}$ & $128_{-23}^{+24}$ & $124_{-29}^{+29}$ & 0.05\\

R4 (East)          & $0.00_{-0.00}^{+0.29}$ & $7.44_{-0.73}^{+0.72}$ & $0.94_{-0.13}^{+0.13}$ & $17.7_{-4.0}^{+2.4}$ & $26.5_{-3.1}^{+3.0}$ & $25.7_{-4.7}^{+4.7}$ & 0.09\\

R5 (North)         & $0.68_{-0.52}^{+0.71}$ & $5.14_{-1.07}^{+1.41}$ & $0.45_{-0.12}^{+0.12}$ & $5.65_{-3.22}^{+4.40}$& $12.4_{-4.2}^{+4.2}$ & $12.0_{-4.4}^{+4.4}$ & 0.05\\

R2+R4              & $2.68_{-0.17}^{+0.19}$ & $3.21_{-0.13}^{+0.12}$ & $10.3_{-0.9}^{+0.9}$ & $31.3_{-7.1}^{+7.7}$ & $240_{-23}^{+24}$ & $232_{-40}^{+40}$ & 0.16\\

R3+R5              & $1.65_{-0.25}^{+0.29}$ & $3.83_{-0.31}^{+0.33}$ & $2.94_{-0.37}^{+0.37}$ & $18.3_{-5.8}^{+7.1}$ & $72.4_{-10.8}^{+11.6}$ & $70.2_{-14.9}^{+14.9}$ & 0.08\\

R2 to R5 (Outer Regions)    & $2.82_{-0.14}^{+0.15}$ & $3.21_{-0.10}^{+0.11}$ & $17.5_{-1.2}^{+1.2}$ & $50.4_{-9.1}^{+9.9}$ & $410_{-31}^{+32}$ & $397_{-64}^{+64}$ & 0.22\\

\hline
\end{tabular}
\footnotetext{$^{a}$ see Sect.\,\ref{sec:specana}} 
\label{tab:detect_gashe2-10}
\end{minipage}
\end{table*}


\begin{figure*}
\centering
\includegraphics[width=15cm]{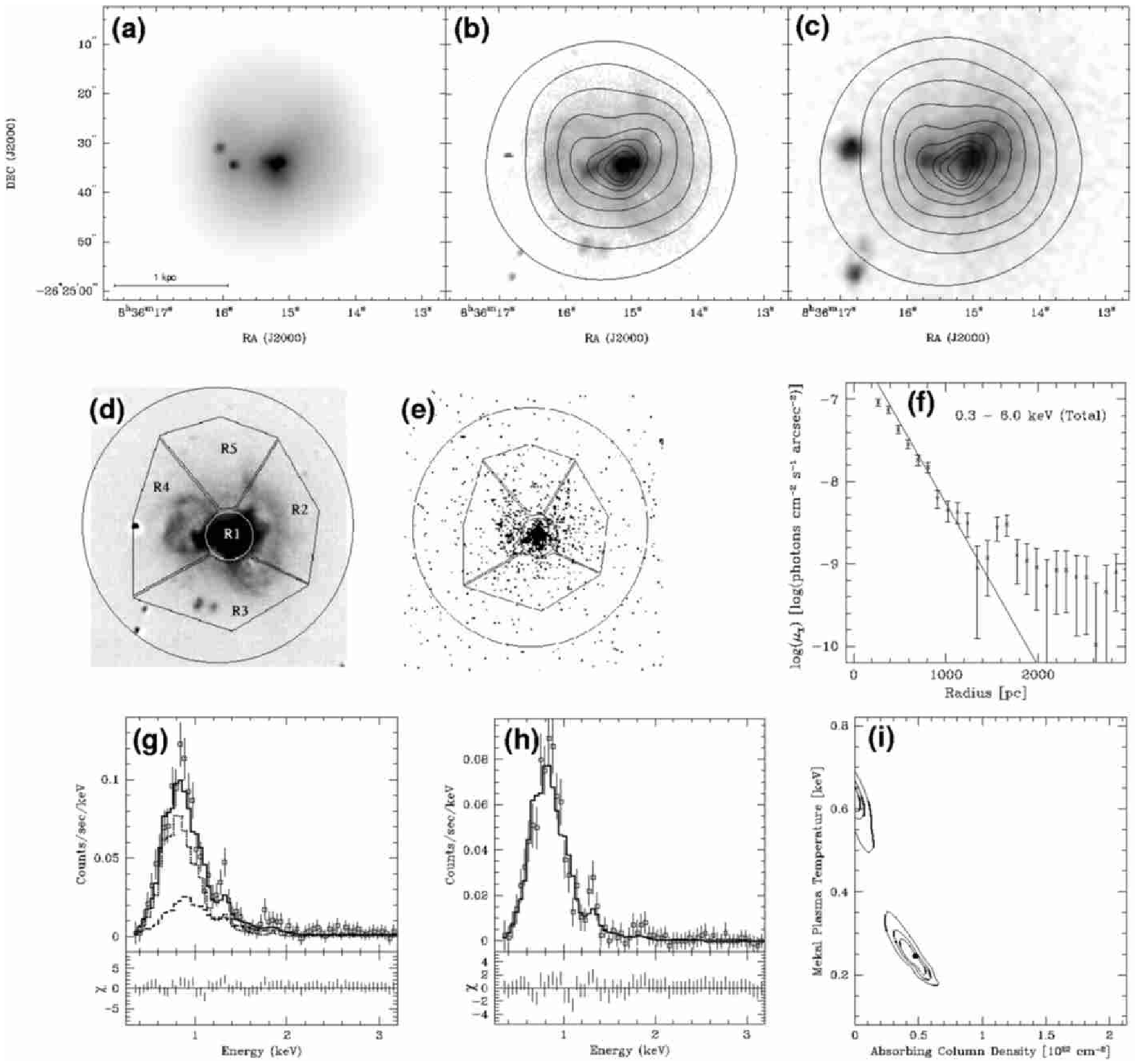}
\caption
{Images and spectra for the X-ray emission of He\,2-10. See the
caption of Fig.\,\ref{fig:olay1569} for details. The contours of the
logarithmic X-ray emission in panels (b) and (c) are spaced by
$0.1\log(peak flux)$ and start at $0.2\log(peak flux)$ level
(continuum subtracted \ha\ image taken from \citealt{mar98}; the
optical $R$--band 1.1m ESO Schmidt image is from \citealt{lau89}).}
\label{fig:olayhe2-10}
\end{figure*}

\subsubsection{Previous X-ray observations and results}

\citet{ste98} analysed 9.2\,ks of {\it ROSAT} PSPC observations of
He\,2-10. They detected a point-like source and fitted a single
temperature RS plasma with a temperature of $\sim 5\times 10^{6}$\,K
and a luminosity of $10^{40}$\,erg~s$^{-1}$ to its spectrum.
\citet{men99} published a multi-wavelength analysis of He\,2-10
including the {\it ROSAT} PSPC and new HRI (38\,ks) observations. Based on a
correlation analysis with H$\alpha$ maps they conclude that hot gas
fills a bipolar superbubble. Furthermore, they speculate that there is
an outflow of hot gas and that part of it can escape the galaxy's
gravitational potential.

\subsubsection{{\it Chandra} observations}

\begin{figure*}
  \centering \includegraphics[width=14cm]{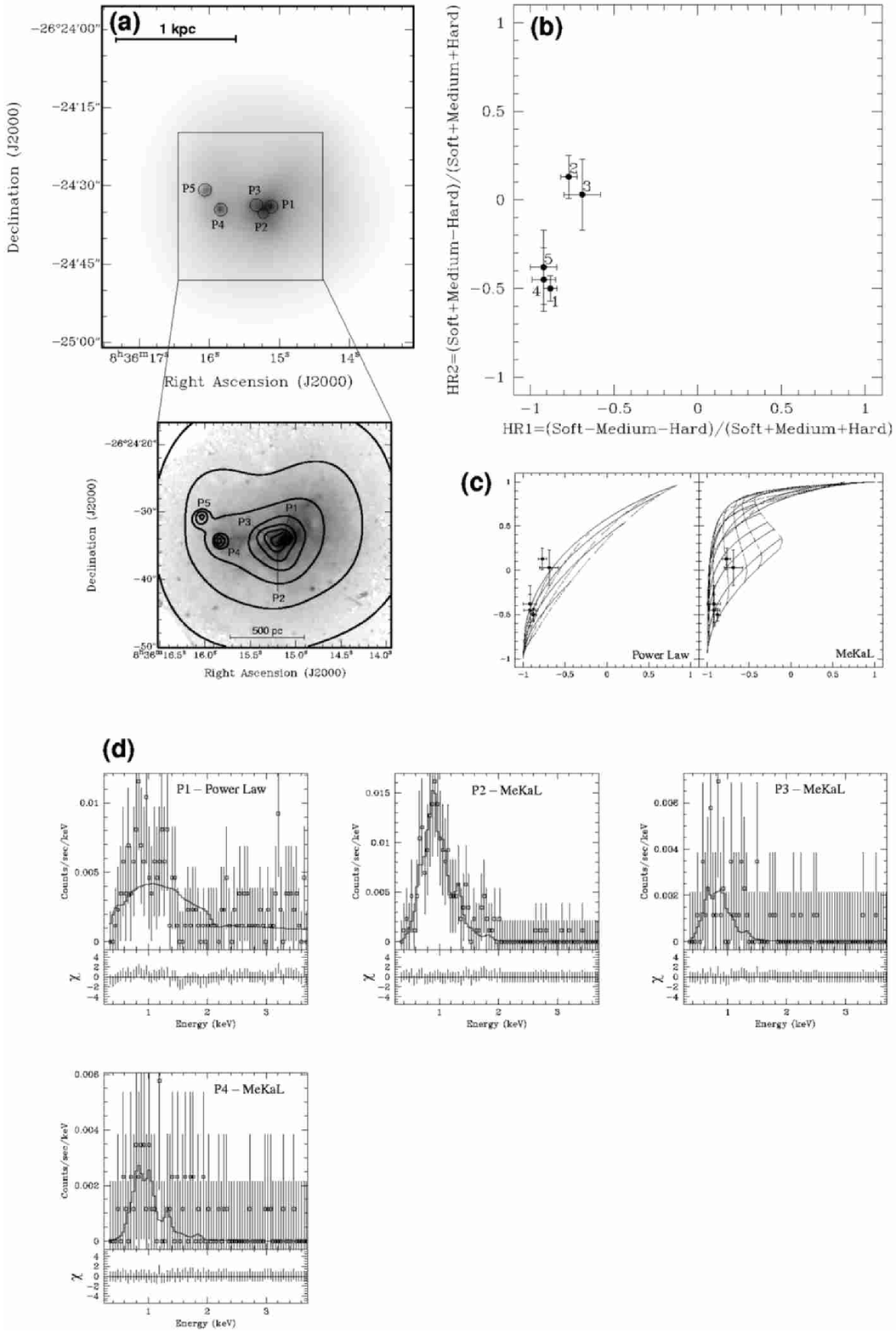}
\caption
{Locations and spectra of the point sources in He\,2-10. See caption
  of Fig.\,\ref{fig:point1569} for details. The count rates,
  hardness ratios and fit parameters are listed in
  Tables\,\ref{tab:pointhe2-10a} and \ref{tab:pointhe2-10b}. }
\label{fig:pointhe2-10}
\end{figure*}

The total X-ray emission of He\,2-10 is approximately circularly
shaped and has a diameter of $1\farcm7$ (4.4\,kpc). Within the diffuse
X-ray emission and the optical and \ha\ counterpart we detect five
point sources: three of them are at the same position as the central
star forming complex (P1, P2, P3) and one is located close to the
eastern complex (P4; see Fig\,\ref{fig:pointhe2-10}[a]). The fifth
source is placed further away to the east and is slightly shifted to
the north (P5). P1 is the strongest of the sources, followed by P2
(see Table\,\ref{tab:pointhe2-10a} for positions, count rates, and
hardness ratios of the point source population). P3 coincides with the
brightest radio continuum source detected by \citet{kob99b} (their
knot 4) and shows a radio spectrum indicative of an optically thick
thermal bremsstrahlung origin. P1 coincides with their knot 3 and
exhibits a flat radio continuum spectrum. A hardness ratio plot of all
X-ray point sources is shown in Figs.\,\ref{fig:pointhe2-10}(b) and
(c) and individual spectra are displayed in
Fig.\,\ref{fig:pointhe2-10}(d). The fitted parameters for point
sources with sufficient counts are listed in
Table\,\ref{tab:pointhe2-10b}. P2, P3, and P4 are best fit by a
collisional thermal plasma, whereas P1 is in better agreement with a
power law function. All fits to P1, however, are unsatisfactory and a
more sophisticated model may be needed to reach a more acceptable
result. Given the large distance to He\,2-10, only the most luminous
sources are detected. They exceed luminosities of $10^{39}$\,\lum\
which puts them in the range of ULXs.

The shape of the underlying diffuse X-ray emission basically follows
the \ha\ morphology of He\,2-10 (see Fig.\,\ref{fig:olayhe2-10}[b]).
The azimuthally averaged X-ray surface brightness profiles are similar
to those of NGC\,5253: the soft emission has a much larger exponential
scale length ($\sim 320$\,pc, Table\,\ref{tab:xprofile_fit}) than the
hard emission ($\sim 150$\,pc).  In addition, at large radii the
profile levels off (see Fig.\,\ref{fig:olayhe2-10}[f]). The central
temperature of the hot gas ($T\sim 7.6\times 10^{6}$\,K) is higher
than that of the outer regions ($T\sim 3.5\times 10^{6}$\,K, see
Fig.\,\ref{fig:olayhe2-10}[d] and [e] for the definitions of the
regions and Table\,\ref{tab:detect_gashe2-10} for the fit results).
Based on the fitted absorbing column densities, the orientation of the
disc of He\,2-10 is such that the north and east of the disc is
directed away from us (the hot gas being in front of the disc) while
the south and the west is more absorbed and the disc is therefore
pointing toward the observer. The total X-ray luminosity of the
diffuse emission is equal to $2\times 10^{40}$\,\lum. Adding up the
diffuse plus the point source X-ray luminosities yields
$8.1\times10^{40}$\,\lum. The thermal plasma results agree within a
factor of $\sim 2$ with the {\it ROSAT} results.

\begin{table*}
\begin{minipage}{170mm}
\centering
\caption
{Positions, count rates and hardness ratios of the point sources in
He\,2-10. The Right Ascension and Declination are given in $^{h}$
$^{m}$ $^{s}$ and $\degr$ $\arcmin$ $\arcsec$ (J2000), respectively.
The count rates of the total, soft, medium, and hard emission are in
units of $10^{-5}$\,cts\,s$^{-1}$. See Fig.\,\ref{fig:pointhe2-10} for
the locations of the point sources coinciding with this galaxy as well
as for their distribution in a hardness ratio plot.}

\begin{tabular}{@{}lllrrrrrr@{}}
\hline

No.&\multicolumn{1}{c}{RA}&\multicolumn{1}{c}{DEC}&\multicolumn{1}{c}{Total}&\multicolumn{1}{c}{Soft}&\multicolumn{1}{c}{Medium}&\multicolumn{1}{c}{Hard}&\multicolumn{1}{c}{HR1}&\multicolumn{1}{c}{HR2}\\

\hline

P1  & 08~36~15.105 & $-$26~24~33.56 & $1209\pm 84$    & $74.47\pm 22.41$ & $225.6\pm 42.1$ & $908.8\pm 69.2$ & $-0.88\pm 0.04$ & $-0.50\pm 0.07$\\
P2  & 08~36~15.252 & $-$26~24~34.67 & $853.5\pm 71.6$ & $97.54\pm 25.04$ & $384.7\pm 49.4$ & $371.3\pm 45.4$ & $-0.77\pm 0.05$ & $+0.13\pm 0.12$\\
P3  & 08~36~15.380 & $-$26~24~33.56 & $260.5\pm 39.6$ & $39.81\pm 16.23$ & $94.14\pm 25.73$& $126.5\pm 25.3$ & $-0.69\pm 0.11$ & $+0.03\pm 0.20$\\
P4  & 08~36~15.838 & $-$26~24~34.30 & $266.7\pm 40.7$ & $10.63\pm 9.88$  & $63.26\pm 23.22$& $192.8\pm 31.9$ & $-0.92\pm 0.07$ & $-0.45\pm 0.18$\\
P5 & 08~36~16.048 & $-$26~24~30.98 & $131.6\pm 25.8$ & $5.06\pm 5.06$   & $35.43\pm 13.39$& $91.10\pm 21.47$& $-0.92\pm 0.08$ & $-0.38\pm 0.21$\\
\hline

\end{tabular}
\label{tab:pointhe2-10a}
\end{minipage}
\end{table*}


\begin{table*}
\begin{minipage}{170mm}
\centering
\caption{Parameters of the fitted X-ray emission models to the point
  sources in He\,2-10. See the caption of Table\,\ref{tab:point1569b}
  for details and Fig.\,\ref{fig:pointhe2-10}(d) for the individual
  spectra and fits. }
\begin{tabular}{@{}lllrrrrrr@{}}
\hline

No.&Model &\multicolumn{1}{c}{$N_{H}$} &\multicolumn{1}{c}{$T/\gamma$} &\multicolumn{1}{c}{$Norm/Ampl^{a}$} & \multicolumn{1}{c}{$F_{\rm X}^{\rm abs}$ }&\multicolumn{1}{c}{$F_{\rm X}$} &\multicolumn{1}{c}{$L_{\rm X}$}\\
&&\multicolumn{1}{c}{[$10^{21}$\,\cden]} &\multicolumn{1}{c}{[$10^{6}$\,K/--]} &\multicolumn{1}{c}{[$10^{-5}$]} & \multicolumn{2}{c}{[$10^{-15}$\,\flux]} &\multicolumn{1}{c}{[$10^{37}$\,\lum]}&\\
\hline
P1   & PL  & \multicolumn{1}{c}{0.00 }& \multicolumn{1}{c}{1.00 }& \multicolumn{1}{c}{0.90  }& \multicolumn{1}{c}{147.4 }& \multicolumn{1}{c}{155.4 }& \multicolumn{1}{c}{150.7}\\
P2   & MEK & \multicolumn{1}{c}{8.41 }& \multicolumn{1}{c}{2.06 }& \multicolumn{1}{c}{297.5 }& \multicolumn{1}{c}{23.09 }& \multicolumn{1}{c}{5779  }& \multicolumn{1}{c}{5606 }\\
P3   & MEK & \multicolumn{1}{c}{5.56 }& \multicolumn{1}{c}{2.11 }& \multicolumn{1}{c}{14.18 }& \multicolumn{1}{c}{3.47  }& \multicolumn{1}{c}{279.1 }& \multicolumn{1}{c}{270.7}\\
P4   & MEK & \multicolumn{1}{c}{6.78 }& \multicolumn{1}{c}{3.46 }& \multicolumn{1}{c}{4.63  }& \multicolumn{1}{c}{4.01  }& \multicolumn{1}{c}{110.5 }& \multicolumn{1}{c}{107.2}\\

\hline
\end{tabular}
\footnotetext{$^{a}$ see Sect.\,\ref{sec:specana}} 
\label{tab:pointhe2-10b}
\end{minipage}
\end{table*}


\subsection{Metallicities and $\alpha/Fe$ Abundances of the hot, coronal gas}
\label{sec:metals}

\begin{figure}
\centering
\includegraphics[width=8cm]{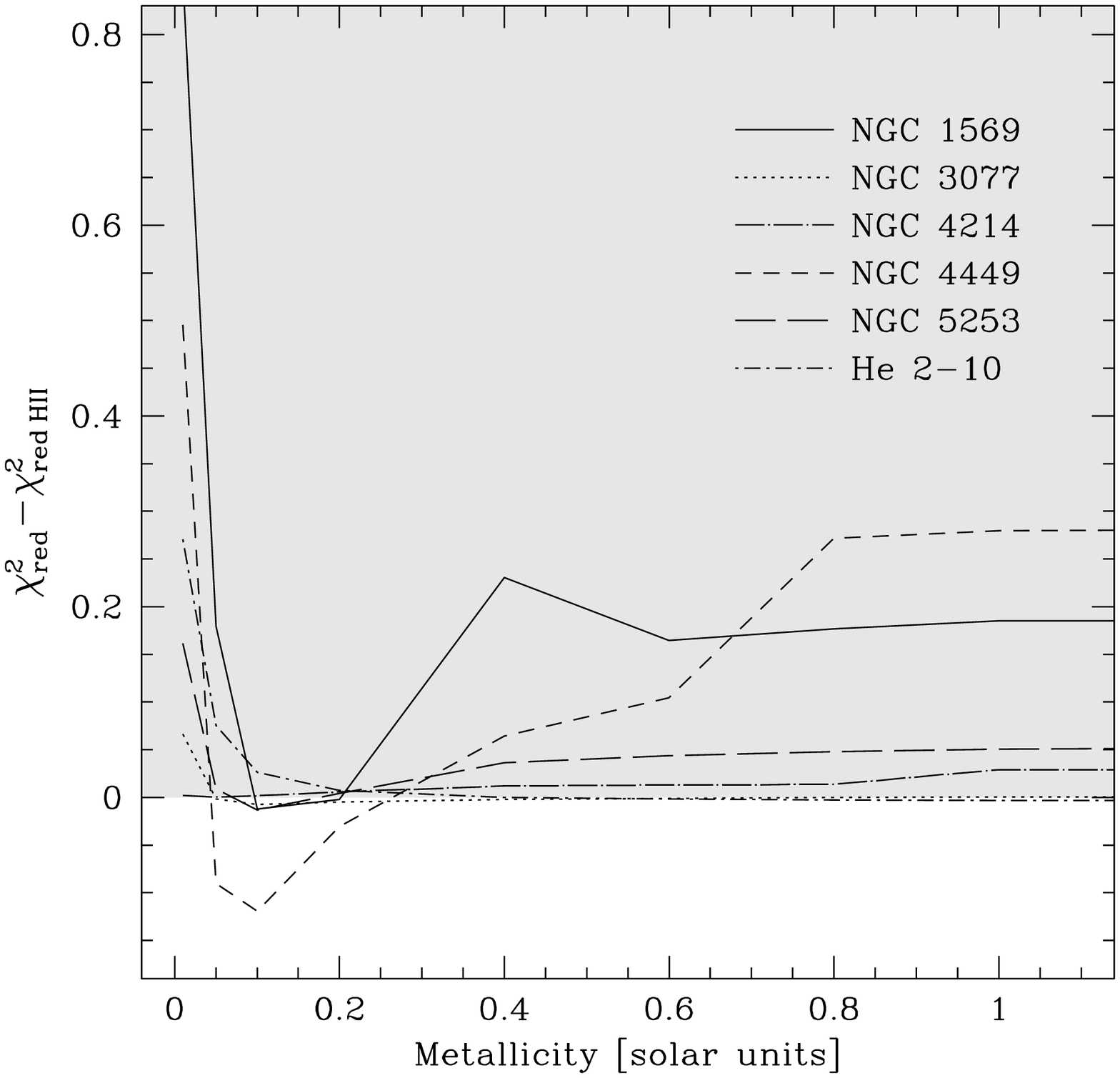}
\caption
{The goodness--of--fit indicator $\chi^{2}_{\rm red}$ of MeKaL fits
with varying metallicity relative to fits with the galaxies'
metallicities ($\chi^{2}_{\rm red HII}$; metallicities derived from
oxygen abundances of \hii\ regions within the objects; see
Table\,\ref{tab:sample_new}). Whenever the graph is within the grey
area the fits are worse than those using the galaxies' \hii\ region
metallicities, i.e., $\chi^{2}_{\rm red}>\chi^{2}_{\rm red HII}$; in
the white area the fits are improved ($\chi^{2}_{\rm
red}<\chi^{2}_{\rm red HII}$).}
\label{fig:metals}
\end{figure}

In order to investigate how sensitive the model fits are to a change
in metallicity \citep[but preserving a solar element mixture as given
by][]{and89} we ran MeKaL fits to the emission of the targets as a
whole and varied the metallicity from 0.01 to 10 times solar. The
absorbing column densities, temperatures, and normalisations of the
hot thermal plasmas were free parameters for these fits. The
goodness-of-fit measure $\chi^{2}_{red}$ as a function of metallicity
relative to the metallicities of the galaxies are plotted in
Fig.\,\ref{fig:metals} (galaxy metallicities are derived from oxygen
abundances measured for \hii\ regions within the objects
[Table\,\ref{tab:sample_new}]). Except for NGC\,4449 all fits
deteriorate for any metallicity other than that of their \hii\
regions. High--metallicity fits may still be acceptable for all
galaxies but NGC\,1569 and NGC\,4449. Metallicities below $\sim 5$\%
solar as, e.g., found in the massive starburst galaxy NGC\,253
\citep{str02} can be excluded for all objects but NGC\,4214 and
NGC\,3077. Any real improvement over the fits with the \hii\ region
metallicity are only obtained for NGC\,4449 which is fitted best
assuming a metallicity of $\sim 10$\% solar.

Theoretical models predict that the element {\it mixture} of galactic
winds differs from that in the solar neighbourhood. In general, the
metal enrichment of the interstellar/intergalactic medium is mainly
provided by SNe. The main contributors of $\alpha$ elements are type
II SNe whereas Fe elements are overwhelmingly synthesised in type Ia
SNe. The evolution of the stars involved in a SN Ia causes a
substantial time delay \citep[see, e.g., ][]{mat01} between type II
and type Ia SNe. For this reason, the hot gas observed in a starburst
should exhibit an $\alpha$ to Fe element abundance ratio ($\alpha/Fe$)
larger than unity in solar units
\citep[see, e.g., theoretical calculations by ][]{sil01b}. The
determination of this ratio is difficult as the results of the
spectral fitting process have quite large error bars. In spite of this
challenge, we performed MeKaL fits to determine $\alpha/Fe$ in four
ways: (a) all parameters ($\alpha$ and Fe element abundances, $N_{\rm
H}$, $T$, and $Norm$) were free, (b) the $\alpha$ element abundances
were fixed to the oxygen abundances measured on the
\hii\ regions of the galaxies, (c) freezing $N_{\rm H}$, $T$, and
$Norm$ to the values obtained for an $\alpha/Fe$ of unity but leaving
the $\alpha$ and Fe element abundances free, and (d) only treating the
Fe element abundances as a free parameter. The results as well as the
corresponding $\Delta\chi^{2}_{\rm red}=\chi^{2}_{\rm
red}-\chi^{2}_{\rm red HII}$ (reduced $\chi^{2}$ relative to the fixed
element mixture based on \hii\ region metallicity) are shown in
Table\,\ref{tab:metals}. The improved parameters resulting from
methods (a) and (b) are listed in Table\,\ref{tab:metalsimprove}. In
Table\,\ref{tab:metals} the results of F--tests are also provided
which were computed as models for (a) to (d) being nested (i.e., they
have the same but fewer free parameters) in the model with an
$\alpha/Fe=1$ and overall metallicity fixed to the galaxies' \hii\
region oxygen abundances. Using the 95\% criterion indicates that all
fits with `F--test'$<0.05$ are better than those with
$\alpha/Fe=1$. Method (d) led to $\alpha/Fe$ of about unity which is
understandable given that the fixed parameters were obtained with
exactly this assumption. With very few exceptions, $\alpha/Fe$ is
determined to be in the range 1.3--2.7 in agreement with $\alpha$
element enrichment. The exceptions are NGC\,3077 and NGC\,4214. As
discussed in \citet{ott03}, the hot gas in NGC\,3077 is still stored
in hot superbubbles whereas all the other galaxies in the sample show
outflow features (see Sect.\,7 in Paper\,II). This might be reflected
in its $\alpha/Fe$ value of 1 but would contradict the reduced
mass-loading of surrounding material (with a solar $\alpha/Fe$) which
is derived for NGC\,3077 in Sect.\,5.7 of Paper\,II. The number of
detected X-ray photons in NGC\,4214 is relatively low (only 10\,ks of
integration time) and methods (c) and (d) did not converge. Methods
(a) and (b) led to results with very large uncertainties for
NGC\,4214, the best results of which are actually $\alpha/Fe<1$. All
fits were conducted using MeKaL thermal plasma models. Very similar
results were obtained when we applied RS and {\sc APEC} \citep{smi01}
models. Again, NGC\,4214 is very poorly fitted, the $\alpha/Fe$ ratio
of the hot gas stored in NGC\,3077 results in about unity and the
ratios for all the other galaxies are $\sim 2$. In summary, regarding
the $\Delta\chi^{2}_{red}$ and the F--test results, an $\alpha$
element enhancement is significant for NGC\,1569, NGC\,4449, and
He\,2--10, but are not statistically supported for NGC\,3077,
NGC\,4214, and NGC\,5253. The fits result in $\alpha/Fe\sim2$ for
NGC\,1569, NGC\,4449, NGC\,5253, and He\,2--10, and $\alpha/Fe\sim1$
for NGC\,3077. The data quality of NGC\,4214 does not allow us to
set any constraints.


\begin{table*}
\renewcommand{\arraystretch}{1.2}
\caption{Fitted parameters for the determination of $\alpha/Fe$ with different parameters being fixed and left free.}
\begin{tabular}{@{}lllllll@{}}

\hline
Method           & \multicolumn{3}{c}{(a)} & \multicolumn{3}{c}{(b)} \\
\hline
Fixed parameters  &\multicolumn{3}{c}{---}&\multicolumn{3}{c}{$\alpha$}\\

Free parameters &\multicolumn{3}{c}{$\alpha, Fe, N_{\rm H}, T, Norm^{a}$}&\multicolumn{3}{c}{$Fe, N_{\rm H}, T, Norm^{a}$}\\
\hline
Galaxy &  \multicolumn{1}{c}{$\alpha/Fe$}&$\Delta\chi^{2}_{\rm red}$&F-test&\multicolumn{1}{c}{$\alpha/Fe$}&$\Delta\chi^{2}_{\rm red}$&F--test\\
\hline
NGC\,1569 &  \phn0.51/0.19 = 2.71 & $-0.20$ & $2\times10^{-9}$& 0.20/0.11 = 1.82 & $-0.15$ &$9\times10^{-8}$ \\
NGC\,3077 &  \phn0.25/0.23 = 1.09 & $\pm0.00$ & 0.67 & 1.00/0.98 = 1.02 & $\pm0.00$ & 0.96 \\
NGC\,4214 &  \phn0.03/0.05 = 0.60 & $\pm0.00$ & 0.01 & 0.25/3.10 = 0.08 & $\pm0.00$ &0.03 \\
NGC\,4449 &  \phn0.19/0.08 = 2.38 & $-0.20$ & $9\times10^{-9}$& 0.25/0.11 = 2.27 & $-0.81$ & $1\times10^{-8}$\\
NGC\,5253 &  \phn0.20/0.09 = 2.22 & $\pm0.00$ & 0.24 & 0.20/0.11 = 1.82 & $\pm0.00$ & 0.08 \\
He\,2-10 & 11.03/4.52 = 2.44 & $-0.05$ & $1\times10^{-7}$& 1.00/0.58 = 1.72 & $-0.03$ & $2\times10^{-5}$\\
\hline

Method & \multicolumn{3}{c}{(c)} & \multicolumn{3}{c}{(d)} \\
\hline
Fixed parameters  &\multicolumn{3}{c}{$N_{\rm H}, T, Norm^{a}$}&\multicolumn{3}{c}{$\alpha$, $N_{\rm H}, T, Norm^{a}$}\\ 

Free parameters &\multicolumn{3}{c}{$\alpha, Fe$}&\multicolumn{3}{c}{$Fe$}\\ 
\hline
Galaxy &  $\alpha/Fe$&$\Delta\chi^{2}_{\rm red}$&F--test&$\alpha/Fe$&$\Delta\chi^{2}_{\rm red}$&F--test\\
\hline
NGC\,1569 &   0.39/0.18 = 2.17 & $-0.15$  & $2\times10^{-7}$ & 0.20/0.21 = 0.95 & $+0.15$ & $7\times10^{-6}$\\ 
NGC\,3077 &   1.00/1.00 = 1.00 & $\pm0.00$  & 0.99 & 1.00/1.00 = 1.00 & $\pm0.00$ & 0.43\\
NGC\,4214 &   \nodata          & \nodata & \nodata & \nodata          & \nodata & \nodata\\
NGC\,4449 &   0.27/0.23 = 1.17 & $+0.03$  & 0.02  & 0.25/0.25 = 1.00 & $+0.06$ & 0.01\\
NGC\,5253 &   0.23/0.19 = 1.21 & $+0.01$  &0.14  & 0.20/0.21 = 0.95 & $+0.02$ & 0.07\\
He\,2-10 &  1.12/0.88 = 1.28 & $-0.01$    & 0.06& 1.00/0.97 = 1.03 & $-0.01$ & 0.31\\

\hline
\end{tabular}
\footnotetext{$^{a}$ see Sect.\,\ref{sec:specana}} 
\label{tab:metals}
\end{table*}

\begin{table*}
\renewcommand{\arraystretch}{1.1}
\caption{Improved fitted parameters for models (a) and (b) listed in Table\,\ref{tab:metals}}
\begin{tabular}{@{}lcccccc@{}}
\hline
Method           & \multicolumn{3}{c}{(a)} & \multicolumn{3}{c}{(b)} \\
\hline
& $N_{\rm H}$  & $T$ &  $Norm^{a}$ &  $N_{\rm H}$  & $T$ &  $Norm^{a}$ \\
& [$10^{21}$\,cm$^{-2}$] & [$10^{6}$\,K] & [$10^{-5}$]& [$10^{21}$\,cm$^{-2}$] & [$10^{6}$\,K] & [$10^{-5}$]\\
\hline
NGC\,1569&\phn1.67 & 7.07 &  63.95 &  \phn2.14 & 7.04 & 112.9\\
NGC\,3077&\phn4.23 & 2.52 & 113.3 & \phn4.70 & 2.33 & 50.07\\
NGC\,4214&11.15 & 1.92 & 1269 &15.12 & 1.62 & 1188\\
NGC\,4449&\phn4.26 & 3.74 & 246.4&\phn3.48 & 3.94 & 184.8\\
NGC\,5253&\phn3.15 & 4.28 & 40.98&\phn3.04 & 4.31 & 41.71\\
He\,2-10&\phn0.00 & 7.06 & 1.04&\phn0.15 & 7.03 & 8.63\\
\hline
\end{tabular}
\footnotetext{$^{a}$ see Sect.\,\ref{sec:specana}} 
\label{tab:metalsimprove}
\end{table*}

\section{Summary}
\label{sec:summary}

We present a sample of eight dwarf starburst galaxies (I\,Zw\,18,
VII\,Zw\,403, NGC\,1569, NGC\,3077, NGC\,4214, NGC\,4449, NGC\,5253,
and He\,2-10) observed with the {\it Chandra} X-ray observatory. The
unique combination of high angular resolution and large collecting
area of the {\it Chandra} X-ray observatory has proven indispensable
for a proper analysis of the unresolved X-ray point sources within
each field, as well as for their removal from the diffuse X-ray
component. We have performed an in-depth analysis of all Chandra
datasets, using the exact same analysis/methods for each dwarf galaxy and
find the following:

\begin{itemize}

\item All galaxies in the sample had previous detections with the {\it
      ROSAT} or {\it EINSTEIN} satellites. The {\it Chandra}
      observations show that in addition to unresolved sources,
      diffuse X-ray emission due to hot (coronal) gas is present in
      six galaxies (NGC\,1569, NGC\,3077, NGC\,4214, NGC\,4449,
      NGC\,5253, and He\,2-10) with sizes of $1-10$\,kpc. In the case
      of I\,Zw\,18 and VII\,Zw\,403 only an unresolved X-ray source is
      detected and an upper limit to the diffuse component is
      determined.

\item In total 55 X-ray point sources are detected in these galaxies.
  Optical counterparts were only identified for a minority of
  sources. In general, sources with sufficient counts (35 sources) are
  well described by power law (14), thermal plasma (18), and black
  body (3) spectra. Ten of the sources in our sample exceed X-ray
  luminosities of $10^{39}$\,\lum\ and hence are ultraluminous X-ray
  sources (ULX).
  
\item Power law indices of the corresponding X-ray point sources
  sources range from 1 to 4 and those sources have X-ray luminosities
  of up to $\sim5\times10^{39}$\,\lum. Temperatures for X-ray point
  sources for which the best fit was provided by a thermal plasma
  model are in the range of $\sim6-40\times10^{6}$\,K with X-ray
  luminosities of up to $6\times 10^{40}$\,\lum. Sources which are fit
  by a black body spectrum have temperatures of $\sim 0.3-0.9\times
  10^{6}$\,K with a maximum luminosity of $\sim6\times10^{40}$\,\lum.
  
\item  For those galaxies with extended emission, most of the X-ray photons
  (typically $60-80$ per cent) are emitted by the diffuse
  component. Photons from X-ray point sources only equal the diffuse
  emission in the case of NGC\,4214. After the removal of the point
  sources, a single temperature thermal plasma model with a
  metallicity close to that measured for \hii\ regions is in good
  agreement with the data. However, the temperatures vary for
  different regions defined within the objects. In particular, the
  temperatures in the centre of the galaxies ($\sim 2-7\times
  10^{6}$\,K) are in general larger than those of their outskirts ($\sim
  2-3\times 10^{6}$\,K). The integrated X-ray luminosities of the
  diffuse component are in the range of $\sim 4\times
  10^{38}-\,2\times 10^{40}$\,\lum\ and therefore span about two
  orders of magnitude. Absorbing column densities are in the range of
  $2-10\times 10^{21}$\,\cden.

\item The metallicities of the hot gas cannot be reliably constrained
  but are compatible with the metallicities measured for \hii\
  regions. The $\alpha/Fe$ ratio is compatible with a value of $\sim
  2$ (except for NGC\,3077: $\alpha/Fe\sim1$) which agrees with a
  scenario in which type II SNe are responsible for the ejected
  material stored in the superwinds.

\item Azimuthally averaged X-ray surface brightness profiles are well
  described by exponential laws with scale lengths of $\sim
  100-600$\,pc. At large radii the profiles may level off. For most
  galaxies the scale length of the soft X-ray emission is larger than
  that of the hard emission.

\item Fitted absorbing column densities are used to derive the
  orientation of the discs of the galaxies assuming that the soft
  X-rays are absorbed by the cooler gas stored in the discs.

\end{itemize}

In Paper\,II we present a comparison of the X-ray data with
observations at other wavelengths, a discussion on the state of the
ISM, correlations with star formation tracers, and the development of
superwinds which can lead to outflows.

\section*{Acknowledgements}
We would like to thank Dominik Bomans, John Cannon, Deidre Hunter,
Chip Kobulnicky, and Vince McIntyre for providing optical and \ha\
images of the galaxies. In particular, we thank Crystal Martin for
providing some optical images as well as for valuable discussions on
NGC\,3077 and He\,2-10. We are also grateful to David Strickland for
the critical reading of the manuscript. JO acknowledges generous
support from the Graduiertenkolleg 118 'The Magellanic System, Galaxy
Interaction, and the Evolution of Dwarf Galaxies' of the Deutsche
Forschungsgemeinschaft (DFG). EB is grateful to CONACyT for financial
support through grant Nr.\ 27606-E. This research has made use of the
NASA/IPAC Extragalactic Database (NED) and the NASA/IPAC Infrared
Science Archive, which are maintained by the Jet Propulsion
Laboratory, Caltech, under contract with the National Aeronautics and
Space Administration (NASA), NASA's Astrophysical Data System Abstract
Service (ADS), NASA's SkyView, and the astronomical database SIMBAD,
provided by the 'Centre de Donn\'ees astronomiques de Strasbourg'
(CDS).

\bsp

\label{lastpage}

\end{document}